\documentclass[a4paper,11pt]{article}
\usepackage{jheppub}
\usepackage[utf8]{inputenc}
\usepackage{bbold}
\usepackage{bbm}
\usepackage{bm}
\usepackage{caption}
\usepackage{subcaption}
\usepackage{hyperref}
\usepackage{mathtools,slashed}
\usepackage{mathtools}
\usepackage{xcolor}
\usepackage{appendix}
\usepackage{amsmath,amssymb}
\usepackage{graphicx}
\usepackage[T1]{fontenc}

\newcommand{\sign}{\text{sign}}

\DeclarePairedDelimiterX\braket[2]{\langle}{\rangle}{#1 \delimsize\vert #2}
\DeclarePairedDelimiterX\brakets[3]{\langle}{\rangle}{#1 \delimsize\vert #2 \delimsize\vert #3}

\graphicspath{{./images/}}

\title{Effective field theories for dark matter pairs in the early universe: Debye mass effects}

\author[a]{S.~Biondini,}
\author[b,c,d]{N.~Brambilla,}
\author[e]{A.~Dashko,}
\author[b]{G.~Qerimi}
\author[b]{and A. Vairo}

\affiliation[a]{School of Science and Technology, University of Camerino,\\
Via Madonna delle Carceri 9, 62032 Camerino, Italy}
\affiliation[b]{Technical University of Munich, TUM School of Natural Sciences, Physics Department,\\
James-Franck-Str. 1, 85748 Garching, Germany}
\affiliation[c]{Institute for Advanced Study, Technical University of Munich,\\
Lichtenbergstrasse 2a, 85748 Garching, Germany}
\affiliation[d]{Munich Data Science Institute, Technische Universit\"{a}t M\"{u}nchen,\\
Walther-von-Dyck-Strasse 10, 85748 Garching, Germany}
\affiliation[e]{Deutsches Elektronen-Synchrotron DESY,\\
Notkestr.~85, 22607 Hamburg, Germany}
\emailAdd{simone.biondini@unicam.it}
\emailAdd{nora.brambilla@tum.de}
\emailAdd{andrii.dashko@desy.de}
\emailAdd{gramos.qerimi@tum.de}
\emailAdd{antonio.vairo@tum.de}

\preprint{DESY-24-154, TUM-EFT 179/23}

\abstract{In some scenarios for the early universe, non-relativistic thermal dark matter chemically decouples from the thermal environment once the temperature drops well below the dark matter mass. The value at which the energy density freezes out depends on the underlying model. In a simple setting, we provide a comprehensive study of heavy fermionic dark matter interacting with the light degrees of freedom of a dark thermal sector whose temperature $T$ decreases from an initial value close to the freeze-out temperature. Different temperatures imply different hierarchies of energy scales. By exploiting the methods of non-relativistic effective field theories at finite $T$, we systematically determine the thermal and in-vacuum interaction rates. In particular, we address the impact of the Debye mass on the observables and ultimately on the dark matter relic abundance. We numerically compare the corrections to the present energy density originating from the resummation of Debye mass effects with the corrections coming from a next-to-leading order treatment of the bath-particle interactions. We observe that the fixed-order calculation of the inelastic heavy-light scattering at high temperatures provides a larger dark matter depletion, and hence an undersized yield for given benchmark points in the parameter space, with respect to the calculation where Debye mass effects are resummed.
}

\begin{document}

\maketitle
\flushbottom

\section{Introduction}
One of the main open challenges across particle physics and cosmology is to reveal the nature of dark matter (DM).
The evidence for such non-luminous and non-baryonic components of the universe is compelling and longstanding \cite{Zwicky:1933gu,Rubin:1980zd,Clowe:2006eq},
see e.g.~\cite{Feng:2010gw,Bertone:2016nfn} for reviews.
The existence of some form of dark matter is rather convincing  because it manifests (gravitationally) at many different scales, ranging from the size of a galaxy to the entire universe.
In the standard paradigm of a cosmological model with cold dark matter and a cosmological constant,
the amount of dark matter in the universe can be inferred from the power spectrum of  the cosmic microwave background (CMB).
The Planck collaboration provides us with an accurate measurement of the dark matter energy density $\Omega_{\textrm{DM}}h^2 = 0.1200 \pm 0.012$~\cite{Planck:2018nkj},
where $h$ is the reduced Hubble constant.  

Particle dark matter is an appealing option and is a strong call for theoretical and experimental particle physicists in the search of physics beyond the Standard Model.
Indeed,  the known constituents of ordinary matter cannot account for the elusive dark matter particle.
The diversified effort in nailing down a particle dark matter involves complementary experimental searches, model building activities and robust predictions for its cosmological abundance.
The latter aspect boils down to connecting the masses and couplings of a given dark sector with the corresponding dark matter energy density.
This is achieved by assuming a production mechanism for dark matter that had to be at play in the early universe.
Thermal \emph{freeze-out} has been the leading paradigm for many well-motivated dark matter candidates, most notably for models that predict a Weakly Interacting Massive Particle (WIMP).

Without dwelling upon the quite diverse realizations of a WIMP dark matter (see e.g.~\cite{Bertone:2016nfn,Roszkowski:2017nbc,Bertone:2010at}),
for the scope of our work, it suffices to recall the main features of dark matter freeze-out. 
The main assumption is that dark massive particles interact efficiently with the other constituents of the thermal environment and are kept in chemical equilibrium for $T \gg M$, where $T$ is the temperature of the bath and $M$ is the mass of the dark matter.
Chemical equilibrium is maintained by number-changing processes, most notably dark-matter pair annihilation into Standard Model particles or lighter degrees of freedom of a given dark sector.
With the cooling of the expanding universe, dark matter particles get progressively diluted
and their abundance gets (almost) frozen when the pair-annihilation process becomes slower than the Hubble rate.
This happens for temperatures of about $T \approx M/25$,
which qualifies the typical kinematic regime for the freeze-out of massive particles as \emph{non-relativistic}. 

The central quantity that enters the evolution equations of the dark matter number density and ultimately enables a prediction for the dark matter cosmological abundance
is the thermally averaged annihilation cross section \cite{Gondolo:1990dk,Griest:1990kh}.
In many concrete situations that comprise simplified dark matter models as well as ultraviolet complete theories, the annihilating dark matter particles interact with lighter states.
Whenever this happens at a small relative velocity, one needs to include the effect of repeated soft-momentum/lighter-state exchange between the heavy particles.
The reason is that, on the one hand, the Sommerfeld factor modifies the annihilation of dark matter pairs in scattering states (or above-threshold states) \cite{Sommerfeld:1931qaf,Hisano:2004ds},
and, on the other hand, bound-state formation triggers an additional channel to deplete dark matter pairs in the form of bound states (or below-threshold states) \cite{Feng:2009mn,vonHarling:2014kha}. 

Recently, there has been quite an effort in the scrutiny of the (thermal) dynamics of bound-state formation, dissociation and bound-to-bound transitions, and the corresponding impact on the predicted dark matter energy density
\cite{vonHarling:2014kha,Beneke:2014hja,Petraki:2016cnz,Kim:2016kxt,Beneke:2016ync,Mitridate:2017izz,Biondini:2017ufr,Harz:2018csl,Biondini:2018pwp,Biondini:2018xor,Biondini:2018ovz,Binder:2018znk,Oncala:2018bvl,Oncala:2019yvj,Harz:2019rro,Binder:2019erp,Biondini:2019zdo,Biondini:2019int,Binder:2020efn,Garny:2021qsr,Biondini:2021ycj,Binder:2021vfo,Biondini:2021ccr,Binder:2023ckj,Biondini:2023ksj,Biondini:2023zcz,Biondini:2024aan}.
Many different and complementary aspects have been considered and addressed,
that range from the identification of the processes responsible for bound-state formation to the inclusion of excited bound states in a network of Boltzmann equations.
In this work, we focus on and assess the effect of the {\it Debye mass} of the force mediator on the bound-state formation cross section, and bound-state dissociation and transition widths.

The Debye mass of the force mediator, $m_\text{D}$,  can be understood as the inverse of the chromoelectric screening length and, at weak coupling, it is $m_\text{D} \sim gT$,
where $g$ is the coupling between the mediator and the other light degrees of freedom of the thermal bath.
Together with the temperature, it enriches the set of thermodynamical scales in the system.
Although the appearance of a Debye mass is model-dependent, it is expected to affect several setups:
{\it (i)} the force mediator between dark matter particles is a Standard Model particle,
e.g.~the Higgs boson \cite{Lahanas:1999uy,Baer:2003bp,Harz:2019rro} or electroweak gauge bosons \cite{Cirelli:2005uq,Beneke:2012tg};
{\it (ii)} the dark matter particles are charged under a non-abelian dark gauge group \cite{Cirelli:2005uq,Carone:2018eka};
{\it (iii)} coannihilating partners of the actual dark matter particles are charged under some, or all, the Standard Model gauge groups \cite{Edsjo:1997bg,Ellis:2001nx,Berggren:2015qua,Garny:2015wea,Arina:2020udz}.
In the latter case, the coannihilating states also freeze out in the non-relativistic regime and contribute to the actual dark matter abundance. 

The Debye mass is often associated with the interaction between the force mediator and other light degrees of freedom.\footnote{
Dark matter fermions or scalars with non-abelian force mediators are an exception.
Indeed, gluon self-interactions produce a Debye mass without the need of further light degrees of freedom.} For example, in the case of a dark photon as a force mediator, additional light fermionic or scalar degrees of freedom are responsible
for quantum corrections to the dark photon propagator at $T \ll M$, whose pole develops a real and an imaginary part.
The real part introduces a screening Debye mass of order $gT$ for the temporal dark photon, whereas the imaginary part of the pole originates
from $2 \to 2$ scatterings with plasma constituents, also referred to as {\it Landau damping} \cite{Braaten:1991gm,Laine:2006ns,Laine:2007qy,Beraudo:2007ky,Brambilla:2008cx,Escobedo:2008sy,Escobedo:2010tu}.
Hence, in a thermal environment, there is an additional process for bound-state formation and dissociation that is different from {\it photo-dissociation}, which is the bound-state formation via radiative emission~\cite{vonHarling:2014kha}.

In this paper, we adopt the framework of {\it non-relativistic effective field theories} (NR\-EFTs)
to deal with the various energy scales that characterize heavy pairs in a thermal environment.
In-vacuum scales, which are generated by the relative dynamics of the heavy pairs,
are hierarchically ordered  as $M \gg M v_{\hbox{\scriptsize rel} } \gg Mv_{\hbox{\scriptsize rel} }^2$,
where $v_{\hbox{\scriptsize rel} }$ is the relative velocity of the particles in the pair.
The corresponding hierarchy  of energy scales for bound states, whenever they are (nearly) Coulombic, is $M \gg M \alpha \gg M \alpha^2$,  
where $\alpha=g^2/(4\pi)$ and $g$ is the coupling between the heavy dark matter particle and the force mediator.
The thermodynamical scales are the temperature of the early universe, the scale of the kinetically equilibrated dark matter particles $\sqrt{MT}$ and the Debye mass.
Contributions coming from the different energy scales may be disentangled and computed in a systematic manner by means of non-relativistic effective field theories~\cite{Brambilla:2004jw,Brambilla:2008cx,Escobedo:2010tu}.
In this work, we follow up on the NREFTs for dark matter freeze-out presented in refs.~\cite{Biondini:2023zcz,Biondini:2024aan} by adding the Debye mass. 

Former applications of NREFTs in the context of dark matter freeze-out,
where the effect of the Debye-screened Yukawa potential and bound-state dissociation via Landau damping has been included for the temperature regime $T \gg M\alpha$,
can be found in refs.~\cite{Kim:2016kxt,Biondini:2017ufr,Biondini:2018pwp}.
For the complementary temperature regime, $T \ll M \alpha$,
the scrutiny of the bound-state formation processes for a dark QED model with additional light states has been put forward in \cite{Binder:2020efn},
however without accounting for a resummation of the Debye mass.
Our aim is to close this gap and explore the effects of the Debye mass for smaller temperatures, $T \ll M \alpha$.
In so doing, we shall improve upon the temperature window where bound-state effects are expected to be efficient.  

The structure of the paper is as follows.
In section \ref{sec:NREFTs_model}, we introduce an abelian DM model,
and in sections \ref{sec:NREFT_part} and \ref{sec:pNREFT_part},
its low-energy effective field theories upon integrating out modes carrying energies and momenta of order $M$, $M v_{\hbox{\scriptsize rel}}$, and $M\alpha$.
Then, we address the calculation of the heavy-pair self-energy in section~\ref{sec:el_dipole_transitions},
which is the main ingredient for the extraction of the bound-state formation cross section in section \ref{sec:bsf},
and for the bound-state dissociation and bound-state to bound-state transition widths in section \ref{sec:bsd}.
Section~\ref{sec:DMevolution} is devoted to a numerical study of the dark matter energy density 
from the obtained thermal rates.
Conclusions are in section \ref{sec:concl}, and supplementary material is collected in the appendices~\ref{sec:app_A}-\ref{sec:app_C}.     

\section{NREFTs for an abelian dark matter model} 
\label{sec:NREFTs_model}
Following the strategy of our former papers \cite{Biondini:2023zcz,Biondini:2024aan}, we consider a simple model for the dark sector.
More specifically, the dark matter is a dark Dirac fermion $X$ with mass $M$ and it is charged under an abelian gauge group U(1)$_\text{DM}$.
We denote the corresponding dark photon with $\gamma$~\cite{Feldman:2006wd,Fayet:2007ua,Goodsell:2009xc,Morrissey:2009ur,Andreas:2011in}.
The choice of such a dark sector is also customary in the literature in order to scrutinize near-threshold effects in a simple, yet meaningful, setting \cite{vonHarling:2014kha,Binder:2018znk,Binder:2020efn}. 

At variance with our former studies, we consider in this work the effects of additional light degrees of freedom that couple to
the dark photon. As anticipated in the introduction,  
quantum corrections to the dark photon propagator will appear, cf.~section~\ref{sec:el_dipole_transitions}.
In the following, we take $n_f$ dark Dirac fermion species $f_i$, $i=1,...,n_f$ as light degrees of freedom, which are also charged under the same gauge group U(1)$_\text{DM}$.
They are assumed to be relativistic, i.e. with $m_i \ll T$, at the freeze-out and later stages of the temperature evolution.
The light fermions, together with the dark photons, make up for the dark sector thermal bath in the form of radiation.  

The Lagrangian density reads
\begin{equation}
\mathcal{L}=\bar{X} (i \slashed {D} -M) X -\frac{1}{4} F_{\mu \nu} F^{\mu \nu} + \sum_{i=1}^{n_f}\bar{f}_i(i \slashed{D}-m_i) f_i + \mathcal{L}_{\textrm{portal}} \, ,
\label{lag_mod_0}
\end{equation}
where the covariant derivative is $D_\mu=\partial_\mu + i g A_\mu$, with $A_\mu$ the dark photon field and $F_{\mu \nu} = \partial_\mu A_\nu - \partial_\nu A_\mu$;
we define the fine structure constant as $\alpha \equiv g^2/(4 \pi)$, and assume a weakly-coupled system with $\alpha \ll 1$.
In practice, in our calculations, we put $m_i=0$ for all the light dark fermions.\footnote{
In ref.~\cite{Escobedo:2010tu}, the photon self-energy in QED for a finite electron mass is derived.
For the purpose of our work, it is not crucial to retain the light fermion masses in the calculation.}
The portal interactions connect the dark sector with the Standard Model.  
A rather common choice is  realized via a kinetic mixing with the neutral components of the SM gauge fields \cite{Holdom:1985ag, Foot:1991kb}.
Typically rather small mixing-induced couplings are needed to keep the two sectors in thermal equilibrium, see e.g.~\cite{Evans:2017kti}.
Therefore, we assume that the dark sector has the same temperature as the Standard Model
and, at the same time, neglect the effect of portal interactions when computing the relevant cross sections and decay widths as triggered by the dark gauge coupling.  

We are especially interested in capturing the dynamics of dark matter pairs during the thermal freeze-out and at later stages, where annihilations may still occur and affect the relic energy density.
The decoupling from chemical equilibrium happens at around $T/M \approx 1/25$, which puts the DM particles in a non-relativistic regime.
Kinetic equilibrium lasts instead for longer and the momenta of kinetically equilibrated DM heavy particles are of order $\sqrt{MT}$.
In this regime, the dark matter pairs are close to threshold, i.e. they move with non-relativistic velocities $v_{\hbox{\scriptsize rel}} \sim \sqrt{T/M} \ll 1$.
This establishes a hierarchy of energy scales, which are the {\it hard} momentum scale $M$, the scale $M v_{\hbox{\scriptsize rel}}$ and the scale $M v_{\hbox{\scriptsize rel}}^2$.
Near threshold, a heavy fermion and an antifermion can interact in the form of either a Coulombic scattering state or a bound state.  
For Coulombic bound states in the center-of-mass frame, the typical relative momentum is of the order of the inverse Bohr radius, $M\alpha$, we call this scale {\it soft}, whereas the
binding energy is of the order of $M\alpha^2$, we call this scale and any scale smaller than $M\alpha $ {\it ultrasoft}. 
Hence, Coulombic bound states realize a further hierarchy of energy scales. 

Energy scales due to the thermal environment, namely the temperature $T$ and the dynamically generated Debye mass $m_\text{D}$, add up to the in-vacuum scales.
In this work, we assume the following hierarchy of energy scales, which is realized to a good extent at the thermal freeze-out and after,
\begin{equation}
M \gg \sqrt{MT} \gg M\alpha  \gg T \,.
\label{scale_arrang}
\end{equation}
Moreover, we assume 
\begin{equation}
T \gg m_\text{D}\,,
\label{scale_arrang2}
\end{equation}
which qualifies the medium as a {\it weakly coupled plasma}.
The novelty of this work consists in not assuming any special hierarchy between the Debye mass and the bound-state energy $M\alpha^2$, although we examine limiting cases.
Debye mass effects need to be resummed under the condition that the temperature is much larger than the energy of the photon emitted or absorbed by the DM pair, 
which we examine in detail in section~\ref{sec:bsf_resummed}. 
For temperatures comparable to or smaller than the energy of the photon emitted or absorbed by the DM pair, 
thermal effects can be computed in fixed-order perturbation theory.
The next-to-leading order computation relevant for bound-state formation can be found in appendix~\ref{sec:app_C}.\footnote{
The case $M\alpha^2 \gtrsim T$ has also been extensively studied in~\cite{Biondini:2023zcz}.
Although interactions with light dark fermions were not considered there, for small $T$ their contributions to the rates are suppressed.}
Cross sections and widths are more conveniently computed by replacing the relativistic model \eqref{lag_mod_0} with suitable non-relativistic effective field theories obtained by integrating out the high energy scales of the system.
We pursue this approach in the following.

\subsection{\texorpdfstring{NRQED$_{\textrm{DM}}$}{NRQED DM}}
\label{sec:NREFT_part}
At energies much smaller than $M$, the effective degrees of freedom are non-relativistic dark fermions and antifermions, low energy dark photons and the light fermions $f_i$.  
The effective field theory that follows from \eqref{lag_mod_0} by integrating out dark photons and fermions of energy or momentum of order $M$ has the form of NRQED~\cite{Caswell:1985ui}.
It is organized as an expansion in $1/M$ and $\alpha$ and its Lagrangian density up to $\mathcal{O}(1/M^2)$ reads\footnote{
At order $1/M^2$, we do not display four-fermion operators made of light quarks as they do not contribute to the annihilation of heavy dark fermion-antifermion pairs.}
\begin{eqnarray}
  \mathcal{L}_{\textrm{NRQED}_{\textrm{DM}}}\!\!&=&\!
        \mathcal{L}^{\text{bilinear}}_{\textrm{NRQED}_{\textrm{DM}}}
        - \frac{1}{4} F^{\mu \nu}F_{\mu \nu}
       +\frac{d_2}{M^2} F^{\mu \nu} \bm{{\rm{D}}}^2 F_{\mu \nu} \nonumber
       \\
       \!&\hspace{0.5cm}+&\!\frac{d_s}{M^2} \psi^\dagger \chi \, \chi^\dagger \psi
       +\frac{d_v}{M^2} \psi^\dagger \, \bm{\sigma} \, \chi \cdot \chi^\dagger \, \bm{\sigma} \, \psi
       + \sum_{i=1}^{n_f}\bar{f}_i i \slashed{D} f_i \, ,
\label{NREFT_lag}
\end{eqnarray}
where we have set the light fermion masses to zero.
Here, $\psi$ is the two-component Pauli spinor that annihilates a dark matter fermion, and $\chi^\dagger$ is the Pauli spinor that annihilates an antifermion.
We have not explicitly displayed the terms in the Lagrangian density that are bilinear in the fields $\psi$ and $\chi$, i.e. $\mathcal{L}^{\text{bilinear}}_{\textrm{NRQED}_{\textrm{DM}}}$;
they can be found in the original reference \cite{Caswell:1985ui} and are discussed in the context of the abelian DM model in reference \cite{Biondini:2023zcz}.

\begin{figure}[ht!]
    \centering
    \includegraphics[scale=0.98]{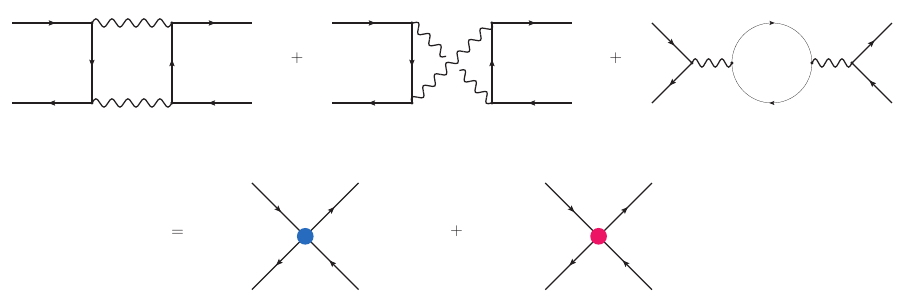}
    \caption{Matching between annihilation diagrams in the relativistic theory at one loop (upper three diagrams)
  and the corresponding four-fermion interactions in $\textrm{NRQED}_{\textrm{DM}}$ (lower two diagrams).
  The associated imaginary parts of the matching coefficients $d_s$ and $d_v$ at order $\alpha^2$ are given in \eqref{matching_coeff}.
  The thick solid lines denote the incoming and outgoing heavy DM particle and antiparticle, whereas wiggly lines stand for dark photons and thin solid lines for light dark fermions.}
    \label{fig:ann}
\end{figure}

We are interested in describing the annihilation of heavy dark matter pairs. 
Annihilation processes happen at the scale $2M$ and the corresponding energy modes are integrated out in the non-relativistic EFT.
The local four-fermion operators shown in eq.~\eqref{NREFT_lag} encode the annihilation of S-wave fermion-antifermion pairs.\footnote{
Higher-dimensional four-fermion operators capture the annihilation of fermion-antifermion pairs with non-vanishing orbital angular momentum.
For example, dimension eight four-fermion operators encode the annihilation of P-wave fermion-antifermion pairs.
We do not consider dimension-eight or higher operators in this work.}
The leading order contribution to the imaginary part of the dimension-six operators comes from the annihilation processes $X \bar{X} \to \gamma \gamma$ and $X \bar{X} \to f \bar{f}$,
see figure~\ref{fig:ann};
the imaginary part of the matching coefficients $d_s$ and $d_v$ in eq.~\eqref{NREFT_lag} may be obtained by cutting the loop diagrams along the photon propagators
and the light fermion propagators~\cite{Bodwin:1994jh,Braaten:1996ix}. 
At the lowest order in the coupling, the heavy $X \bar{X}$ pair in an S-wave annihilates either into two dark photons or into a light fermion-antifermion pair,
respectively for a spin-singlet and spin-triplet configuration. 
The imaginary parts of the four-fermion matching coefficients $d_s$ and $d_v$ at order $\alpha^2$ read 
\begin{equation}
\text{Im}[d_s] = \pi \alpha^2  \, , \quad 
\text{Im}[d_v] =  \frac{n_f}{3} \pi \alpha^2 \, .
\label{matching_coeff}
\end{equation}
The matching procedure can be systematically improved by adding higher orders in the coupling.
The four-fermion Wilson coefficients are known up to next-to-next-to leading order, see refs.~\cite{Petrelli:1997ge,Pineda:1998kj} and the review~\cite{Vairo:2003gh}.

\subsection{\texorpdfstring{pNRQED$_{\textrm{DM}}$}{pNRQED DM}}
\label{sec:pNREFT_part}
Following the hierarchy of energy scales~\eqref{scale_arrang}, the next relevant scales to be integrated out are $\sqrt{MT}$ and the soft scale $M\alpha$. At the lowest order, integrating out photons of momentum $\sqrt{MT}$ and energy $T$ is equivalent to integrating out photons with energy or momentum of order $M\alpha$. 
At higher orders, momenta of order $\sqrt{MT}$ may affect the real and imaginary part of the potential between the DM heavy fermion-antifermion pair; for a QCD analogue where the role of the temperature is played by $\Lambda_\text{QCD}$ see~\cite{Brambilla:2003mu}.
These effects are beyond our present accuracy. 

After integrating out dark photons with energy or momentum  of order $M \alpha$, 
the resulting effective field theory takes the form of potential NRQED (pNRQED)~\cite{Pineda:1997bj,Pineda:1997ie} and we denote it as pNRQED$_\textrm{DM}$ \cite{Biondini:2023zcz}.
The dynamical degrees of freedom are heavy dark fermions and antifermions with energies $T$ and $M\alpha^2$ and momenta $\sqrt{MT}$ and $M\alpha$,
photons with energies and momenta of order $T$ and $M \alpha^2$, and light dark fermions and antifermions with energies and momenta of order $T$.
The case of pNRQED at finite temperature has been studied in refs.~\cite{Escobedo:2008sy,Escobedo:2010tu} for the hydrogen atom and the muonic hydrogen,
whereas its application to dark matter freeze-out in the early universe can be found in refs.~\cite{Biondini:2023zcz,Biondini:2024aan}.

According to the assumed hierarchy of scales, the matching can be carried out by setting to zero energy scales smaller than $M\alpha$, 
which are the ultrasoft scales.
The temperature is one of the ultrasoft scales, hence, the matching between NRQED$_{\textrm{DM}}$ and pNRQED$_{\textrm{DM}}$ is in practice as in-vacuum.   
The building blocks of pNRQED$_{\textrm{DM}}$ are fermion-antifermion pairs,
hence it is customary to project the EFT on the fermion-antifermion space 
and express it in terms of gauge singlet fermion-antifermion bilocal fields $\phi(t,\bm{r},\bm{R})$,
where $\bm{r} \equiv \bm{x}_1-\bm{x}_2$ is the distance between a fermion located at $\bm{x}_1$
and an antifermion located at $\bm{x}_2$ and $\bm{R}\equiv(\bm{x}_1+\bm{x}_2)/2$ is the center of mass coordinate.
Fermion-antifermion pairs above the threshold form scattering states of positive energy and fermion-antifermion pairs below the threshold form bound states of negative energy.
In order to ensure that the photons are ultrasoft, photon fields are multipole expanded in $\bm{r}$.
Hence, the pNRQED$_\textrm{DM}$ Lagrangian density for the dark matter theory \eqref{lag_mod_0} is organized as an expansion in $1/M$, 
inherited from NRQED$_\textrm{DM}$, $\bm{r}$ and $\alpha$ (at weak coupling):
\begin{eqnarray}
  \mathcal{L}_{\textrm{pNRQED}_{\textrm{DM}}}&=&   \int d^3r \; \phi^\dagger(t,\bm{r},\bm{R})
             \, \left[ i \partial_0 -H(\bm{r},\bm{p},\bm{P},\bm{S}_1,\bm{S}_2)  + g \, \bm{r} \cdot \bm{E}(t,\bm{R})\right] \phi (t,\bm{r},\bm{R})  + \dots  \nonumber \\
&&-\frac{1}{4} F_{\mu \nu} F^{\mu \nu}   + \sum_{i=1}^{n_f}\bar{f}_i i \slashed{D} f_i  \, ,
\label{pNREFT_1}
\end{eqnarray}
where $\bm{E}$ is the (dark) electric field, $E^i=F^{i0}$, and
\begin{eqnarray}
 &&H(\bm{r},\bm{p},\bm{P},\bm{S}_1,\bm{S}_2) = 2M + \frac{\bm{p}^2}{M}+\frac{\bm{P}^2}{4M} - \frac{\bm{p}^4}{4M^3} +  V (\bm{r},\bm{p},\bm{P},\bm{S}_1,\bm{S}_2) + \ldots\, , 
 \label{ham_pNRQED}\\
  &&V (\bm{r},\bm{p},\bm{P},\bm{S}_1,\bm{S}_2)= V^{(0)} + \frac{V^{(1)}}{M} + \frac{V^{(2)}}{M^2} + \ldots \, ,
 \label{pot_pNRQED}    
\end{eqnarray}
with $\bm{S}_1=\bm{\sigma}_1/2$ and $\bm{S}_2=\bm{\sigma}_2/2$ the spin operators acting on the fermion and antifermion, respectively.
The static potential $V^{(0)}$ is just the Coulomb potential $-\alpha/r$.
At order $r$, the term $\phi^\dagger(t,\bm{r},\bm{R}) \, \bm{r} \cdot \bm{E}(t,\bm{R}) \, \phi(t,\bm{r},\bm{R})$ in the pNRQED$_{\textrm{DM}}$ Lagrangian describes
the \emph{electric dipole interaction} of the dark fermion-antifermion pair with ultrasoft dark photons.
The electric-dipole vertex is at the origin of the ultrasoft transitions triggering bound-state formation, bound-state dissociation and bound-state to bound-state transitions.
We deal with these processes in section~\ref{sec:el_dipole_transitions}.  
The matching coefficient of the electric dipole interaction has been taken equal to one. 
 
As far as pair annihilations go, both for a scattering state and a bound state,
they are accounted for by the imaginary part of a local potential in pNRQED$_{\textrm{DM}}$. 
At order $1/M^2$ and in the center-of-mass frame, it reads
\begin{equation}
  \delta V^{\textrm{ann}}(\bm{r})
  = - \frac{i}{M^2} \, \delta^3(\bm{r})\, \left[ {2\rm{Im}}(d_s) - \bm{S}^2 \left( {\rm{Im}}(d_s)- {\rm{Im}}(d_v) \right) \right],
\label{pNREFT_2}
\end{equation}
where $\bm{S}=\bm{S}_1+\bm{S}_2$ is the total spin of the pair. 
Such potential describes the annihilation of fermion-antifermion pairs in an S-wave.
One obtains an annihilation cross section when projecting on an above-threshold scattering state and a decay width when projecting on a bound state.
The effect of soft-photon exchanges is contained in the fermion-antifermion wave function.
The annihilation cross section reads\footnote{
We can compute the annihilation cross section in the center-of-mass frame by means of the optical theorem in pNRQED$_{\textrm{DM}}$:
$$
(\sigma_{\hbox{\scriptsize ann}} v_{\hbox{\scriptsize rel}})(\bm{p}) 
= \frac{1}{2} \langle \,  \bm{p}, \bm{0}| \int d^3r \, \phi^\dagger(\bm{r},\bm{R},t)\, 
\left[-\rm{Im} \,\delta V^{\rm ann}(\bm{r})\right] \, \phi(\bm{r},\bm{R},t)\, |  \bm{p},\bm{0} \rangle
\, ,
$$
where $|\bm{p},\bm{P}\rangle$ is the scattering state of an incoming unbound DM pair with relative momentum $\bm{p}$ and center-of-mass momentum $\bm{P}$. A detailed derivation can be found in ref.~\cite{Biondini:2023zcz}.}
\begin{eqnarray}
(\sigma_{\hbox{\scriptsize ann}} v_{\hbox{\scriptsize rel}})(\bm{p}) 
&=&  \frac{{\rm{Im}}(d_s)+3{\rm{Im}}(d_v)}{M^2} \left|\Psi_{\bm{p} 0}(\bm{0})\right|^2
    =  (\sigma^{\hbox{\tiny NR}}_{\hbox{\scriptsize ann}} v_{\hbox{\scriptsize rel}})  \, S_{\hbox{\scriptsize ann}}(\alpha/v_\text{rel})\,,
\label{ann_fact_scat}
\end{eqnarray}
where $v_{\hbox{\scriptsize rel}}$ is the relative velocity of the pair,  
$\Psi_{\bm{p},\ell}(\bm{r})$ is the $\ell$-th partial wave of the scattering state and $S_{\hbox{\scriptsize ann}}\equiv|\Psi_{\bm{p},0}(\bm{0})|^2$ is the {\it Sommerfeld factor} capturing the soft dynamics.
In eq.~\eqref{ann_fact_scat}, $\sigma^{\hbox{\tiny NR}}_{\hbox{\scriptsize ann}} v_{\hbox{\scriptsize rel}}$ is the annihilation cross section of free scatters, and hence it only depends on $\text{Im}(d_s)$ and $\text{Im}(d_v)$ that encode the hard scale effects.
At leading order, using the expressions in~\eqref{matching_coeff}, we find 
\begin{eqnarray}
        \sigma^{\hbox{\tiny NR}}_{\hbox{\scriptsize ann}} v_{\hbox{\scriptsize rel}}=(1+n_f)\frac{\pi\alpha^2}{M^2} \, .
\end{eqnarray}
For bound states below threshold, the decay widths are
\begin{eqnarray}
\Gamma^{n,\hbox{\scriptsize para}}_{\textrm{ann}} &=& \frac{4 {\rm{Im}}(d_s)}{M^2} \frac{|R_{n0}(0)|^2}{4\pi} \,,
\label{ann_para}\\
\Gamma^{n,\hbox{\scriptsize ortho}}_{\textrm{ann}}&=& \frac{4 {\rm{Im}}(d_v)}{M^2} \frac{|R_{n0}(0)|^2}{4\pi} \,,
\label{ann_ortho}
\end{eqnarray}
where $R_{n0}(r)$ is the radial part of an S-wave ($\ell=0$) with principal quantum number $n$, 
and we further distinguish between spin-singlet paradarkonium and spin-triplet orthodarkonium. 
At leading order in the coupling, the decay widths for the para- and orthodarkonium ground state 1S are 
\begin{equation}
    \Gamma^{\text{1S},\hbox{\scriptsize para}}_{\textrm{ann}}=\frac{M\alpha^5}{2} \, , \quad \Gamma^{\text{1S},\hbox{\scriptsize ortho}}_{\textrm{ann}}=\frac{n_f}{3} \frac{M\alpha^5}{2} \, .
\label{eq:Gammannihilation}    
\end{equation} 
The {\it natural} renormalization scale for the coupling appearing in $d_s$ and $d_v$ is 
of the order of $2M$, whereas the one for the coupling in the wave function is of the order of the soft scale $M\alpha$.
In the applications of section~\ref{sec:DMevolution}, we set it at these values.

\section{Dark matter heavy pairs at finite temperature}
\label{sec:el_dipole_transitions}
Together with the annihilation of unbound dark matter pairs,
the formation of bound states and their decays into light degrees of freedom provide another way of depleting the heavy dark matter particles.
More specifically, bound-state formation, dissociation and transitions among bound states may significantly affect the dark matter energy density.   

\begin{figure}[ht]
    \centering
    \includegraphics[scale=0.85]{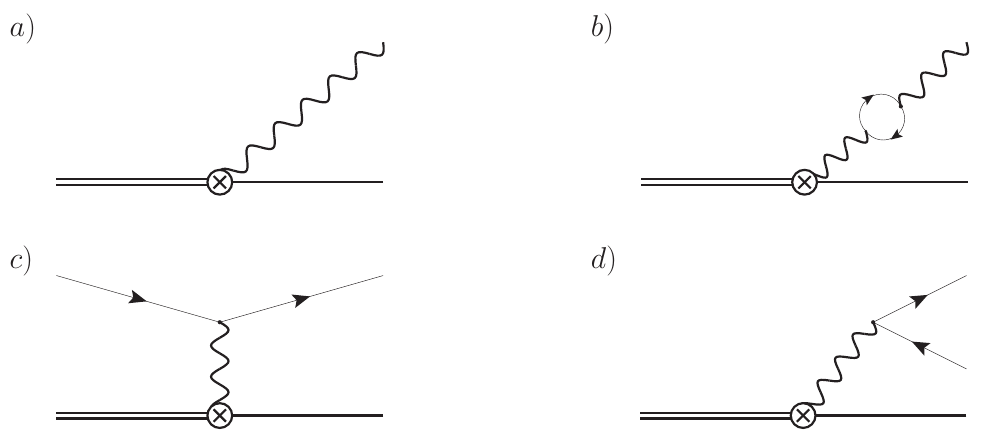}
    \caption{Bound-state formation processes.
      Solid double lines stand for dark matter pairs in a scattering state, 
      solid lines for bound states, wiggly lines for dark photons, arrowed thin lines for light plasma constituents (light fermions in our case), and  
      circled-crossed vertices for the electric dipole interactions in pNRQED$_\text{DM}$.}
    \label{fig:bsf_different}
\end{figure}

In the dark matter model~\eqref{lag_mod_0}, and equivalently in its low-energy version \eqref{pNREFT_1}, bound-state formation may proceed via different mechanisms.
Their relative importance depends on the temperature. 
One finds: 
{\sl a)} bound-state formation via radiative emission of the gauge boson;
process {\sl b)}, which is the same radiative emission process as {\sl a)}, 
however with a quantum correction for the photon;
{\sl c)} bound-state formation via inelastic collisions of a scattering state with a constituent of the thermal bath that turns it into a bound state;
{\sl d)} the decay of the emitted gauge boson into a pair of light fermions.
See figure~\ref{fig:bsf_different} for a diagrammatic representation of the different bound-state formation mechanisms.
The reverse processes correspond to bound-state dissociation, where $a)$ and $b)$ can be referred to as photo-dissociation and $c)$ traces back to the Landau damping in a QED plasma.
Diagrams $b)$, $c)$ and $d)$ of figure~\ref{fig:bsf_different} only contribute if the plasma contains light fermions.
Bound-state to bound-state transitions may also occur with the same topologies, however replacing the scattering state with a bound state.
Bound-state to bound-state transitions comprise both excitations and de-excitation processes. 

For bound-state formation and dissociation, as well as bound-state to bound-state transitions,
the relevant ultrasoft energy scale 
is the energy difference, $\Delta E$, between the incoming and outgoing pair.
At sufficiently high temperatures, it holds that\footnote{If the kinetic energy of the unbound heavy DM fermion pair is of the same order as or smaller than the binding energy, then the size of $\Delta E$ is set by the binding energy.
With respect to the absolute value of the ground state binding energy, $|E_1^b|=M\alpha^2/4$,
the condition $T \gg |E^b_1|$ is realized for $T \gg M/400$ if we choose $\alpha=0.1$.
This condition is fulfilled at freeze-out, where $T \approx M/25$.} 
\begin{equation}
T \gg \Delta E\,.
\label{scale_arrang3}
\end{equation}
We analyze this situation in section~\ref{sec:bsf_resummed}.
Under $T \gg \Delta E$, the dispersion relation for the dark photon in the self-energy gets modified:
interactions between the thermal dark photons and the light dark particles from the medium dynamically generate a thermal mass $m_\text{D}= \sqrt{4\pi n_f \alpha/3}\,T$, called Debye mass, 
see appendix~\ref{sec:app_B}.
The Debye mass depends on the plasma temperature $T$, the gauge coupling $\alpha$ computed at the scale of the temperature and the number $n_f$ of dark light fermions.
In order to satisfy the hierarchy~\eqref{scale_arrang2} between thermal scales, namely $T \gg m_\text{D}$, one has to require $\sqrt{\pi n_f \alpha } \ll 1$.
We may further distinguish between relative arrangements of the lowest-lying scales,
namely whether $m_\text{D}$ is of the order of $\Delta E$, or $m_\text{D} \ll \Delta E$ or $m_\text{D} \gg \Delta E$.\footnote{
The condition  $m_\text{D} \gg |E^b_1|$ is realized for $M/T \ll 8  \alpha^{-3/2} \sqrt{\pi n_f/3} ~(\approx 260$ for $\alpha=0.1$ and $n_f=1$), while the  opposite situation is realized at very low temperatures.}
It is one of the main results of our work to include the resummation of the Debye mass effects in the calculation of the bound-state formation cross section.

If the temperature is of the same order as or smaller than the binding energy and/or the unbound heavy DM fermion pair is in kinetic equilibrium with the plasma, which implies that its kinetic energy is of order $T$, 
then it holds that 
\begin{equation}
T \lesssim \Delta E\,.
\label{scale_arrang4}
\end{equation}
Since the temperature keeps decreasing in the expanding universe,
the condition $T \lesssim \Delta E$ is unavoidably reached in the late stages of its evolution.
Moreover, it is realized in vacuum.
It has been treated in ref.~\cite{Biondini:2023zcz} for the case without light fermions.
For $T \lesssim \Delta E$, the polarization tensor does not generate a Debye mass, and 
bound-state formation and dissociation can be computed at fixed order in perturbation theory.
In ref.~\cite{Binder:2020efn}, they have been computed for the model of eq.~\eqref{lag_mod_0} at next-to-leading order (NLO).
In appendix~\ref{sec:app_C}, we reproduce the NLO calculation of the bound-state formation cross section. 

\subsection{Debye mass contribution to the self-energy}
\label{sec:bsf_resummed}
The bound-state formation cross section can be computed from the imaginary part of the self-energy in pNRQED$_{\textrm{DM}}$.
As we are now dealing explicitly with thermal scales, the computation needs to be performed in the \emph{thermal field theory} version of pNRQED$_{\textrm{DM}}$.
We use the real-time Schwinger--Keldysh formalism~\cite{Bellac:2011kqa,Laine:2016hma,Ghiglieri:2020dpq}.
The real-time formalism necessarily leads to a doubling of the degrees of freedom called type 1 and 2. 
We collect the relevant propagators of the fermion-antifermion field $\phi$ in  pNRQED$_{\textrm{DM}}$, as well as the thermal photon propagator in appendix \ref{sec:app_A} and \ref{sec:app_B}. 
For the heavy dark matter fermion, a rather important simplification can be made. 
As shown in ref.~\cite{Brambilla:2008cx}, the 12 component of a heavy-field propagator vanishes in the heavy-mass limit,
hence the physical heavy fields do not propagate into type 2 fields.
Therefore, the type 2 fermion-antifermion fields decouple and may be ignored in the heavy-mass limit, 
which makes the real-time formalism convenient when dealing with heavy particles in a thermal environment.

\begin{figure}[ht]
    \centering
    \includegraphics[scale=0.85]{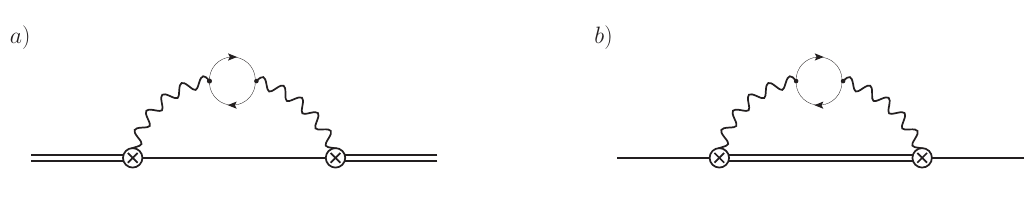}
    \caption{Two-loop self-energy diagrams in pNRQED$_{\textrm{DM}}$ with an initial scattering state and an intermediate bound state on the left, 
    and with an initial bound state and an intermediate scattering state on the right. 
    The represented (massless) fermion loops provide the complete one-loop correction to the photon propagator.
    The imaginary part of the left diagram contributes to the bound-state formation process, while the imaginary part of the right diagram contributes to the bound-state dissociation.}
    \label{fig:bsf_two_loop_fixed_NLO}
\end{figure}

The 11 component of the self-energy diagram shown at two loops in the left or right panel of figure~\ref{fig:bsf_two_loop_fixed_NLO} reads 
\begin{eqnarray}
  \Sigma^{11}(p_0) =   -ig^2 \frac{\mu^{4-D}}{D-1} \int \frac{d^D q}{(2\pi)^D}\, r^i\frac{i}{p_0-q_0-H+i\epsilon}r^i \,[q_0^2 D^{11}_{ii} (q)+{\bm{q}}^2 D^{11}_{00} (q)]\, ,
  \label{general_self_energy}
\end{eqnarray}
where the integral has been regularized in $D=4 -2 \epsilon$ dimensions and $p_0$ is the energy of the incoming pair. 
We compute the in-vacuum photon polarization contribution to the self-energy in appendix~\ref{sec:app_C}, 
where we label it $n_\textrm{F}= 0$, $n_\textrm{F}$ being the Fermi--Dirac distribution of the fermions in the loop. 
We add this contribution to the self-energy in section~\ref{sec:bsf}.

In this section, instead, we compute the contribution to the self-energy~\eqref{general_self_energy} coming from the thermal part of the photon polarization and from the one-loop self energy.
We split the calculation into different momentum and energy regions.
The temperature contribution at NLO, valid under the condition $T \lesssim \Delta E$, can be found  
in appendix~\ref{sec:app_C}.
Its impact on the bound-state formation cross section is analyzed in section~\ref{sec:bsf}.
Here, in section~\ref{sec:TggE}, we compute the contribution from the scale $T$ under the condition $T \gg \Delta E$.
Furthermore, the condition $T \gg \Delta E$ requires accounting for the Debye mass and the resummation of the polarization tensor in the dark photon propagator.
We consider the limiting cases $m_\text{D} \gg \Delta E$ and $\Delta E \gg m_\text{D}$
in sections~\ref{sec:mDggE} and~\ref{sec:mDllE}, respectively, 
and the most general case $m_\text{D} \sim \Delta E$ in section~\ref{sec:mDsimE}.

\subsubsection{Temperature contribution in the hierarchy \texorpdfstring{$T \gg \Delta E$}{T >> delta E}}
\label{sec:TggE}
We start with the energy/momentum modes of the order of the temperature, namely $q_0, |\bm{q}| \sim T$.  
We expand the heavy pair propagator in $\Delta E \equiv p_0 - H \ll q_0 \sim T$ up to leading order, 
\begin{equation}
      \frac{i}{p_0-H-q_0+i\epsilon} = 
         \frac{i}{-q_0+i\epsilon}\left[1+\mathcal{O}\left(\frac{p_0-H}{q_0}\right)\right] \, ,
\label{DM_prop_expanded}
\end{equation}
and insert it into~\eqref{general_self_energy}, that now reads 
\begin{equation}
\begin{aligned}
        \Sigma^{11}_{T}(p_0) &=-ig^2 \frac{\mu^{4-D}}{D-1} \bm{r}^2 \int \frac{d^D q}{(2\pi)^D} \frac{i}{-q_0+i\epsilon}[q_0^2 D^{11}_{ii} (q)+{\bm{q}}^2 D^{11}_{00} (q)] \, .
\label{scattering_self_energy_T}
\end{aligned}
\end{equation}
The subscript $T$ for the self-energy  denotes the corresponding energy region. 
We shall keep a similar notation  when integrating out the subsequent lower scales $m_\text{D}$ and $\Delta E$.
By inserting the leading order expression for the dark photon propagator given in~\eqref{XXpropGT} in eq.~\eqref{scattering_self_energy_T}, we obtain a vanishing integral.\footnote{
In fact, the vacuum part of the free dark photon propagator results in a scaleless integral for all terms in the expansion in~\eqref{DM_prop_expanded}. 
As for the thermal part, contributions up to the third order in the expansion of the heavy pair propagator can be shown to either vanish in dimensional regularization or be real and hence not contribute to the bound-state formation cross section given in~\eqref{bsf_projected}. 
Further terms of higher order in $\Delta E/T$ in~\eqref{DM_prop_expanded} can be omitted 
since they scale like $\alpha r^2 (\Delta E)^4$ or smaller, 
which is beyond our accuracy of interest~\cite{Brambilla:2010vq}.} 
Next, we insert the NLO expression into~\eqref{scattering_self_energy_T}, and, given that $q \sim T \gg m_\text{D}$, 
we can expand the dark photon propagator. 
Since the electric correlator is even in $q_0$, only the even part of the zeroth order expression of the DM pair propagator contributes, 
\begin{equation}
\begin{aligned}
      \Sigma^{11}_{T}(p_0) &=-ig^2 \frac{\mu^{4-D}}{D-1} \bm{r}^2 \int \frac{d^D q}{(2\pi)^D} \pi \delta(q_0){\bm{q}}^2 D^{11,\text{NLO}}_{00} (q) \, ,
\label{scattering_self_energy_T2}
\end{aligned}
\end{equation}
and hence the imaginary part reads
\begin{equation}
\begin{aligned}
        \text{Im}\left[\Sigma^{11}_{T}(p_0)\right] &=-g^2 \frac{\mu^{4-D}}{D-1} \bm{r}^2 \int \frac{d^D q}{(2\pi)^D} \pi \delta(q_0){\bm{q}}^2 ~\text{Re}\left[D^{11,\text{NLO}}_{00} (q)\right] \\
        &=-\frac{g^2}{2} \frac{\mu^{4-D}}{D-1} \bm{r}^2 \int \frac{d^D q}{(2\pi)^D} \pi \delta(q_0){\bm{q}}^2 \frac{-i\Pi_{00}^S(q)}{\bm{q}^4} \, ,
\label{scattering_self_energy_T3}
\end{aligned}
\end{equation}
where we have split the dark photon propagator (and polarization tensor) according to~\eqref{LOprop_coulombgauge} (and~\eqref{symm_pol_tensor_hierarchy1}) into its real symmetric and imaginary antisymmetric parts. 
Only the real part is relevant to our calculation. 
Since in the integrand of \eqref{scattering_self_energy_T3} $q_0=0$, 
it follows that the relevant momentum region is $|\bm{q}|\sim T \gg q_0$; 
using the appropriate symmetric polarization tensor in eq.~\eqref{symm_pol_tensor_hierarchy1}, we obtain~\cite{Brambilla:2008cx}
\begin{equation}
\begin{aligned}
        \text{Im}\left[\Sigma^{11}_{T}(p_0)\right] &=-4g^4n_f T \bm{r}^2 \frac{\mu^{4-D}}{D-1}  \int \frac{d^{D-1} q}{(2\pi)^{D}} \frac{1}{|\bm{q}|^3}\int^{\infty}_{|\bm{q}|/2} d|\bm{k}| |\bm{k}|n_\textrm{F}(|\bm{k}|) \\
        &=\frac{\alpha}{6} \bm{r}^2 T m_\text{D}^2 \left[\frac{1}{\epsilon}+\gamma_\text{E}+\frac{2}{3}-4\ln{2}-2\frac{\zeta'(2)}{\zeta(2)}-\ln\left(\frac{T^2}{\pi \mu^2}\right)\right] \, ,
\label{scattering_self_energy_T4}
\end{aligned}
\end{equation}
where $\zeta$ is Riemann's zeta function and $\zeta(2)=\pi^2/6$. 
The quantity in eq.~\eqref{scattering_self_energy_T4} is infrared divergent and scale-dependent. 
However, at this stage, this is not a concern.  
It is in fact a manifestation, or an artifact, of the separation of energy scales. 
Upon adding the contributions from the lower energy scales, either $m_\text{D}$ or $\Delta E$, the divergence and the renormalization scale $\mu$ indeed cancel out. 
In the next sections, we consider the three possible scale arrangements between $m_\text{D}$ and $\Delta E$. 

Higher order corrections coming from the contributions to the heavy pair propagator beyond the zeroth order expansion are suppressed by a factor $\alpha r^2 m_\text{D}^2 T \times (\Delta E/T)^2 \ll \alpha r^2 m_\text{D}^2 T$. Similarly, corrections to the dark photon propagator are suppressed by $\alpha r^2 m_\text{D}^2 T \times (m_\text{D}/T)^2 \ll \alpha r^2 m_\text{D}^2 T$ and hence beyond the accuracy of eq.~\eqref{scattering_self_energy_T4}.

In a similar manner, one could calculate the real part of the self-energy. 
It gives thermal corrections to the potential of the heavy DM pair, see for instance ref.~\cite{Brambilla:2008cx} for the case of heavy quarkonium in a QGP or refs.~\cite{Escobedo:2008sy,Escobedo:2010tu} for hydrogen and muonic atoms.

\subsubsection{Hierarchy \texorpdfstring{$m_{\rm D} \gg \Delta E$}{mD >> delta E}} 
\label{sec:mDggE}
We consider here the scale arrangement $m_\text{D} \gg \Delta E$. 
Hence, the next relevant scale to contribute in the loop is the energy and momentum of the order of the Debye mass scale, i.e. $q\sim m_\text{D}$. 
Also in this case, we can expand the DM pair propagator as in eq.~\eqref{DM_prop_expanded} because $\Delta E  \ll q_0 \sim m_{\textrm{D}}$, however, the dark photon propagator needs to be resummed because $q_0$, $|\bm{q}| \sim m_{\textrm{D}}$. 
We present the hard thermal loop resummation in appendix~\ref{sec:b3}.
The imaginary part of the self-energy becomes~\cite{Brambilla:2008cx}
\begin{equation}
\begin{aligned}
        \text{Im}\left[\Sigma^{11}_{m_\text{D}}(p_0)\right] &=-\frac{g^2}{2} \frac{\mu^{4-D}}{D-1} \bm{r}^2 \int \frac{d^D q}{(2\pi)^D} \pi \delta(q_0){\bm{q}}^2 D^S_{00}(q) \\
        &=-g^2 \frac{\mu^{4-D}}{D-1}  \bm{r}^2 \int \frac{d^D q}{(2\pi)^D}  \pi \delta(q_0)\bm{q}^2 \pi \frac{T}{|\bm{q}|}\frac{m_\text{D}^2}{(\bm{q}^2+m_\text{D}^2)^2} \\
        &=-\frac{\alpha}{6} \bm{r}^2 T m_\text{D}^2 \left[\frac{1}{\epsilon}-\gamma_\text{E}+\frac{5}{3}-\ln\left(\frac{m_\text{D}^2}{\pi \mu^2}\right)\right] \, ,
\label{scattering_self_energy_mD}
\end{aligned}
\end{equation}
where we have used the longitudinal component of the resummed symmetric dark photon propagator in eq.~\eqref{resummed_symm_prop}. 
The result is ultraviolet divergent and depends on the scale $\mu$. 
As anticipated, by summing eqs.~\eqref{scattering_self_energy_T4} and~\eqref{scattering_self_energy_mD}, the divergences, as well as the renormalization-scale dependence, cancel out.

The last contributing region to the self-energy
is $q\sim \Delta E$. 
At variance with the former situations, we cannot expand the DM pair propagator.
We obtain
\begin{equation}
\begin{aligned}
        \text{Im}&\left[\Sigma^{11}_{\Delta E}(p_0)\right] \\
        &=\text{Im}\left[-ig^2 \frac{\mu^{4-D}}{D-1} \int \frac{d^D q}{(2\pi)^D}\, r^i \frac{i}{p_0 - q_0 - H +i\epsilon} r^i\, 
         [q_0^2 D^{11}_{ii} (q)+{\bm{q}}^2 D^{11}_{00} (q)]\right]  \, .\\
        &=-\frac{2}{3} \alpha r^i (\Delta E)^2 r^iT\, ,
\label{scattering_self_energy_E}
\end{aligned}
\end{equation}
which accounts for the bound-state formation process via the emission of a thermal transverse dark photon. 
The photon propagators are the resummed ones, which can be taken at leading order.
The term proportional to $D^{11}_{00}$ vanishes when performing the integral in $q_0$. 
The result is of order $\alpha r^2 m_\text{D}^2 T \times (\Delta E/m_\text{D})^2$, i.e.  suppressed by a factor $(\Delta E/m_\text{D})^2$ with respect to the contributions from the scales $m_\text{D}$ and $T$. 
Therefore, for the particular hierarchy $T\gg m_\text{D}\gg \Delta E$, the Landau damping phenomenon induced by the scattering with particles in the thermal bath dominates over the thermal photo-emission process~\cite{Brambilla:2008cx}.

The total imaginary part of the heavy pair self-energy is the sum of the contributions from the scale $T$,  $m_\text{D}$ and  $\Delta E$ from eqs.~\eqref{scattering_self_energy_T4}, \eqref{scattering_self_energy_mD} and \eqref{scattering_self_energy_E}:
\begin{equation}
\begin{aligned}
\text{Im}\left[\Sigma^{11}(p_0)\right]&=\text{Im}\left[\Sigma^{11}_{T}(p_0)\right]+\text{Im}\left[\Sigma^{11}_{m_\text{D}}(p_0)\right]+\text{Im}\left[\Sigma^{11}_{\Delta E}(p_0)\right] \\
&=\frac{\alpha}{6} T m_\text{D}^2 r^i\left[2\gamma_\text{E}-1-2\frac{\zeta'(2)}{\zeta(2)}+\ln\left(\frac{m_\text{D}^2}{16T^2}\right)-\frac{(2\Delta E)^2}{m_\text{D}^2}\right]r^i \, ,
\label{scattering_self_energy_hierarchy1}
\end{aligned}
\end{equation}
which is finite and scale-independent. 
We recall that $\Delta E=p_0 -H$ in the formula above is an operator.
In eq. \eqref{scattering_self_energy_hierarchy1}, the parametrically dominant term is $\alpha T m_\text{D}^2 r^i\left[2\gamma_\text{E}-1-2\zeta'(2)/\zeta(2)\right.$ $\left. +\ln\left(m_\text{D}^2/(16T^2)\right)\right]r^i/6$,while we display terms up to relative order $(\Delta E/m_\text{D})^2$.

\subsubsection{Hierarchy \texorpdfstring{$\Delta E \gg m_{\rm D}$}{delta E >> mD}} 
\label{sec:mDllE}
In the reversed hierarchy arrangement $\Delta E \gg m_\text{D}$, we first integrate out energy and momentum modes of the order of $\Delta E$. 
We cannot expand the DM pair propagator in $\Delta E/q_0$, and this results in a richer set of contributions from both longitudinal as well as transverse dark photons. 
We write the longitudinal contribution as
\begin{equation}
\begin{aligned}
        \Sigma^{11,\textrm{long}}_{\Delta E}(p_0)
        &=-ig^2 \frac{\mu^{4-D}}{D-1} \int \frac{d^D q}{(2\pi)^D} r^i \frac{i}{p_0-H-q_0+i\epsilon} r^i \bm{q}^2 D^{11}_{00} (q) \\
        &=-i\frac{g^2}{2} \frac{\mu^{4-D}}{D-1} \int \frac{d^D q}{(2\pi)^D} r^i\frac{i}{p_0-H-q_0+i\epsilon} r^i \bm{q}^2 \left[D^{AS}_{00}(q)+D^{S}_{00}(q)\right]\\
        &\equiv \Sigma^{11,\textrm{long},AS}_{\Delta E}(p_0)+\Sigma^{11,\textrm{long},S}_{\Delta E}(p_0) \, ,
\label{scattering_self_energy_E2_long}
\end{aligned}
\end{equation}
which again can be split into an antisymmetric ($AS$) and symmetric ($S$) part according to the corresponding decomposition of the photon propagator, see appendix~\ref{sec:app_A}. 
For the antisymmetric propagator, one can simplify it as follows
\begin{equation}
    D^{AS}_{00}(q) = \frac{2i\bm{q}^2+i(\Pi_{00}^R(q)+\Pi_{00}^A(q))}{(\bm{q}^2+\Pi_{00}^{R}(q))(\bm{q}^2+\Pi_{00}^{A}(q))}  \approx  2\frac{i}{\bm{q}^2} \, ,
\label{resummed_asymm_prop_E}
\end{equation}
where we used $\Pi_{00}^{R/A}(q) \sim m_\text{D}^2 \ll \bm{q}^2 \sim (\Delta E)^2$. 
It is now possible to show that the antisymmetric part gives a vanishing contribution,  $\Sigma^{11,\textrm{long},AS}_{\Delta E}(p_0)=0$, by integrating over $q_0$ and using the residue theorem. 
NLO corrections in the dark photon propagator contribute to order $\alpha r^2 m_\text{D}^2T \times (\Delta E/T)$ and are beyond our accuracy of interest.

As for the symmetric part, we  use $\Pi^{S,T\neq 0}_{00}(q)$ given in~\eqref{symm_long_pol_tensor_hierarchy2}, which holds for $T\gg q \sim \Delta E$, and we get
\begin{equation}
\begin{aligned}
       \Sigma^{11,\textrm{long},S}_{\Delta E}(p_0) 
        &=-ig^2 \frac{\mu^{4-D}}{D-1} \int \frac{d^D q}{(2\pi)^D} r^i\frac{i}{p_0-H-q_0+i\epsilon}r^i\bm{q}^2\pi T\frac{m_\text{D}^2}{|\bm{q}|^5}\theta(-q^2)\, .
\label{scattering_self_energy_E2_long_symm}
\end{aligned}
\end{equation}
Hence, the corresponding imaginary part is~\cite{Brambilla:2010vq}
\begin{equation}
\begin{aligned}
        \text{Im}\left[\Sigma^{11,\textrm{long},S}_{\Delta E}(p_0) \right]
        &=-g^2 \frac{\mu^{4-D}}{D-1} \int \frac{d^D q}{(2\pi)^D} r^i\pi \delta(p_0-H-q_0)r^i \pi T\frac{m_\text{D}^2}{|\bm{q}|^3}\theta(-q^2)\\
        &=\frac{\alpha}{6} T m_\text{D}^2 r^i\left[-\frac{1}{\epsilon}+\gamma_\text{E}-\frac{8}{3}+\ln\left(\frac{(\Delta E)^2}{\pi \mu^2}\right)\right]r^i \, .
\label{scattering_self_energy_E2_long_symm2}
\end{aligned}
\end{equation}
As expected, the ultraviolet divergence cancels the infrared divergence in eq.~\eqref{scattering_self_energy_T4} and the dependence on the renormalization scale also vanishes. 
Higher-order corrections from the dark photon propagator are suppressed. 

We handle the heavy-pair self-energy with transverse dark photons in a similar way. 
For the antisymmetric part, we write
\begin{equation}
\begin{aligned}
       \Sigma^{11,\textrm{trans},AS}_{\Delta E}(p_0) &=-i\frac{g^2}{2} \frac{\mu^{4-D}}{D-1} \int \frac{d^D q}{(2\pi)^D} r^i \frac{i}{p_0-H-q_0+i\epsilon} r^i q_0^2D_{ii}^{AS} \\
        &=-ig^2 \mu^{4-D}\frac{D-2}{D-1} \int \frac{d^D q}{(2\pi)^D} r^i \frac{i}{p_0-H-q_0+i\epsilon} r^i q_0^2~\text{P}\left[\frac{i}{q^2}\right] \, ,
\label{scattering_self_energy_E2_trans_asymm}
\end{aligned}
\end{equation}
where we have used
\begin{equation}
\begin{aligned}
    D^{AS}_{ii}(q) &= D^{R}_{ii}(q) + D^{A}_{ii}(q) \\
    &=\left(\delta_{ij}-\frac{q_iq_j}{\bm{q}^2}\right)\left[\frac{i}{q^2+\Pi^R_{\text{trans}}(q)+i\sign(q_0)\epsilon}  +\frac{i}{q^2+\Pi^A_{\text{trans}}(q)-i\sign(q_0)\epsilon}\right]\\
    &\approx 2(D-2)~\text{P}\left[\frac{i}{q^2}\right] \, ,
\label{resummed_asymm_prop_E_T}
\end{aligned}
\end{equation}
since $\Pi_{ii}^{R/A}(q) \sim m_\text{D}^2 \ll \bm{q}^2 \sim (\Delta E)^2$. 
Writing the principal value term as
\begin{equation}
    \text{P}\left[\frac{i}{q^2}\right] = \frac{i}{q^2 + i\epsilon} - \pi \delta(q^2) \, ,
\end{equation}
the imaginary part of~\eqref{scattering_self_energy_E2_trans_asymm} becomes~\cite{Brambilla:2010vq}
\begin{equation}
\begin{aligned}
        \text{Im}&\left[\Sigma^{11,\textrm{trans},AS}_{\Delta E}(p_0)\right] \\
        &=\text{Im}\left[-ig^2 \mu^{4-D}\frac{D-2}{D-1} \int \frac{d^D q}{(2\pi)^D} r^i \frac{i}{p_0-H-q_0+i\epsilon}r^i q_0^2\left(\frac{i}{q^2 + i\epsilon} - \pi \delta(q^2)\right)\right] \\
        &=-\frac{\alpha}{3}r^i(\Delta E)^3r^i \, .
\label{scattering_self_energy_E2_trans_asymm2}
\end{aligned}
\end{equation}
The calculation of the symmetric part of the scattering-state self-energy with transverse dark photons can be done analogously as in ref.~\cite{Brambilla:2010vq} for heavy quarkonium in a quark-gluon plasma, leading for the imaginary part to the result 
\begin{equation}
\begin{aligned}
        \text{Im}\left[\Sigma^{11,\textrm{trans},S}_{\Delta E}(p_0)\right]
        &= \text{Im}\left[-i\frac{g^2}{2} \frac{\mu^{4-D}}{D-1} \int \frac{d^D q}{(2\pi)^D} r^i \frac{i}{p_0-H-q_0+i\epsilon} r^i q_0^2 D^{11,S}_{ii}(q) \right] \\
        &=-\frac{\alpha}{6}T m_\text{D}^2r^i\left[1-2\ln{2}+\frac{(2\Delta E)^2}{m_\text{D}^2}\right]r^i \, ,
\label{scattering_self_energy_E2_trans}
\end{aligned}
\end{equation}
which is finite and scale-independent. 
We neglect higher-order corrections.

Summing up the longitudinal and transverse terms that contribute at the scale $\Delta E$ (cf.~eqs.~\eqref{scattering_self_energy_E2_long_symm2},~\eqref{scattering_self_energy_E2_trans_asymm2} and~\eqref{scattering_self_energy_E2_trans}), we find
\begin{equation}
\begin{aligned}
        \text{Im}&\left[\Sigma^{11}_{\Delta E}(p_0)\right] \\
        &= -\frac{\alpha}{6}T m_\text{D}^2r^i\left[\frac{1}{\epsilon}-\gamma_\text{E}+\frac{11}{3}-2\ln{2}+\frac{(2\Delta E)^2}{m_\text{D}^2}+2\frac{(\Delta E)^3}{T m_\text{D}^2}-\ln\left(\frac{(\Delta E)^2}{\pi \mu^2}\right)\right]r^i \, .
\label{scattering_self_energy_E_final}
\end{aligned}
\end{equation}
Contributions from the modes of the order of the Debye mass $m_\text{D}$ to the imaginary part of the scattering-state self-energy are at least of order $\alpha r^2 m_\text{D}^2 T \times (m_\text{D}/\Delta E)$ and hence beyond the accuracy of eq.~\eqref{scattering_self_energy_E_final}~\cite{Brambilla:2010vq}.\footnote{
In order to neglect $\alpha r^2 m_\text{D}^2 T \times (m_\text{D}/\Delta E)$ with respect to the smallest term in \eqref{scattering_self_energy_E_final}, we need to further require $g \gg (m_{\text{D}}/\Delta E)^4$.
} 
Therefore, we do not compute $\text{Im}\left[\Sigma^{11}_{m_\textrm{D}}(p_0)\right]$.

Finally, adding up the contributions from the scales $T$ in eq.~\eqref{scattering_self_energy_T4} and $\Delta E$ in eq.~\eqref{scattering_self_energy_E_final}, the total imaginary part of the self-energy of the pair for $T \gg \Delta E \gg m_{\textrm{D}}$ reads
\begin{equation}
\begin{aligned}
\text{Im}&\left[\Sigma^{11}(p_0)\right] \\
&=\frac{\alpha}{6} T m_\text{D}^2 r^i\left[2\gamma_\text{E}-3-2\frac{\zeta'(2)}{\zeta(2)}-\left(\frac{2\Delta E}{m_\text{D}}\right)^2\left(1+\frac{\Delta E}{2T}\right)+\ln\left(\frac{(\Delta E)^2}{4T^2}\right)\right]r^i \, .
\label{scattering_self_energy_hierarchy2}
\end{aligned}
\end{equation}
In eq. \eqref{scattering_self_energy_hierarchy2}, the parametrically dominant term is $-2 \alpha T r^i(\Delta E)^2 r^i/3$, 
while we display terms up to relative order $\Delta E/T$ and $(m_\text{D}/\Delta E)^2$.

\subsubsection{Hierarchy \texorpdfstring{$m_{\rm D} \sim \Delta E$}{mD ~ delta E}} 
\label{sec:mDsimE} 
We complete the section by considering the case where the Debye mass scale is of the order of $\Delta E$, which generalizes the results in the former sections~\ref{sec:mDggE} and~\ref{sec:mDllE}. 
For the hierarchies $m_\text{D}\gg \Delta E$ and $\Delta E \gg m_\text{D}$, we have shown how to obtain analytic results, see eqs.~\eqref{scattering_self_energy_hierarchy1} and~\eqref{scattering_self_energy_hierarchy2}, respectively. 
For the more general case, $m_\text{D}\sim \Delta E$, we shall instead obtain closed-form results that have to be further integrated numerically. 
To the best of our knowledge, the relaxation of the hierarchy between the binding energy and the Debye mass has not been carried out in the context of heavy quarkonium either. 

There are two main complications that arise in the extraction of the self-energy for $m_\text{D}\sim \Delta E$.  
First, we cannot expand the DM pair propagator and hence we must keep both the transverse and longitudinal parts of the electric field correlator. 
Second, the resummation of the loop corrections to the dark photon propagator is essential, since $q\sim m_\text{D}$. 
We find it convenient to compute the 21-component of the scattering-state self-energy, $\Sigma^{>}$, instead of $\text{Im}[\Sigma^{11}]$~\cite{Biondini:2023zcz}.\footnote{
We have explicitly checked that 
computing $\Sigma^{>}$ or $-2i\textrm{Im}\Sigma^{11}$ leads to the same result.} 
The corresponding self-energy reads 
\begin{equation}
\begin{aligned}
        \Sigma^{>}_{m_\text{D} \sim \Delta E}(p_0) &=ig^2 \frac{\mu^{4-D}}{D-1} \int \frac{d^D q}{(2\pi)^D} r^i 2\pi \delta(p_0-H-q_0) r^i [{\bm{q}}^2 D^{>}_{00} (q)+q_0^2 D^{>}_{ii} (q)] \\
        &= \Sigma^{>,\textrm{long}}_{m_\text{D} \sim \Delta E}(p_0)+\Sigma^{>,\textrm{trans}}_{m_\text{D} \sim \Delta E}(p_0)\, .
\label{21_scattering_self_energy}
\end{aligned}
\end{equation}

\subsubsection*{Longitudinal contribution}
First, we compute the longitudinal part. 
We abbreviate the resummed longitudinal retarded/advanced dark photon propagator in~\eqref{resummed_longitudinal_prop} as
\begin{equation}
\begin{aligned}
    D^{R/A}_{00}(q) &=  \frac{i}{\bm{q}^2 + f(q) \pm ig(q)} \, ,
\label{resummed_longitudinal_prop1}
\end{aligned}
\end{equation}
with the functions $f(q)=\text{Re}\left[\Pi^{R/A}_{00}(q)\right]$, cf.~\eqref{Re_long_pol_tensor_hierarchy2}, and $g(q)=\text{Im}\left[\Pi^{R/A}_{00}(q)\right]$, cf.~\eqref{Im_long_pol_tensor_hierarchy2}. 
The resummed longitudinal 21-propagator is then 
\begin{equation}
\begin{aligned}
    D^>_{00}(q) &=[1+n_\text{B}(q_0)]\left[D^R_{00}(q)-D^A_{00}(q)\right] \approx \frac{\pi T}{|\bm{q}|}\frac{m_\text{D}^2\theta(-q^2)}{[\bm{q}^2+f(q)]^2+g(q)^2} \, ,
\label{resummed_longitudinal_prop2}
\end{aligned}
\end{equation}
and the corresponding longitudinal part of $\Sigma^{>}$ reads
\begin{equation}
\begin{aligned}
       \Sigma^{>,\textrm{long}}_{m_\text{D} \sim \Delta E}(p_0)&= ig^2 \frac{\mu^{4-D}}{D-1} \int \frac{d^D q}{(2\pi)^D} r^i 2\pi \delta(p_0-H-q_0) r^i \bm{q}^2 D^{>}_{00} (q)  \\
        &=ig^2 \frac{\mu^{4-D}}{D-1} \int \frac{d^D q}{(2\pi)^D} r^i 2\pi \delta(\Delta E-q_0) r^i  \bm{q}^2\frac{\pi T}{|\bm{q}|}\frac{m_\text{D}^2\theta(-q^2)}{[\bm{q}^2+f(q)]^2+g(q)^2} \, ,
\label{21_scattering_self_energy_long}
\end{aligned}
\end{equation}
where, as in the previous sections, $\Delta E = p_0-H$ is an operator. 
The ultraviolet divergence in~\eqref{21_scattering_self_energy_long} cancels the infrared divergence in $\Sigma^{>}_{T}=-2i~\text{Im}\left[\Sigma^{11}_{T}\right]$, cf.~eq.~\eqref{scattering_self_energy_T4}. 
In order to see the cancellation explicitly, we extract the divergent part out of the self-energy in eq.~\eqref{21_scattering_self_energy_long} by writing
 \begin{equation}
\begin{aligned}
       \Sigma^{>,\textrm{long}}_{m_\text{D} \sim \Delta E}(p_0) &= ig^2 \frac{\mu^{4-D}}{D-1} \int \frac{d^D q}{(2\pi)^D} r^i 2\pi \delta(\Delta E-q_0) r^i \frac{\pi T}{|\bm{q}|^3}m_\text{D}^2\theta(-q^2) \\
        &\hspace{3cm}\times\left[1+\frac{1-[1+f(q)/\bm{q}^2]^2-[g(q)/\bm{q}^2]^2}{[1+f(q)/\bm{q}^2]^2+[g(q)/\bm{q}^2]^2} \right]\\
        &=\Sigma^{>,\textrm{long}, \epsilon}_{m_\text{D} \sim \Delta E}(p_0)+\Sigma^{>,\textrm{long}, \slashed{\epsilon}}_{m_\text{D} \sim \Delta E}(p_0)\, ,
\label{21_scattering_self_energy_long2}
\end{aligned}
\end{equation}
where we have indicated the divergent (finite) part with the superscript $\epsilon$ ($\slashed{\epsilon}$).
The integration of the first term in the square bracket, which gives the divergent part, can be done as in~\eqref{scattering_self_energy_E2_long_symm2}, and we obtain
\begin{equation}
\begin{aligned}
       \Sigma^{>,\textrm{long}, \epsilon}_{m_\text{D} \sim \Delta E}(p_0)
        &=i\frac{\alpha}{3} T m_\text{D}^2 r^i\left[\frac{1}{\epsilon}-\gamma_\text{E}+\frac{8}{3}-\ln\left(\frac{(\Delta E)^2}{\pi \mu^2}\right)\right]r^i \, .
\label{21_scattering_self_energy_long2_div}
\end{aligned}
\end{equation}
The result of the integration of the second term in the square bracket, which is finite, can be written as follows
\begin{equation}
\begin{aligned}
       \Sigma^{>,\textrm{long}, \slashed{\epsilon}}_{m_\text{D} \sim \Delta E}(p_0)
        &=i\frac{2}{3}\alpha 
        T m_\text{D}^2 r^i\mathcal{X}_l\left(\frac{\Delta E}{m_\text{D}}\right)r^i \, ,
\label{21_scattering_self_energy_long2_fin}
\end{aligned}
\end{equation}
where we have introduced the integral
\begin{equation}
        \mathcal{X}_l(x)\equiv \int_1^{\infty} dt \frac{1}{t}\frac{(xt)^4-[(xt)^2+\hat{f}(t)]^2-\hat{g}(t)^2}{[(xt)^2+\hat{f}(t)]^2+\hat{g}(t)^2} \, ,
    \label{dimless_xl}
\end{equation}
over the dimensionless variable $t\equiv |\bm{q}|/\Delta E$ and the auxiliary functions
\begin{equation}
        \hat{f}(t)\equiv 1+\frac{1}{2t}\ln{\left|\frac{1-t}{1+t}\right|} \, , 
        \qquad \qquad \hat{g}(t)\equiv \frac{\pi}{2t} \, .
\end{equation}
Hence, summing the quantities in eqs.~\eqref{21_scattering_self_energy_long2_div} and~\eqref{21_scattering_self_energy_long2_fin}, the longitudinal part of the heavy-pair self-energy reads
 \begin{equation}
\begin{aligned}
        \Sigma^{>,\textrm{long}}_{m_\text{D} \sim \Delta E}(p_0) &= i\frac{\alpha}{3}Tm_\text{D}^2r^i\left[\frac{1}{\epsilon}-\gamma_\text{E}+\frac{8}{3}-\ln\left(\frac{(\Delta E)^2}{\pi \mu^2}\right)+2\mathcal{X}_l\left(\frac{\Delta E}{m_\text{D}}\right)\right]r^i \\
        &=-2i~\text{Im}\left[\Sigma^{11,\textrm{long}}_{m_\text{D} \sim \Delta E}(p_0)\right] \, .
\label{21_scattering_self_energy_long3}
\end{aligned}
\end{equation}
Comparing the result with the longitudinal self-energy in eq.~\eqref{scattering_self_energy_E2_long_symm2}, which comes from integrating out modes of order $\Delta E \gg m_\text{D}$, 
we observe that the difference is due to the additional term proportional to $-2\mathcal{X}_l(\Delta E/m_\text{D})$.
It is a correction due to the less strict relation $\Delta E \sim m_\text{D}$ and, indeed, in the limit $\Delta E/m_\text{D}\to \infty$ the function $\mathcal{X}_l$ vanishes.

\subsubsection*{Transverse contribution}
We consider now the finite transverse part in~\eqref{21_scattering_self_energy}:
\begin{equation}
\begin{aligned}
      \Sigma^{>,\textrm{trans}}_{m_\text{D} \sim \Delta E}(p_0)&= i\frac{g^2}{3}  \int \frac{d^4 q}{(2\pi)^4} r^i 2\pi \delta(p_0-H-q_0) r^i q_0^2 D^{>}_{ii} (q) \, .
\label{21_scattering_self_energy_trans}
\end{aligned}
\end{equation}
Starting from the 21-component of the photon propagator
\begin{equation}
\begin{aligned}
    D^>_{ii}(q) &=[1+n_\text{B}(q_0)]\left[D^R_{ii}(q)-D^A_{ii}(q)\right] \, ,
\label{resummed_trans_prop1}
\end{aligned}
\end{equation}
and using the resummed retarded/advanced propagators in eq.~\eqref{resummed_transversal_prop}, we obtain 
(expanding in $\Delta E/T$ and keeping the leading order term)
\begin{equation}
\begin{aligned}
         \Sigma^{>,\textrm{trans}}_{m_\text{D} \sim \Delta E}(p_0)&= i\frac{2}{3}g^2 Tr^i \Delta E \int \frac{d^3 q}{(2\pi)^3} \left[\frac{i}{(\Delta E)^2 - \bm{q}^2 +h(\Delta E,\bm{q}) + ik(\Delta E,\bm{q}) + i\epsilon} \right. \\
        &\hspace{3cm} \left. - \frac{i}{(\Delta E)^2 - \bm{q}^2 +h(\Delta E,\bm{q}) - ik(\Delta E,\bm{q}) - i\epsilon}\right] r^i \, ,
\label{21_scattering_self_energy_trans1}
\end{aligned}
\end{equation}
where we have defined the functions $h(q)=\text{Re}\left[\Pi^{R/A}_{\rm trans}(q)\right]$, cf.~\eqref{Re_trans_pol_tensor_hierarchy2}, and $\pm k(q)=\text{Im}\left[\Pi^{R/A}_{\rm trans}(q)\right]$, cf.~\eqref{Im_trans_pol_tensor_hierarchy2}. 
Accordingly, we write eq.~\eqref{21_scattering_self_energy_trans1} as 

\begin{equation}
       \Sigma^{>,\textrm{trans}}_{m_\text{D} \sim \Delta E}(p_0) = i\frac{2}{3}\alpha
       Tm_\text{D}^2 r^i\mathcal{X}_t\left(\frac{\Delta E}{m_\text{D}}\right)r^i \\ =-2i~\text{Im}\left[\Sigma^{11,\textrm{trans}}_{m_\text{D} \sim \Delta E}(\Delta E)\right] \, ,
\label{21_scattering_self_energy_trans2}
\end{equation}
where we have defined the finite integral
\begin{equation}
\begin{aligned}
        \mathcal{X}_t(x)&\equiv \frac{2}{\pi}x^4\int_0^{\infty} dt ~t^2\left[\frac{i}{(1-t^2)x^2-\hat{h}(t)+i\hat{k}(t)+i\epsilon}-\frac{i}{(1-t^2)x^2-\hat{h}(t)-i\hat{k}(t)-i\epsilon}\right] \\
        &= \frac{2}{\pi}x^4\left[\int_0^1 dt~t^2 2\pi \delta[(1-t^2)x^2-\hat{h}(t)] + \int_1^\infty dt~t^2\frac{2\hat{k}(t)}{[(1-t^2)x^2-\hat{h}(t)]^2+\hat{k}(t)^2} \right]\, ,
\label{dimless_integral}
\end{aligned}
\end{equation}
over the dimensionless integration variable $t\equiv |\bm{q}|/\Delta E$ and the functions
\begin{equation}
        \hat{h}(t)\equiv \frac{1}{2t^2}\left[1-\frac{1}{2t}(1-t^2)\ln{\left|\frac{1+t}{1-t}\right|}\right] \, , 
        \qquad\qquad \hat{k}(t)\equiv \frac{\pi}{4t^3}(t^2-1)\theta(t^2-1) \, .
\label{hhatkhat}        
\end{equation}
In the last equality in eq.~\eqref{dimless_integral}, we have split the integral into two integration regions that can be associated with distinct physical processes contributing to heavy-pair transitions.
By considering bound-state formation, the first term in eq.~\eqref{dimless_integral} contributes to the radiative formation via the emission of a 
time-like dark photon with momentum $|\bm{q}| < \Delta E$, namely through the process $(X \bar{X})_p \to (X \bar{X})_n + \gamma^*$  shown in diagram $b)$ of figure~\ref{fig:bsf_different}.\footnote{
The labels $p$ and $n$ identify scattering and bound states, respectively.
} 
The second term in eq.~\eqref{dimless_integral} contributes to the $2 \to 2$ scattering process with light dark fermions from the plasma, the process being $(X \bar{X})_p + f \to (X \bar{X})_n +f$ shown in diagram $c)$ of  figure~\ref{fig:bsf_different}.
In this case, the interaction is mediated by a space-like dark photon with momentum $|\bm{q}| > \Delta E$. 
The $1\to 3$ process shown in diagram $d)$ of figure~\ref{fig:bsf_different} is kinematically forbidden in the temperature regime that we consider here, $T\gg \Delta E$. 
This is signaled by the theta function in the expression of $\hat{k}$ in \eqref{hhatkhat} 
originating from $\theta (-q^2)$ in eq.~\eqref{Im_trans_pol_tensor_hierarchy2}, 
which explicitly excludes the region $|\bm{q}| < \Delta E$, 
where the production of two thermal fermions, i.e. with total momentum $|\bm{q}| \sim T$, would happen.

Summarizing the result for the total self-energy, the sum of eqs.~\eqref{21_scattering_self_energy_long3} and~\eqref{21_scattering_self_energy_trans2} leads to
 \begin{equation}
\begin{aligned}
        \Sigma^{>}_{m_\text{D} \sim \Delta E}&(p_0) \\
        &= i\frac{\alpha}{3}Tm_\text{D}^2r^i\left[\frac{1}{\epsilon}-\gamma_\text{E}+\frac{8}{3}-\ln\left(\frac{(\Delta E)^2}{\pi \mu^2}\right)+2\mathcal{X}_l\left(\frac{\Delta E}{m_\text{D}}\right)+2\mathcal{X}_t\left(\frac{\Delta E}{m_\text{D}}\right)\right]r^i \\
        &=-2i~\text{Im}\left[\Sigma^{11}_{m_\text{D} \sim \Delta E}(p_0)\right] \, ,
\label{21_scattering_self_energy_mD_E}
\end{aligned}
\end{equation}
and adding the contribution in~\eqref{scattering_self_energy_T4} coming from the scale $T$, we end up with
\begin{equation}
\begin{aligned}
\text{Im}\left[\Sigma^{11}(p_0)\right]
&=\frac{\alpha}{3} T m_\text{D}^2 r^i\left[\gamma_\text{E}-1-\frac{\zeta'(2)}{\zeta(2)}+\ln\left(\frac{\Delta E}{4T}\right)-\mathcal{X}_l\left(\frac{\Delta E}{m_\text{D}}\right)-\mathcal{X}_t\left(\frac{\Delta E}{m_\text{D}}\right)\right]r^i \, .
\label{scattering_self_energy_hierarchy3}
\end{aligned}
\end{equation}
Having completed the calculation of the self-energy of a heavy pair in various hierarchies of scales, we proceed in the next sections by selecting particular combinations of the incoming and outgoing pairs. 
In doing so, we compute the observables of interest, namely the bound-state formation cross section, bound-state dissociation and bound-state to bound-state transition widths.

\subsection{Bound-state formation} 
\label{sec:bsf}
The bound-state formation (bsf) cross section can be computed from the imaginary part of the time-ordered scattering-state self-energy using the optical theorem, see eq.~\eqref{bsf_projected}. 
At one loop, it is the sum of the contribution coming from the in-vacuum photon polarization, 
which has been computed in eq.~\eqref{bsf_final_NLO}, 
and the contribution coming from the thermal part of the photon polarization.

If $T \lesssim \Delta E$, the photon polarization in the self-energy diagram does not develop a Debye mass and the computation can be done in a strict perturbative expansion.
For the hierarchy $T \lesssim \Delta E$, the bound-state cross section at NLO has been computed in appendix~\ref{sec:app_C} and can be read off eq.~\eqref{bsf_final_fixed_NLO}, which includes in vacuum and thermal contributions. 
The situation changes if $T \gg \Delta E$ because then the photon polarization develops a Debye mass that becomes a relevant scale of the system.

For the hierarchy  $T \gg m_\text{D}\gg \Delta E$,
adding eq.~\eqref{bsf_final_NLO} to the contribution coming from the thermal part of the photon polarization given in eq.~\eqref{scattering_self_energy_hierarchy1} leads to
\begin{equation}
\begin{aligned}
(\sigma_{\text{bsf}} \, &v_{\hbox{\scriptsize rel}})_{m_\text{D}\gg \Delta E}(\bm{p}) \\
&=
\sum_n (\sigma^n_{\text{bsf}} \, v_{\hbox{\scriptsize rel}})^{\text{LO}}(\bm{p})\; 
\left\{
\frac{n_f}{3\pi}\alpha
\left[\ln{\left(\frac{4(\Delta E^p_n)^2}{\mu^2}\right)}-\frac{10}{3}\right] \right.\\
&\hspace{2cm}
\left. +\left(\frac{m_\text{D}}{2\Delta E^p_n}\right)^2\left[1-2\gamma_\text{E}+2\frac{\zeta'(2)}{\zeta(2)}-\ln\left(\frac{m_\text{D}^2}{16T^2}\right)  + \left(\frac{2 \Delta E^p_n}{m_\text{D}}\right)^2\right]\right\}
\, ,
\label{bsf_final_hierarchy1}
\end{aligned}
\end{equation}
where $(\sigma^n_{\text{bsf}} \, v_{\hbox{\scriptsize rel}})^{\text{LO}}(\bm{p})$
is the bound-state formation cross section for a given bound state $|n\rangle$ with specific quantum numbers $n$ due to the emission of one real photon (diagram $a)$ of figure~\ref{fig:bsf_different}).
It can be read off eq.~\eqref{bsf_final_LO}.
The energy $\Delta E^p_n$ is the difference between the energy of the incoming unbound dark heavy fermion  pair with relative momentum $\bm{p}$ and the outgoing bound state, i.e. $\Delta E^p_n \equiv \bm{p}^2/M + M\alpha^2/(4n^2)$.\footnote{
In the expression of the self-energy the operator $\Delta E = p_0-H$ becomes the energy difference $\Delta E^p_n \geq 0$ upon inserting a complete set of bound states, $\displaystyle \mathbb{1} =\sum_n|n\rangle \langle n|$.}

Similarly, for the hierarchy $T \gg \Delta E \gg m_\text{D}$, the Debye mass resummation results in the contribution to the self-energy given in~eq.~\eqref{scattering_self_energy_hierarchy2}.
Adding it to eq.~\eqref{bsf_final_NLO}, the bound-state formation cross section reads
\begin{equation}
\begin{aligned}
(\sigma_{\text{bsf}} \, &v_{\hbox{\scriptsize rel}})_{\Delta E \gg m_\text{D}}(\bm{p}) \\
=&\sum_n (\sigma^n_{\text{bsf}} \, v_{\hbox{\scriptsize rel}})^{\text{LO}}(\bm{p})
\left\{
\frac{n_f}{3\pi}\alpha
\left[\ln{\left(\frac{4(\Delta E^p_n)^2}{\mu^2}\right)}-\frac{10}{3}\right] \right.\\
&\hspace{0cm}
\left. +\left(\frac{m_\text{D}}{2\Delta E^p_n}\right)^2\left[3-2\gamma_\text{E}+2\frac{\zeta'(2)}{\zeta(2)}+\left(\frac{2\Delta E^p_n}{m_\text{D}}\right)^2\left( 1+ \frac{\Delta E^p_n}{2T}\right)
-\ln\left(\frac{(\Delta E^p_n)^2}{4T^2}\right)\right]\right\}
\, .
\label{bsf_final_hierarchy2}
\end{aligned}
\end{equation}
Note that this is the only situation where $(\sigma_{\text{bsf}} \, v_{\hbox{\scriptsize rel}})^{\text{LO}}_{n}(\bm{p})$ is indeed the leading contribution to the cross section.

For the general case $T \gg m_\text{D} \sim \Delta E$,
the contribution to the self-energy coming from the thermal part of the photon polarization is given in eq.~\eqref{scattering_self_energy_hierarchy3}.
Adding it to eq.~\eqref{bsf_final_NLO}, the bound-state formation cross section results in 
\begin{equation}
\begin{aligned}
&(\sigma_{\text{bsf}} \, v_{\hbox{\scriptsize rel}})_{m_\text{D} \sim \Delta E}(\bm{p}) \\
&=\sum_n (\sigma^n_{\text{bsf}} \, v_{\hbox{\scriptsize rel}})^{\text{LO}}(\bm{p})
\left\{\frac{n_f}{3\pi}\alpha
\left[\ln{\left(\frac{4(\Delta E^p_n)^2}{\mu^2}\right)}-\frac{10}{3}\right] \right.\\
&\hspace{1cm}
\left. +\left(\frac{m_\text{D}}{2\Delta E^p_n}\right)^2\left[2-2\gamma_\text{E}+2\frac{\zeta'(2)}{\zeta(2)}-\ln\left(\frac{(\Delta E^p_n)^2}{16T^2}\right)+2\mathcal{X}_l\left(\frac{\Delta E^p_n}{m_\text{D}}\right)+2\mathcal{X}_t\left(\frac{\Delta E^p_n}{m_\text{D}}\right)\right]\right\}
\, .
\label{bsf_final_hierarchy3}
\end{aligned}
\end{equation}

The NLO terms in the second lines of eqs.~\eqref{bsf_final_hierarchy1},~\eqref{bsf_final_hierarchy2} and~\eqref{bsf_final_hierarchy3} are suppressed by $n_f\alpha/\pi$ with respect to the LO radiative photon emission. 
As shown in appendix~\ref{sec:app_C}, the $\mu$ dependence in the NLO terms cancels against the $\mu$ dependence of $\alpha$ in $(\sigma^n_{\text{bsf}} \, v_{\hbox{\scriptsize rel}})^{\text{LO}}(\bm{p})$.
In the case of eq.~\eqref{bsf_final_hierarchy3}, we have verified the cancellation of the $\mu$ dependence numerically, 
as $(\sigma^n_{\text{bsf}}\, v_{\hbox{\scriptsize rel}})^{\text{LO}}(\bm{p})$ is contained in the term originating from the transverse photon, proportional to~$\mathcal{X}_t$.
In the bound-state formation cross section formulas, the {\it natural} renormalization scale for the coupling associated with the electric dipole interaction is of the order of the ultrasoft scale $M \alpha^2$, while for the one contained in the wave function and in the binding energy it is of the order of the soft scale $M \alpha$.
In the following applications, we set the coupling at these values.

Next, we perform the thermal average of $v_{\hbox{\scriptsize rel}}$ times the bound-state formation cross section with respect to the relative velocity of the incoming heavy DM pair.\footnote{
We thermally average in the laboratory frame and since we omit the center-of-mass motion in this work, the relative velocity in the laboratory frame coincides with the one in the center-of-mass frame, see \cite{Gondolo:1990dk,Biondini:2024aan}.} 
Thermally averaged cross sections enter the evolution equations, cf. section~\ref{sec:DMevolution}. 
Since we assume kinetically equilibrated non-relativistic DM pairs, the distribution of the incoming particle and antiparticle is the Maxwell--Boltzmann distribution~\cite{Gondolo:1990dk}.

\begin{figure}[ht]
    \centering
    \includegraphics[scale=0.78]{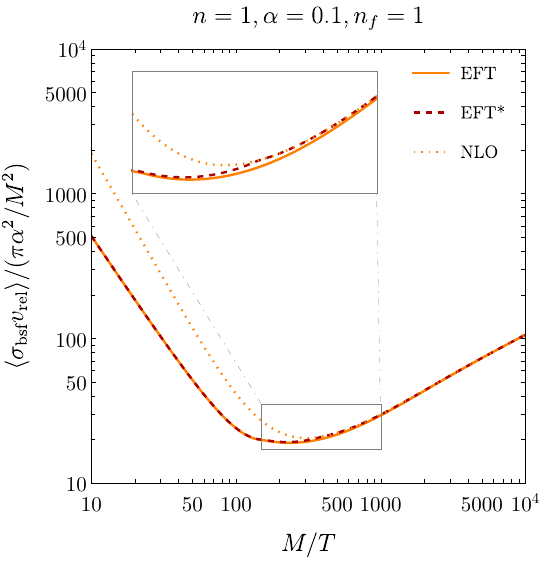}
    \hspace{0.5cm}
    \includegraphics[scale=0.78]{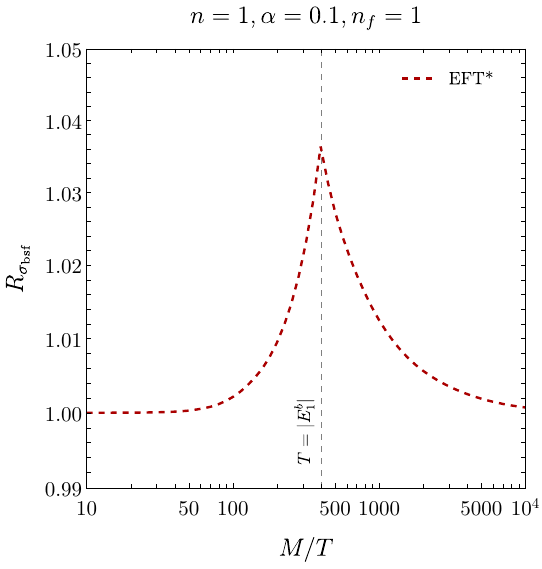}
    \caption{(Left)  Thermally averaged bound-state formation cross section for the ground state up to next-to-leading order, eq.~\eqref{bsf_final_fixed_NLO}, (orange dotted line, labeled NLO), 
    with Debye mass resummation, eq.~\eqref{bsf_final_hierarchy3}, (orange solid line, labeled EFT) 
    and with thermal average split into the region $T<\Delta E^p_1$, where the cross section is given by~\eqref{bsf_final_fixed_NLO}, and into the region $T \ge \Delta E^p_1$, where the cross section is given by~\eqref{bsf_final_hierarchy3}, (red dashed line, labeled EFT$^*$) as a function of $M/T$ for coupling $\alpha(2M)=0.1$ and $n_f=1$. (Right) The ratio of the thermally averaged cross sections labeled EFT$^*$
    and EFT of the previous plot.}
    \label{fig:bsf_cut_vrel}
\end{figure}

The thermal average then reads
\begin{equation}
        \langle \sigma_{\text{bsf}} \, v_{\hbox{\scriptsize rel}}\rangle = \sqrt{\frac{2}{\pi}}\left(\frac{M}{2T}\right)^{3/2}\int_0^\infty dv_{\hbox{\scriptsize rel}} \, v_{\hbox{\scriptsize rel}}^2 \, e^{-\frac{Mv_\text{rel}^2}{4T}} \sigma_{\text{bsf}} \, v_{\hbox{\scriptsize rel}} \, ,
\label{Boltzmann_distr}
\end{equation}
where we have changed variable from $p$ to $v_{\hbox{\scriptsize rel}}$ according to $p = M v_{\hbox{\scriptsize rel}}/2$. 
Because $v_{\hbox{\scriptsize rel}}$ is just a rescaling of the momentum, it can assume all positive values.
Nevertheless, it coincides with the relative velocity in the non-relativistic limit.
Since the integral in \eqref{Boltzmann_distr} goes over all $v_{\hbox{\scriptsize rel}}$, we should split it in a region where $v_{\hbox{\scriptsize rel}}$ is such that $T \ge \Delta E^p_n$ and the bound-state formation cross section is given by the resummed expression of eq.~\eqref{bsf_final_hierarchy3}, 
and in a region where $v_{\hbox{\scriptsize rel}}$ is such that $T<\Delta E^p_n$
and the bound-state formation cross section is given by the NLO expression of eq.~\eqref{bsf_final_fixed_NLO}.
Since for $T \ge \Delta E^p_n$ we take the expression in eq.~\eqref{bsf_final_hierarchy3}, which is valid for any hierarchy between $m_\text{D}$ and $\Delta E^p_n$, 
we do not need to further distinguish between the $v_{\hbox{\scriptsize rel}}$ region for which $m_\text{D} \ge \Delta E^p_n$ and the one for which $m_\text{D} < \Delta E^p_n$.
The effect of splitting the integral in~\eqref{Boltzmann_distr} into the two $v_{\hbox{\scriptsize rel}}$ regions $T \ge \Delta E^p_n$ and $T < \Delta E^p_n$ is, however, tiny compared to using the bound-state formation cross section \eqref{bsf_final_hierarchy3} over the whole range of relative velocities.
We show this for the 1S state in figure~\ref{fig:bsf_cut_vrel}.
In the left plot, we compare the thermal average of the bound-state formation cross section obtained from  \eqref{bsf_final_hierarchy3} over the whole $v_{\hbox{\scriptsize rel}}$ range (orange line) with the one obtained splitting the $v_{\hbox{\scriptsize rel}}$ integral and taking for $T<\Delta E^p_n$ the cross section from \eqref{bsf_final_fixed_NLO} (red dashed line).
In the right plot, we show the ratio of these two determinations. 
The effect is at most 4\% in the region around $T=|E^b_1|$ and fades away rather rapidly.
The effect becomes even smaller in the effective cross section, see figure~\ref{fig:effcsnew}, and completely negligible in the relic density.
The reason is that the LO result is already a very good approximation of the full result at low temperatures, 
and the LO result is included in \eqref{bsf_final_hierarchy3}.
For these reasons and in order to keep more transparent the comparison between the resummed expressions and the 
fixed order ones, we have chosen to present our results for the thermal average of the bound-state formation cross section using the expression \eqref{bsf_final_hierarchy3} over the whole range of relative velocities 
and analogously for the bound-state dissociation width, unless differently specified. 
Moreover, we use those results to compute the DM relic density.

\begin{figure}[ht]
    \centering
    \includegraphics[scale=0.78]{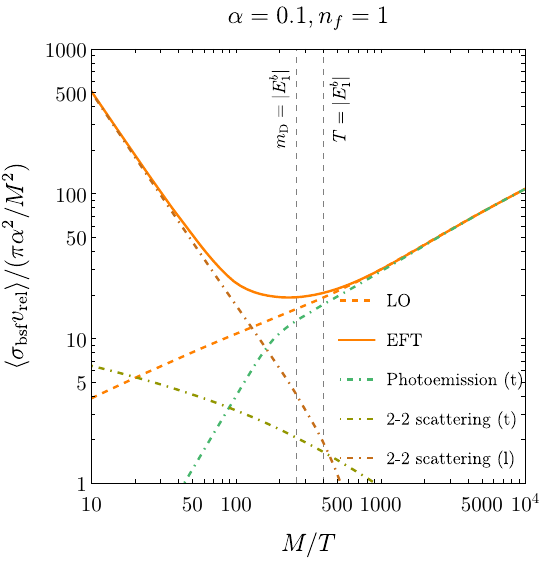}
    \hspace{0.5cm}
    \includegraphics[scale=0.78]{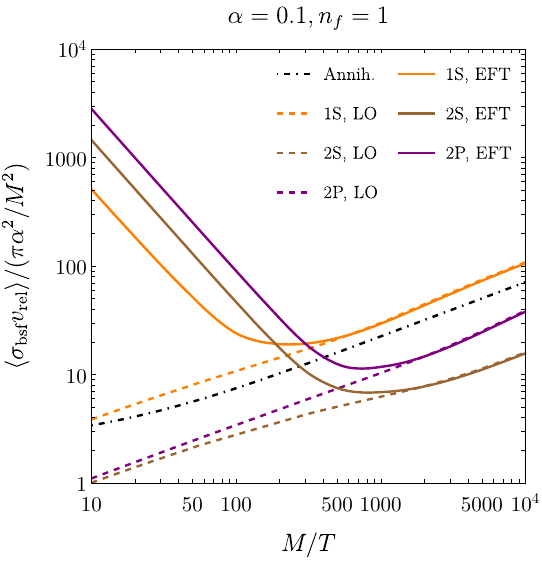}
    \caption{(Left) Thermally averaged bound-state formation cross section for the 1S state computed from eq.~\eqref{bsf_final_hierarchy3} (orange solid line) plotted as a function of $M/T$ for $n_f=1$ and running coupling with starting value $\alpha(2M)=0.1$.
    The dash-dotted cyan, green and red lines denote the individual contributions accounting for photo-emission and bath-particle scattering via transverse and longitudinal dark photons, respectively. 
    The orange dashed line depicts the cross section at LO. 
    The cross sections are normalized by the free annihilation cross section, $\pi \alpha^2/M^2$. 
    The vertical dashed lines mark the positions where $m_\text{D} = |E^b_1|$ and $T = |E^b_1|$, respectively, where $E^b_n=-M\alpha^2/(4n^2)$ are the Bohr levels.
    (Right) Thermally averaged annihilation (black dash-dotted line) and bound-state formation cross sections for the 1S (orange lines), 2S (brown lines) and 2P state (purple lines). 
    Dashed lines are for the results at LO, cf.~\eqref{bsf_final_LO}, whereas solid lines are obtained from~\eqref{bsf_final_hierarchy3}.}
    \label{fig:bsf_individual_contr}
\end{figure}

Let us consider the formation of the ground state 1S. 
We thermally average the expression in eq.~\eqref{bsf_final_hierarchy3} for the 1S state, where $\Delta E^p_1=\bm{p}^2/M +M\alpha^2/4$, and plot it together with partial contributions in figure~\ref{fig:bsf_individual_contr} left, normalized by the free annihilation cross section $\pi \alpha^2/M^2$, as a function of $M/T$ and for the coupling $\alpha(2M)=0.1$. 
The coupling runs at one loop and we consider only one light dark fermion species in the bath, i.e. we set $n_f=1$. 
The cyan dash-dotted line displays the contribution from the first term in the bracket in eq.~\eqref{dimless_integral}, which corresponds to the bound-state formation process via on-shell emission of a thermal transverse dark photon. 
It increases with decreasing temperature, eventually approaching the orange dashed line representing the thermally averaged bound-state formation cross section at leading order, cf.~eq.~\eqref{bsf_final_LO}, which exclusively accounts for the thermal photo-emission process. 
The contribution from the second term in the bracket in eq.~\eqref{dimless_integral} (green dash-dotted line) corresponds to a $2 \to 2$ scattering process with a dark light fermion from the bath due to a transverse off-shell dark photon exchange. 
Instead, the bath-particle scattering via longitudinal dark photons (red dash-dotted line), coming from the first five terms in the bracket in the third line of eq.~\eqref{bsf_final_hierarchy3}, is the dominant process for large temperatures. 
Overall, the $2 \to 2$ scatterings between heavy pairs and light thermal degrees of freedom, which originate the Landau damping phenomenon, are the more important contributions to the bound-state formation in the regime $m_\text{D} > |E^b_1|$, while bound-state formation via photon emission dominates for  $m_\text{D} < |E^b_1|$.
All of the partial contributions are either monotonically increasing (thermal photo-emission process) or decreasing (longitudinal/transverse bath-particle scattering) functions with decreasing~$T$. 

In the right panel of figure~\ref{fig:bsf_individual_contr}, we add to the 1S state (orange lines) also the results for the first excited bound states: 2S (brown lines) and 2P state (purple lines), where for the latter we have summed over the magnetic quantum numbers $m=-1,0,1$. 
Dashed lines denote the bound-state formation as obtained from the leading-order expression in~eq.~\eqref{bsf_final_LO}.
The solid lines depict the results from eq.~\eqref{bsf_final_hierarchy3}, 
where Debye mass effects have been resummed.
Although the curves for the different bound states share the same behavior as a function of~$T$, there is still a significant difference: 
for small $T$ the dominant process is the formation of the ground state, but, for large $T$ the bound-state formation processes of the excited states dominate. 
Hence, we observe the Landau damping phenomenon to be more significant for excited bound states.
For comparison, we also plot the thermally and spin-averaged annihilation cross section  (black dash-dotted line). 

\begin{figure}[ht]
    \centering
    \includegraphics[scale=0.78]{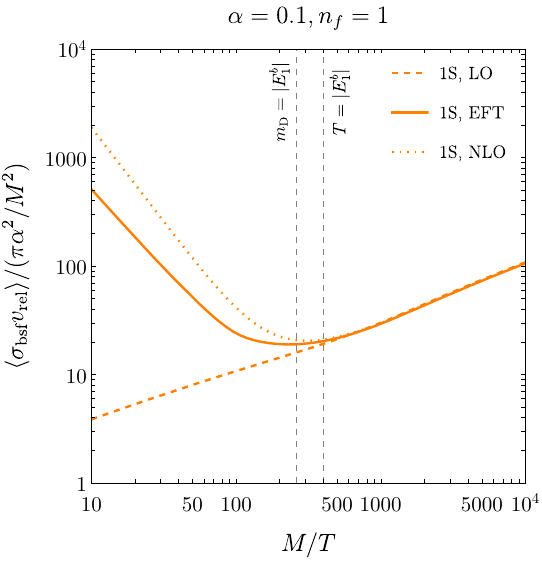}
    \hspace{0.5cm}
    \includegraphics[scale=0.745]{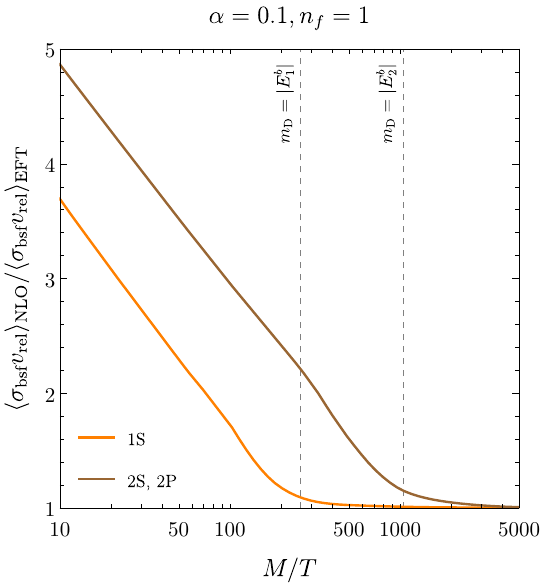}
    \caption{(Left) Thermally averaged bound-state formation cross section for the 1S state at LO (orange dashed line, cf.~\eqref{bsf_final_LO}), up to NLO (orange dotted line, cf.~\eqref{bsf_final_fixed_NLO}) and with Debye mass resummation (orange solid line, cf.~\eqref{bsf_final_hierarchy3}), plotted as a function of $M/T$ for $n_f=1$ and running coupling with starting value $\alpha(2M)=0.1$, normalized by the free annihilation cross section $\pi \alpha^2/M^2$. 
    The vertical dashed lines mark the positions, from left to right, where $m_\text{D} =  |E^b_1|$ and $T=  |E^b_1|$.
    (Right) The ratio of the thermally averaged bound-state formation cross section up to NLO, cf.~\eqref{bsf_final_fixed_NLO}, to the one obtained by resumming the Debye mass, cf.~\eqref{bsf_final_hierarchy3}, for the ground state (orange solid line) and 2S, 2P states (brown solid line). The vertical dashed lines mark the positions where $m_\text{D} = |E^b_1|$ and $m_\text{D} = |E^b_2|$.
    }
    \label{fig:bsf_comparison}
\end{figure}

Next, we compare the thermally averaged bound-state formation cross section obtained upon resummation of the Debye mass, cf.~eq.~\eqref{bsf_final_hierarchy3}, with the corresponding expressions at fixed order, eq.~\eqref{bsf_final_fixed_NLO}. 
In figure~\ref{fig:bsf_comparison} left, we plot the thermally averaged bound-state formation cross section for the ground state at leading order (orange dashed line, cf.~\eqref{bsf_final_LO}), the bound-state formation cross section at next-to-leading order (orange dotted line, cf.~\eqref{bsf_final_fixed_NLO}), and with resummed Debye mass (orange solid line) as functions of $M/T$. 
The last two are also in the left panel of figure~\ref{fig:bsf_cut_vrel}.
We choose again $n_f=1$ and $\alpha=0.1$ at the hard scale $2M$, and run the coupling at one loop. 
The orange solid and dotted lines approach the orange dashed line with decreasing temperature. 
This indeed meets expectations, since at small $T$ the dominant bound-state formation process is via on-shell emission of a thermal dark photon. 
However, for large temperatures, where the thermal scale $m_\text{D}$ becomes important and eventually needs to be resummed, we observe a corresponding smaller cross section (orange solid line) with respect to the cross section at NLO (orange-dotted curve). 
In the right panel of figure~\ref{fig:bsf_comparison}, we plot the ratio between the two cross sections for the ground state (orange solid line) and the first excited states 2S and 2P (brown solid line).\footnote{
Since the excited states 2S and 2P have the same binding energy, the ratio 
is the same for the two states.}
We observe an overestimation of the bound-state formation cross section at NLO compared to the cross section obtained from the EFT treatment by a factor of up to five in the large $T$ region. 
The reduction factor that is expected from the resummation of the Debye mass was already mentioned in ref. \cite{Binder:2021otw}, where it was estimated by a logarithmic factor originally introduced in~\cite{Burnier:2010rp}. 
Here we have computed explicitly the two different predictions for the bound-state formation cross section. 
As we show in section~\ref{sec:DMevolution}, the discrepancy impacts the dark matter density evolution in the early universe. 
When the scale $m_\text{D}$ approaches and gets larger than the one associated with $\Delta E$, the HTL resummation becomes essential to describe accurately the formation of bound states. 
With the EFT approach and the Debye mass resummation, we are able to estimate more reliably the bound-state formation process up to temperatures $T \lesssim M\alpha$.

One additional comment:  In \cite{Binder:2020efn}, it was remarked that the next-to-leading order contribution becomes significant for strong couplings and/or a large number of light dark fermions. 
In this situation, it may happen that $\sqrt{\pi n_f\alpha} \sim 1$, 
for which $T\sim m_\text{D}$ and the plasma behaves as a {\it strongly coupled plasma}.
The treatment of a strongly coupled plasma requires a dedicated study, which is beyond the scope of the present work.

\subsection{Bound-state dissociation and bound-state to bound-state transitions}
\label{sec:bsd}
The dissociation process of bound states into unbound pairs (bsd), either through absorption of a thermal dark photon from the bath (called photo-dissociation) or through scatterings with the other constituents from the bath,  can be studied in a similar manner as the bound-state formation process. 
The dissociation width may be computed from the self-energy diagram in figure~\ref{fig:bsf_two_loop_fixed_NLO} (right).
The dissociation width, $\Gamma^n_\text{bsd}$, follows from projecting the self-energy onto bound states with quantum numbers $n$ and employing the optical theorem, 
\begin{equation}
\Gamma^n_\text{bsd} = -2\,\langle n| {\rm{Im}}[\Sigma^{11}(p_0)] |n\rangle = \langle n|[-i\Sigma^{21}(p_0)]|n\rangle \, .
  \label{bsd_projected}
\end{equation}
Written in terms of the 21 component of the dark electric-field correlator and projecting on intermediate scattering states with relative momentum $\bm{p}$, it reads, similarly to~\eqref{bsf_final_form}, 
\begin{equation}
\begin{aligned}
\Gamma^n_\text{bsd} = g^2 \frac{\mu^{4-D}}{D-1}\int\frac{d^3p}{(2\pi)^3}|\langle n|\bm{r}|\bm{p}\rangle|^2 \int \frac{d^{D-1} q}{(2\pi)^{D-1}}  \langle \bm{E}  \bm{E} \rangle^{>}(-\Delta E^p_n , \bm{q})  \, .
\end{aligned}
\label{bsd_final_form}
\end{equation}
Note that the energy argument in the electric correlator comes with a relative sign difference with respect to the energy argument in the bound-state formation cross section. 

The leading order photo-dissociation width comes from taking the photon propagator without light fermion loop corrections. 
At next-to-leading order, following closely the derivation of the bound-state formation cross section in appendix \ref{sec:app_C}, we obtain the bound-state dissociation width
\begin{equation}
\begin{aligned}
\Gamma_\text{bsd}^{n, \scalebox{0.6}{\text{LO}+\text{NLO}}}
&= 2\int\frac{d^3p}{(2\pi)^3}\sigma^{n,\scalebox{0.6}{\text{LO}}}_{\hbox{\scriptsize bsd}}(\bm{p})\bigg\{1+\frac{n_f}{\pi}\alpha \left[\mathcal{X}_1(\Delta E^p_n,\mu_{\text{us}})+\mathcal{X}_2(\Delta E^p_n/T)\right]\bigg\} \, ,
\end{aligned}
\label{bsd_final_fixed_NLO}
\end{equation}
where the dimensionless functions $\mathcal{X}_1$  and $\mathcal{X}_2$ can be inferred from \eqref{bsf_final_NLO_adding}, \eqref{bsf_final_fixed_NLO} and \eqref{dimless_x2}.\footnote{
Note that the functions $\mathcal{X}_1(x)$ and $\mathcal{X}_2(x)$ are symmetric for 
$x \to -x$.}
The expression at the lowest order, i.e. accounting for the photo-dissociation process at  leading order in $\alpha$ only, reads
\begin{equation}
\Gamma^{n, \scalebox{0.6}{\text{LO}}}_{\textrm{bsd}} =2\int\frac{d^3p}{(2\pi)^3}\, \sigma^{n,\scalebox{0.6}{\text{LO}}}_{\hbox{\scriptsize bsd}}(\bm{p})=
\frac{4}{3}\alpha \int \frac{d^3p}{(2\pi)^3} \, (\Delta E^p_n)^3 \, n_{\textrm{B}}(\Delta E^p_n) \, |\langle n|\bm{r}|\bm{p}\rangle|^2 \, ,
\label{bsd_width_LO}
\end{equation} 
which implicitly defines $\sigma^{n,\scalebox{0.6}{\text{LO}}}_{\hbox{\scriptsize bsd}}(\bm{p})$.
As elaborated in the previous sections, a resummation of the Debye mass scale is required once the temperature makes the Debye mass comparable with the scale associated to $\Delta E$.
The result for the dissociation width in the case $T\gg m_\text{D} \sim \Delta E$, which incorporates the special limits $m_\text{D} \gg \Delta E$, cf. section \ref{sec:mDggE}, and $\Delta E \gg m_\text{D}$, cf. section \ref{sec:mDllE}, reads
\begin{equation}
\begin{aligned}
&\left(\Gamma^{n}_{\textrm{bsd}}\right)_{m_\text{D} \sim \Delta E} \\
&=2\int \frac{d^3p}{(2\pi)^3}\, \sigma^{n, \scalebox{0.6}{\text{LO}}}_{\textrm{bsd}}(\bm{p}) \, \bigg\{\frac{n_f}{3\pi}\alpha
\left[\ln{\left(\frac{4(\Delta E^p_n)^2}{\mu^2}\right)}-\frac{10}{3}\right] \\
&\hspace{1cm}+\left(\frac{m_\text{D}}{2\Delta E^p_n}\right)^2\left[2-2\gamma_\text{E}+2\frac{\zeta'(2)}{\zeta(2)}-\ln\left(\frac{(\Delta E^p_n)^2}{16T^2}\right)+2\mathcal{X}_l\left(\frac{\Delta E^p_n}{m_\text{D}}\right)+2\mathcal{X}_t\left(\frac{\Delta E^p_n}{m_\text{D}}\right)\right]\bigg\}
\, ,
\label{bsd_final_hierarchy3}
\end{aligned}
\end{equation}
where the dimensionless functions $\mathcal{X}_l$ and $\mathcal{X}_t$ are defined in eqs.~\eqref{dimless_xl} and \eqref{dimless_integral}, respectively.
The explicit renormalization scale dependence in the second line of \eqref{bsd_final_hierarchy3}
cancels against the one of the coupling in $\sigma^{n, \scalebox{0.6}{\text{LO}}}_{\textrm{bsd}}(\bm{p})$.
The {\it natural} renormalization scale for the coupling associated to the electric dipole interaction is of the order of the ultrasoft scale $M \alpha^2$, 
while for the one contained in the wave function and in the binding energy it is of the order of the soft scale $M \alpha$.
In the following applications, we set the coupling at these values.

\begin{figure}[ht]
    \centering
    \includegraphics[scale=0.78]{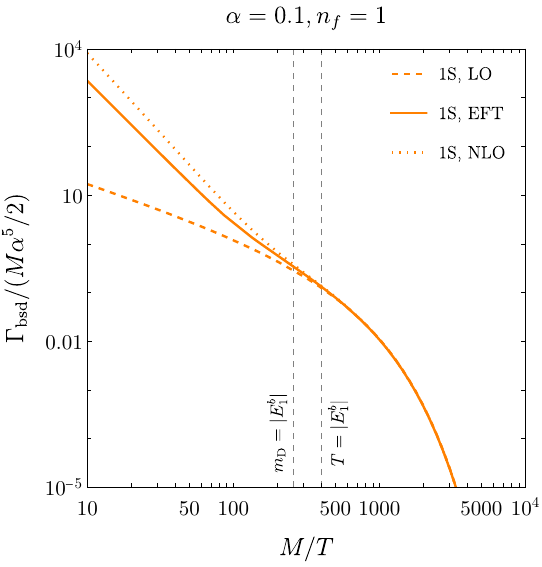}
    \hspace{0.5cm}
    \includegraphics[scale=0.78]{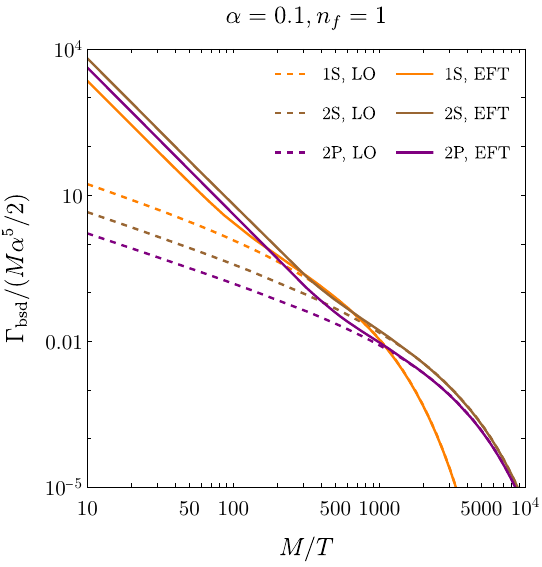}
    \caption{(Left) Bound-state dissociation width, normalized by the leading-order paradarkonium decay width $M\alpha^5/2$, for the 1S state at leading order (orange dashed line, cf.~\eqref{bsd_width_LO}), up to next-to-leading order (orange dotted line, cf.~\eqref{bsd_final_fixed_NLO}) and with Debye mass resummation (orange solid line, cf.~\eqref{bsd_final_hierarchy3}), plotted as a function of  $M/T$ for $n_f=1$ and running coupling with starting value $\alpha(2M)=0.1$. 
    The vertical dashed lines mark the positions, from left to right, where $m_\text{D} =  |E^b_1|$ and $T =  |E^b_1|$.
    (Right) Bound-state dissociation widths for the 1S (orange lines), 2S (brown lines) and 2P state (purple lines). 
    Dashed lines represent the results at LO, cf.~\eqref{bsd_width_LO}, solid lines the ones from~\eqref{bsd_final_hierarchy3}.
    }
    \label{fig:bsd}
\end{figure}

In figure \ref{fig:bsd} left, we plot the bound-state dissociation width of the ground state divided by the leading order paradarkonium decay width, $M\alpha^5/2$, at leading order (orange dashed line), including next-to-leading order corrections (orange dotted line) and with Debye mass resummation (orange solid line).
As in the case of the bound-state formation and for the same reasons discussed there, 
we do not distinguish between the momentum region where $T \ge \Delta E^p_1$ and $T<\Delta E^p_1$.
We do not make this distinction also in the rest of the section.

We set $n_f =1$ and let the coupling run at one loop from the starting value $\alpha(2M)=0.1$. 
At low temperatures, of the order of the ultrasoft scale or smaller, the curves approach each other.
The dominant process is via photo-dissociation. 
At larger $T$, the bath-particle scattering starts being relevant, eventually becoming the dominant process and enhancing the width by about two orders of magnitude for the largest temperatures that we consider. 
As in the case of the bound-state formation process, see  figure~\ref{fig:bsf_comparison}
left, the width at NLO leads to an overestimation by a factor of up to five compared to the width obtained taking into account the Debye mass scale. 
In the right plot of figure \ref{fig:bsd}, we add the widths of the first excited states 2S (brown lines) and 2P (purple lines), where we have averaged over the magnetic quantum number of the incoming bound state. 
Dashed lines are for the width at LO, only accounting for the photo-dissociation process, solid lines are when incorporating the Debye mass resummation. 
At low $T$, the widths of the excited states are larger than the one of the ground state because it is more likely to dissociate a broader Coulombic bound state.
Since at larger temperatures, the widths of the excited states are still larger than the 1S width, contrary to the results at LO, 
we conclude that the resummation of the Debye mass has a greater impact on the excited states than on the ground state, as we already observed in the reversed process.

\begin{figure}[ht]
    \centering
    \includegraphics[scale=0.78]{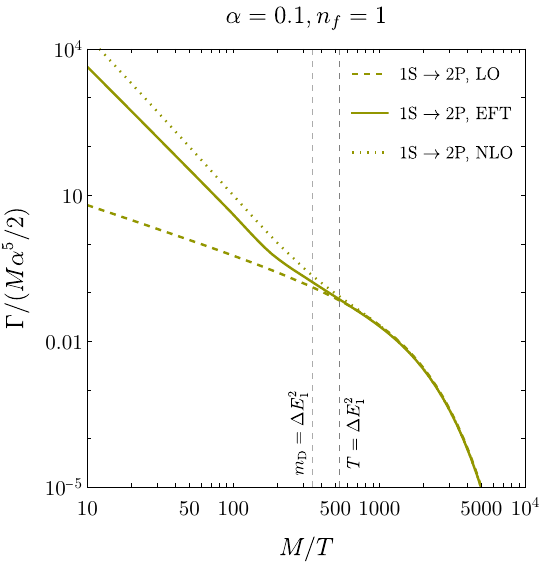}
    \hspace{0.5cm}
    \includegraphics[scale=0.78]{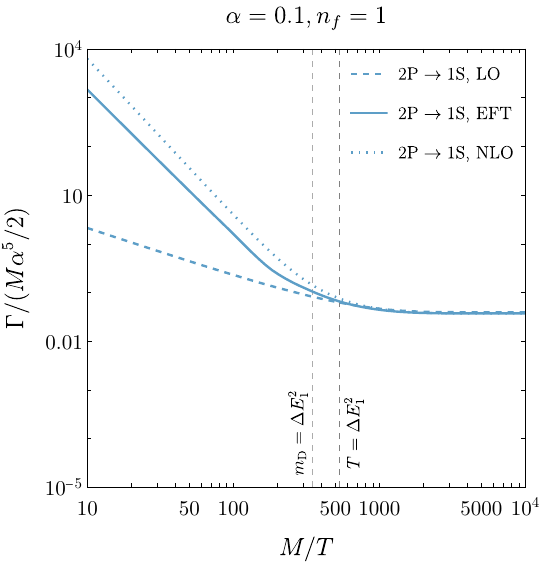}
    \caption{Ratios of excitation (left plot) and de-excitation (right plot) widths
    over the paradarkonium annihilation width at leading order for the 1S $\leftrightarrow$ 2P process, at leading order (dashed lines), up to next-to-leading order (dotted lines) and with Debye mass resummation (solid lines), for $n_f=1$ and running coupling with $\alpha(2M)=0.1$. 
    The vertical dashed lines mark the positions, from left to right, where $m_\text{D} =  \Delta E^2_1$ and $T =  \Delta E^2_1$. 
    }
    \label{fig:btob}
\end{figure}

The computation of the bound-state to bound-state transition width goes like the computation of the bound-state formation cross section done in section \ref{sec:bsf_resummed} and the bound-state dissociation width. 
The incoming or outgoing scattering states in figure~\ref{fig:bsf_two_loop_fixed_NLO} need to be replaced by bound states. 
The result for the de-excitation width up to NLO and with Debye mass resummation equals the expressions for the bound-state formation cross section at NLO, eq.~\eqref{bsf_final_fixed_NLO}, and with Debye mass resummation, eq.~\eqref{bsf_final_hierarchy3}, respectively, but with the energy difference  $\Delta E^p_{n}$ and $(\sigma^n_{\text{bsf}} \, v_{\hbox{\scriptsize rel}})^{\text{LO}}(\bm{p})$ replaced by $\Delta E^n_{n'} = E_n-E_{n'} =(M\alpha^2/4)\left(1/{n'}^2-1/n^2\right)$ and
\begin{equation}
\Gamma^{n, \scalebox{0.6}{\text{LO}}}_{\textrm{de-ex.}} =\frac{4}{3}\alpha \sum_{n', E_{n'}<E_n}\left(\Delta E_{n'}^{n}\right)^3 \, \left(1+n_{\text{B}}\left(\Delta E_{n'}^{n}\right)\right) \, \left| \langle n'|  \bm{r}  | n \rangle \right|^2 \, ,
\label{gamma_de-excitation1}
\end{equation}
respectively. 
Similarly, the excitation width follows from the result for the bound-state dissociation  width at NLO, eq.~\eqref{bsd_final_fixed_NLO}, and the one with Debye mass resummation, eq.~\eqref{bsd_final_hierarchy3}, by replacing $\Delta E^p_{n}$ and $\displaystyle \int \frac{d^3p}{(2\pi)^3}\sigma^{n, \scalebox{0.6}{\text{LO}}}_{\textrm{bsd}}(\bm{p})$ by $\Delta E^n_{n'}$
and $\Gamma^{n, \scalebox{0.6}{\text{LO}}}_{\textrm{ex.}}$, respectively, where
\begin{equation}
\Gamma^{n, \scalebox{0.6}{\text{LO}}}_{\textrm{ex.}} =\frac{4}{3}\alpha \sum_{n', E_{n'}>E_n}\, (\Delta E_{n}^{n'})^3  \, n_{\text{B}}(\Delta E_{n}^{n'}) \, \left| \langle n' | \bm{r} | n \rangle \right|^2 \, .
\label{gamma_excitation1}
\end{equation}
In figure \ref{fig:btob}, left, we plot the excitation width for the 1S $\rightarrow$ 2P process at leading order (green dashed line), including next-to-leading order corrections (green dotted line) and with Debye mass resummation (green solid line). 
In the right panel, we plot the results for the de-excitation process 2P $\rightarrow$ 1S. 
We normalize by the paradarkonium decay width at LO, and choose $n_f=1$ and running coupling at one loop with $\alpha(2M)=0.1$. 
The results are comparable to the ones for the bound-state dissociation in figure~\ref{fig:bsd}, left, i.e., at large temperatures, we observe an overestimation of the NLO widths by a factor of up to five with respect to the widths that account for the resummation of the Debye mass.

\section{Dark matter energy density}
\label{sec:DMevolution}
In this section, we study the evolution of kinetically equilibrated dark matter pairs 
as determined by the annihilation and near-threshold thermal rates computed in section~\ref{sec:pNREFT_part} and~\ref{sec:el_dipole_transitions}, respectively. 
The evolution of the sum of the dark matter particle and antiparticle number densities, $n = n_X + n_{\bar{X}} = 2 n_X$, as well as of the para- and orthodarkonium bound-state number densities $n_n^{\textrm{para}}$ and $n_n^{\textrm{ortho}}$ is governed by a network of semi-classical Boltzmann equations. 
In~\cite{Biondini:2023zcz}, one can find the coupled equations for the 1S state without the inclusion of bound-state to bound-state transitions. 
With respect to that work, here the spin-triplet dark matter bound states are treated at the same level as the spin-zero bound states due to the orthodarkonium and paradarkonium decaying at the same order in the coupling, see~\eqref{eq:Gammannihilation}.\footnote{
Without light fermion species, the orthodarkonium decay occurs at a higher order in the coupling, as it requires three dark photons in the final state.} 

If the condition $H \ll \Gamma^{1\textrm{S},\hbox{\scriptsize para\,(ortho)}}_{\textrm{ann}},  \Gamma^{n}_{\textrm{bsd}}$ is fulfilled, 
where $H$ is the Hubble rate, and the bound states are kept close to equilibrium, which is the case in the considered cosmological scenario at the thermal freeze-out, then the coupled Boltzmann equations reduce to a single evolution equation for the DM density $n$~\cite{Ellis:2015vaa}. 
Upon neglecting the contributions from the (de-)excitation processes among bound states, it reads
\begin{equation}
    (\partial_t + 3H) n = - \frac{1}{2}\langle \sigma_{\textrm{eff}} \, v_{\textrm{rel}} \rangle (n^2-n^2_{\textrm{eq}}) \,,
    \label{Boltzmann_eq_eff}
\end{equation}
where $n_\textrm{eq} = 2n_{X,\textrm{eq}}=4(MT/2 \pi)^{3/2}e^{-M/T}$ and the thermally averaged {\it effective cross section} is defined as
 \begin{equation}
   \langle  \sigma_{\textrm{eff}} \, v_{\textrm{rel}} \rangle  =
   \langle \sigma_{\textrm{ann}} v_{\textrm{rel}} \rangle
   + \sum_{n} \left(\frac{1}{4}\langle   \sigma^n_{\textrm{bsf}} \, v_{\textrm{rel}} \rangle \, \frac{\Gamma^{n,\hbox{\scriptsize para}}_{\textrm{ann}}}{\Gamma^{n,\hbox{\scriptsize para}}_{\textrm{ann}}+\Gamma_{\textrm{bsd}}^n} +\frac{3}{4}\langle   \sigma^n_{\textrm{bsf}} \, v_{\textrm{rel}} \rangle \, \frac{\Gamma^{n,\hbox{\scriptsize ortho}}_{\textrm{ann}}}{\Gamma^{n,\hbox{\scriptsize ortho}}_{\textrm{ann}}+\Gamma_{\textrm{bsd}}^n} \right) .
    \label{Cross_section_eff}
 \end{equation}
If, instead, we keep the transitions between bound states, then a more general expression for the effective cross section has to be used~\cite{Garny:2021qsr}.
We use that implementation when including excited states in our work; 
we refer to it as {\it beyond the non-transition approximation}.

There are two asymptotic regimes, the limits of large and small temperatures, for which the effective cross section in eq.~\eqref{Cross_section_eff} can be further simplified. 
For $T \gg \Delta E$, $\Gamma^n_{\textrm{bsd}} \gg \Gamma^{n,\hbox{\scriptsize para\,(ortho)}}_{\textrm{ann}}$ and using the detailed balance condition at equilibrium between the bound-state formation and dissociation process
\begin{equation}
\begin{aligned}
&\frac{1}{16} \langle \sigma^{n}_{\hbox{\scriptsize bsf}} v_{\hbox{\scriptsize rel}} \rangle \, n^2_{\textrm{eq}} =
\Gamma^{n}_{\textrm{bsd}} \, n_{n,\textrm{eq}}^{\textrm{para}} \, , \\
&\frac{3}{16}\langle \sigma^{n}_{\hbox{\scriptsize bsf}} v_{\hbox{\scriptsize rel}} \rangle \, n^2_{\textrm{eq}} =
\Gamma^{n}_{\textrm{bsd}} \, n_{n,\textrm{eq}}^{\textrm{ortho}} \, ,
\label{detailedbalance}
\end{aligned}
\end{equation}
where $n_{n,\textrm{eq}}^{\textrm{ortho}} = 3n_{n,\textrm{eq}}^{\textrm{para}}$
and $n_{n,\textrm{eq}}^{\textrm{para}} = (2\ell+1)(MT/\pi)^{3/2}e^{-E_n/T}$ with $E_n = 2M + E_n^b$, equation \eqref{Cross_section_eff} becomes
\begin{equation}
    \langle  \sigma_{\textrm{eff}} \, v_{\textrm{rel}} \rangle \approx \langle \sigma_{\textrm{ann}} v_{\textrm{rel}} \rangle + \frac{4}{n^2_{\textrm{eq}}}\sum_{n} \left(\Gamma^{n,\hbox{\scriptsize para}}_{\textrm{ann}} n_{n,\textrm{eq}}^{\textrm{para}} + \Gamma^{n,\hbox{\scriptsize ortho}}_{\textrm{ann}} n_{n,\textrm{eq}}^{\textrm{ortho}} \right) \, ,
\label{effxsection_ion_eq}
\end{equation}
which is independent of both the bound-state formation cross section and the bound-state dissociation width. 
In the region of large temperatures, sometimes referred to as \textit{ionization equilibrium} \cite{Binder:2018znk}, the rapid close-to-threshold processes keep the bound and unbound pairs in detailed balance. 
With decreasing temperature, eventually approaching values lower than the ultrasoft scale ($T < \Delta E$), the thermal dissociation width becomes very small, see figure~\ref{fig:bsd}, such that $\Gamma^n_{\textrm{bsd}} \ll \Gamma^{n,\hbox{\scriptsize para\,(ortho)}}_{\textrm{ann}}$, and we obtain
\begin{equation}
    \langle  \sigma_{\textrm{eff}} \, v_{\textrm{rel}} \rangle \approx \langle \sigma_{\textrm{ann}} v_{\textrm{rel}} \rangle + \sum_{n} \langle \sigma^{n}_{\hbox{\scriptsize bsf}} v_{\hbox{\scriptsize rel}} \rangle \, .
\label{effxsection_smallT}
\end{equation}
At small $T$, the bound states that are formed decay much faster than they are ionized into unbound pairs. 

\begin{figure}[ht]
    \centering
    \includegraphics[scale=0.78]{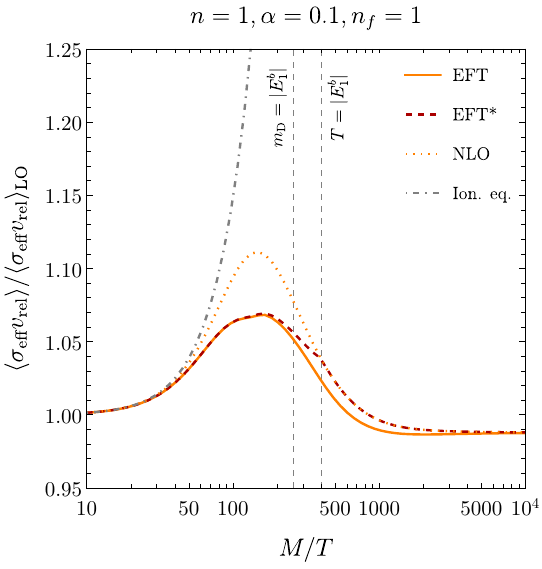}
    \hspace{0.5cm}
    \includegraphics[scale=0.78]{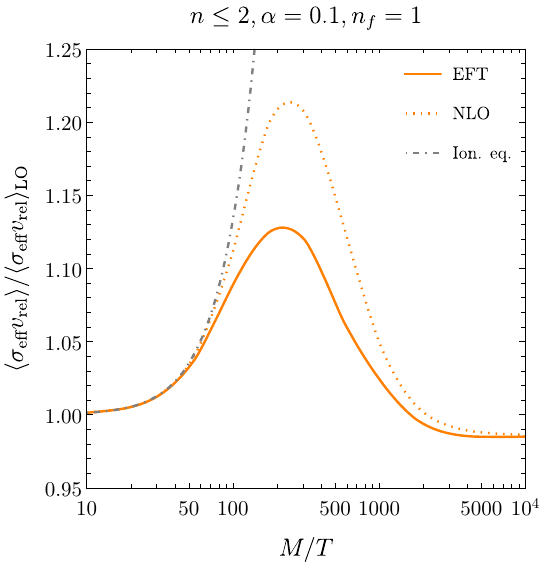}
    \caption{(Left) Thermally averaged effective cross section, cf. \eqref{Cross_section_eff}, with the inclusion of next-to-leading order corrections in the rates (orange dotted line, labeled NLO) or with full resummation of the Debye-mass effects (orange solid line, labeled EFT) for the 1S state, normalized by the same quantity where the thermal rates are evaluated at leading order. 
    The red dashed line, labeled EFT$^*$, represents the effective cross section thermally averaged by splitting the $v_{\hbox{\scriptsize rel}}$ or $p$ integrals into the two regions $T \ge \Delta E^p_1$ and $T < \Delta E^p_1$ and taking for them the Debye mass resummed and NLO rates, respectively, 
    in accordance with the discussion in section~\ref{sec:bsf}.
    The cross section is normalized as the previous ones.
    The plot shows the dependence on $M/T$ for $n_f=1$ and one-loop running coupling with $\alpha(2M)=0.1$. 
    The vertical dashed lines mark the positions, from left to right, where $m_\text{D} =  |E^b_1|$ and $T =  |E^b_1|$. 
    The gray dot-dashed line, labeled Ion.~eq., is for the ratio in ionization equilibrium \eqref{effxsection_ion_eq}. 
    (Right) Thermally averaged effective cross section with the inclusion of excited states up to 2P and with 1S $\leftrightarrow$ 2P transitions and next-to-leading order corrections (orange dotted line) or HTL resummation effects (orange solid line), normalized by the same quantity with the rates evaluated at leading order. 
    }
    \label{fig:effcsnew}
\end{figure}

In the following, we quantify the impact of the resummation of the Debye mass on the thermally averaged effective cross section and eventually on the DM relic abundance. 
As a first step, we consider the evolution of unbound DM pairs and the ground state, and omit the bound-state to bound-state transitions. 
In figure \ref{fig:effcsnew} left, we plot the thermally averaged effective cross section~\eqref{Cross_section_eff} with next-to-leading order corrections included in the thermal rates for the electric dipole transition processes (orange dotted line), the thermally averaged effective cross section when incorporating the Debye mass in the thermal rates (orange solid line), 
and the thermally averaged effective cross section computed from the Debye mass resummed and 
NLO rates in the two regions $T \ge \Delta E^p_1$ and $T < \Delta E^p_1$, respectively, as discussed in section~\ref{sec:bsf} (red dashed line), with all cross sections normalized with respect to the effective cross section at leading order in the rates.
We choose $n_f=1$ and $\alpha$ runs at one loop with $\alpha(2M)=0.1$. 
As expected, at large $T$ the curves approach the gray dot-dashed line representing the region where ionization equilibrium holds. 
The ratio is close to one, reflecting the fact that \eqref{effxsection_ion_eq} only depends on the annihilation rates that are not affected by thermal effects. 
As the temperature drops, ionization equilibrium is lost and the ratios deviate from the gray dot-dashed line. 
The peak values for the deviations from the leading order effective cross section are about 5\% and 10\% at $T \approx M/100$ for the resummed and NLO cases, respectively. 
The treatment of thermal effects at NLO results in an overestimation of the effective cross section, 
which originates from a systematic overestimation of the individual rates (see figures~\ref{fig:bsf_comparison} and \ref{fig:bsd}). 
At small temperatures, the orange solid and dotted lines eventually approach each other and stay constant at a value close to one. 
Here, the processes involving a thermal incoming/outgoing dark photon are dominant and are largely accounted for by the individual rates at leading order. 
The deviation of about 1\% from one in the ratios is due to the suppressed bath-particle scatterings, which are absent in the leading-order expressions. 
Finally, as the red dashed line shows, splitting the momentum or $v_{\hbox{\scriptsize rel}}$ integrals in the rates in the two regions $T \ge \Delta E^p_1$ and $T < \Delta E^p_1$ and using Debye mass resummed rates in the first one and NLO rates in the second one does not lead to a significant change in the thermally averaged cross section, 
as the already modest changes in the individual rates (see figure~\ref{fig:bsf_cut_vrel} for the bound-state formation cross section) tend to compensate each other in the effective cross section.
There is an increase with respect to the cross section computed from the Debye mass resummed rates only (orange solid line) by about 1\% in the region $T\sim |E_1^b|$.

In the right panel of figure \ref{fig:effcsnew}, we plot the ratio of thermally averaged effective cross sections when adding bound states up to 2P and the transitions 1S $\leftrightarrow$ 2P. 
The orange dotted line represents again the ratio of the next-to-leading order cross section over the LO one, while the orange solid line is the ratio with the Debye mass resummed cross section.
The cross section increases when adding more bound states.  
However, the overestimation of the NLO expression remains about a factor 2 around the peak. 
We remark that including bound-state to bound-state transitions and bound-state formation/dissociation via thermal bath scattering in the effective cross section is implemented here for the first time. 

As already discussed in the former sections, at large temperatures the Landau damping phenomenon is relatively important and by far the dominant process, resulting in an enhancement of about two orders of magnitude of the bound-state formation cross section (figure~\ref{fig:bsf_comparison}) and bound-state dissociation width (figure~\ref{fig:bsd}) for the Debye mass resummed cross section; 
an even larger enhancement is observed for the NLO bound-state formation cross section. However, at the level of the effective cross section we observe an enhancement that ranges from 5\% to 10\% (figure~\ref{fig:effcsnew}, left panel). 
This is due to an increase of the bound-state formation cross section \emph{and} bound-state dissociation width, which both enter the effective cross section and largely compensate each other. 
The physical reason is that, while bound states are more efficiently formed at high temperatures via frequent $2 \to 2$ scatterings, it is also true that they are more likely to be dissociated by collisions with light fermions. 

Next, we study the effects on the dark matter relic abundance. 
We rewrite equation~\eqref{Boltzmann_eq_eff} in terms of the yield $Y\equiv n/s$, $s$ being the entropy density, and solve numerically the so-obtained Boltzmann equation. 
The solution is inserted in the expression of the present-day DM relic density $\Omega_{\textrm{DM}} = M s_0 Y_0/\rho_{\textrm{crit},0}$,
where $Y_0, s_0$ and $\rho_{\textrm{crit},0}$ are the present yield, entropy density and critical density, respectively. 
We take the values of $s_0$ and  $\rho_{\textrm{crit},0}$ from~\cite{ParticleDataGroup:2022pth}, such that $\Omega_{\textrm{DM}} h^2=(M/\textrm{GeV)}\, Y_0 /(3.645 \times 10^{-9})$, where $h$ is the reduced Hubble constant.
Eventually, the obtained values for $\Omega_{\hbox{\tiny DM}}$ in terms of the parameters of the DM model can be matched to the present dark matter relic abundance $\Omega_{\hbox{\tiny DM}} h^2 = 0.1200 \pm 0.0012$~\cite{Planck:2018nkj}, 
which leads to constraints on the possible values for the coupling $\alpha$ and DM mass~$M$. 
In the radiation-dominated era, the Hubble rate takes the form $H=T^2\sqrt{4 \pi^3 g_{\textrm{eff}}(T)/45}/M_{\textrm{Pl}}$, where $g_{\textrm{eff}}(T)$ are the effective number of relativistic degrees of freedom and $M_{\textrm{Pl}} \approx 1.2 \times 10^{19} \textrm{GeV}$ is the Planck mass. 
In addition to the Standard Model degrees of freedom, and by assuming thermal equilibrium between the dark sector and the Standard Model, we add the dark photon and the $n_f$ dark light fermions to $g_{\textrm{eff}}(T)$.

\begin{figure}[ht]
    \centering
    \includegraphics[scale=0.79]{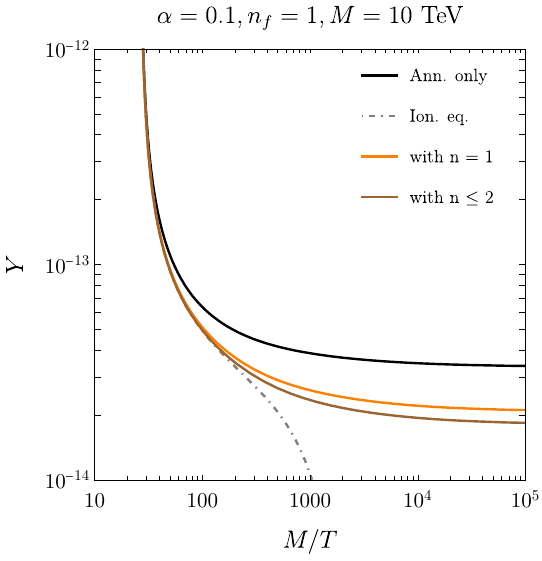}
    \hspace{0.5cm}
    \includegraphics[scale=0.76]{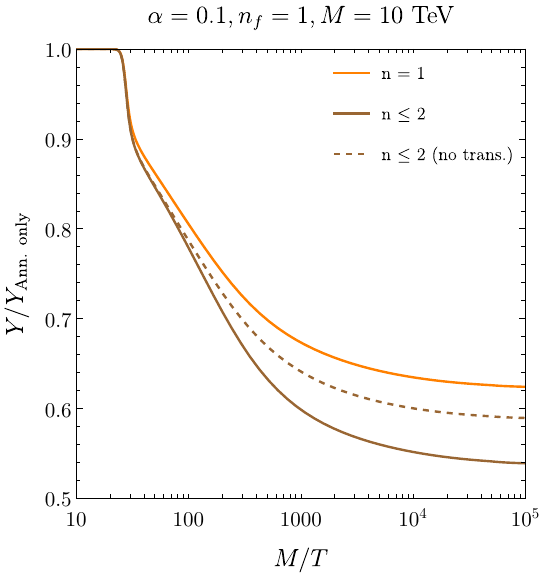}
    \caption{(Left) DM yield as a function of $M/T$ for $n_f=1$, $M=10$ TeV and $\alpha$ running at one loop with $\alpha(2M)=0.1$. 
    The black solid line is for annihilations only, the black dash-dotted line when including the bound-state effects for the 1S state in ionization equilibrium, the orange solid line for the same situation beyond ionization equilibrium and the brown solid line when adding 2S and 2P states beyond the no-transition approximation. 
    (Right) $Y$ normalized by the yield obtained only via Sommerfeld enhanced annihilations (black line in the previous plot).} 
    \label{fig:boltz}
\end{figure}

In figure \ref{fig:boltz} left, we show the dark matter yield $Y$ as a function of $M/T$ for $n_f=1$ dark light fermions and with one-loop running coupling starting at $\alpha(2M)=0.1$. 
We fix the DM mass to be $M=10$ TeV. The black solid line represents the solution when neglecting the bound-state effects and hence only considering the annihilation term in eq. \eqref{Cross_section_eff}. 
The solution that accounts for the ground state in ionization equilibrium is shown by the black dash-dotted line; 
we remind the reader that this solution is only valid at large temperatures and for the dissociation width much larger than the annihilation width,
cf.~eq.~\eqref{effxsection_ion_eq}. 
By solving the effective Boltzmann equation with the most general effective cross section for the ground state in eq.~\eqref{Cross_section_eff}, 
we obtain the result shown by the orange solid line that holds for the whole range of $T$. 
Finally, adding excited states up to 2P and bound-state transitions between them, i.e. solving the effective evolution equation with the effective cross section given in \cite{Garny:2021qsr}, 
the yield, shown by the brown line, is depleted by around 15\% compared to the yield obtained only with the ground state.  
The yields obtained by considering the 1S state or adding excited states up to 2P (with and without taking into account transitions between bound states) normalized by the yield obtained from annihilations are given by the orange solid and brown lines shown in the right plot of figure~\ref{fig:boltz}. 
Larger couplings or more light dark particles would enhance the effective cross section, and hence reduce the corresponding yields even more.

\begin{figure}[ht]
    \centering
    \includegraphics[scale=0.78]{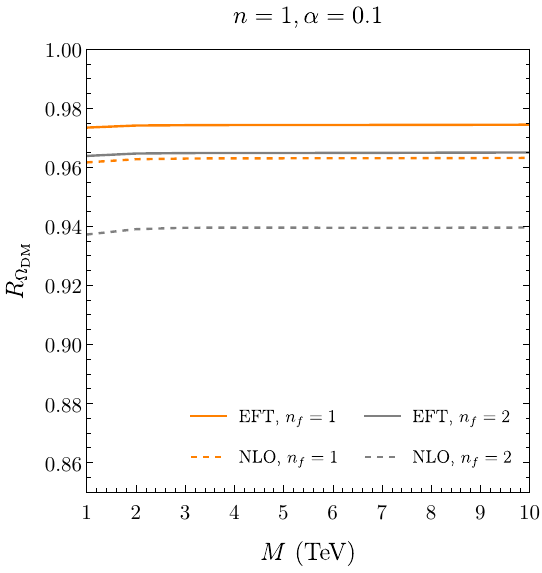}
    \hspace{0.5cm}
    \includegraphics[scale=0.78]{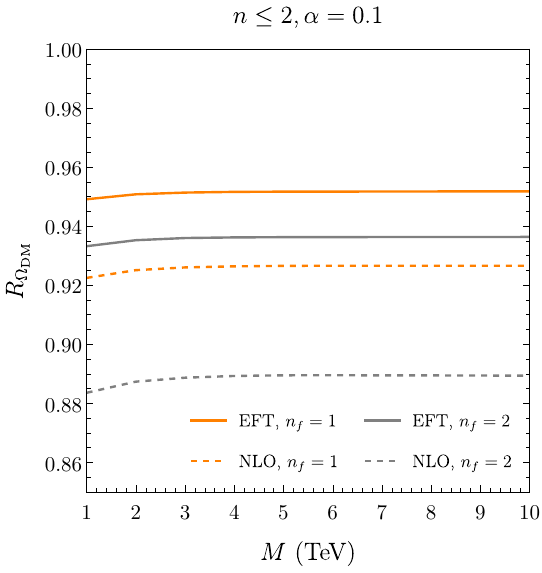}
    \caption{(Left) Present-day relic density normalized by $\Omega_\text{DM}h^2$ obtained 
    including only the 1S state with rates at LO, 
    as a function of the DM mass in TeV and with one-loop running coupling starting at  $\alpha(2M)=0.1$. 
    Solid lines include the effects of the Debye mass resummation, while dashed lines 
    include only NLO effects. 
    Orange lines refer to the $n_f=1$ case and gray lines to the $n_f=2$ case.
    (Right) The ratio when including excited states up to 2P and bound-to-bound transitions.}
    \label{fig:boltznew}
\end{figure}

In figure~\ref{fig:boltznew}, we quantify the corrections to the dark matter relic density coming from the thermal processes that are captured by the NLO rates and the rates with Debye mass resummation. 
We plot the ratios of the dark matter energy densities obtained from NLO or Debye mass resummed cross sections and widths, and the dark matter relic density obtained from LO electric dipole transition rates. 
We indicate the ratios with $R_{\Omega_{\textrm{DM}}}$.  
First, we include the ground state only. 
In the left plot of figure~\ref{fig:boltznew}, we display $R_{\Omega_{\textrm{DM}}}$ by including NLO corrections to the rates (dotted lines) or by resumming the Debye mass (solid lines), for a DM mass range from 1 TeV to 10 TeV and for one-loop running coupling starting at $\alpha(2M)=0.1$. 
Orange lines are for the $n_f=1$ case, whereas gray lines refer to a model with two dark light fermion species. 
The next-to-leading order effects reduce $\Omega_\text{DM}h^2$ by about 4\% for $n_f=1$ and 6\% for $n_f=2$. 
When accounting for Debye mass resummation, one sees a smaller impact and  the energy density is reduced by 2.5\% and 3.5\% for $n_f=1$ and $n_f=2$, respectively. 
The relic density changes more drastically upon including bound-state to bound-state transitions. 
In the right panel of figure~\ref{fig:boltznew}, we show the same ratios for the energy density, however now including the first excited states 2S and 2P beyond the no-transition limit, i.e. with (de-)excitation processes included. 
The Debye mass resummation depletes the relic density by about 5\% and 6\% for $n_f=1$ and $n_f=2$, whereas the NLO correction by about 8\% and 11\%, respectively. 
Increasing the number of dark light fermion species gives origin to larger effects both for the NLO rates and the resummed rates. 
However, the larger number of light fermion species impacts the fixed NLO rates more than the resummed ones.  

\begin{figure}[ht]
    \centering
    \includegraphics[scale=0.78]{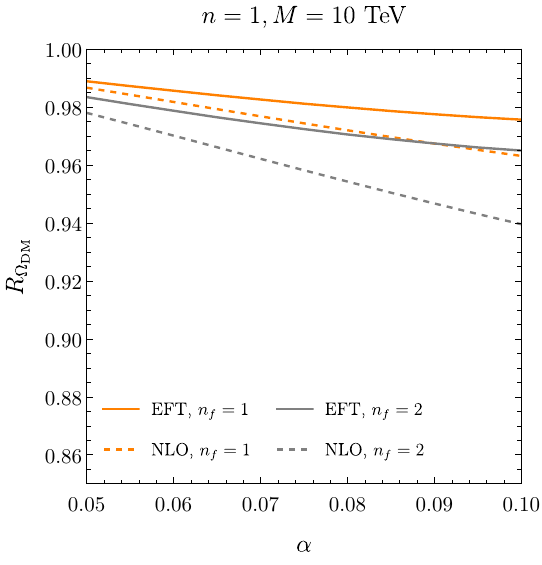}
    \hspace{0.5cm}
    \includegraphics[scale=0.78]{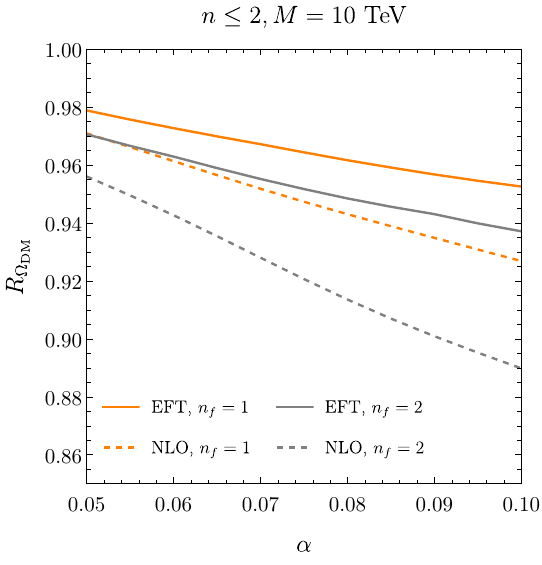}
    \caption{(Left) 
     Present-day relic density normalized by $\Omega_\text{DM}h^2$ obtained 
    including only the 1S state with rates at LO, as a function of $\alpha \equiv \alpha(2M)$ and fixed $M=10$ TeV. 
    Solid orange and gray lines are for the Debye mass resummation case and $n_f=1$ and $n_f=2$, respectively. 
    Dashed lines are for the NLO case. 
    (Right) The ratio when including excited states up to 2P and bound-to-bound transitions. }
    \label{fig:boltznew_alpha}
\end{figure}

Finally, in figure \ref{fig:boltznew_alpha}, we vary the ratio of energy densities as a function of $\alpha(2M)$ from 0.05 to 0.1, for a fixed DM mass of 10 TeV and $n_f=1$ and $n_f=2$.  
We observe that the increase of the coupling corresponds to larger effects of higher order corrections on the depletion of dark matter (both in the NLO and Debye mass resummed case). 
Since the HTL Debye mass resummation performed in this work relies on the condition $T \gg m_\text{D}$, which is valid as long as $\sqrt{\pi n_f \alpha } \ll 1$, we do not investigate couplings larger than $\alpha(2M)=0.1$ and do not include further species of dark light fermions.

Using in the Boltzmann equation the thermally averaged effective cross section computed by splitting the momentum or $v_{\hbox{\scriptsize rel}}$ integrals in the rates in the two regions $T \ge \Delta E^p_1$ and $T < \Delta E^p_1$ and inserting Debye mass resummed rates in the first one and NLO rates in the second one (red dashed line in figure~\ref{fig:effcsnew}) leads to a correction of less than 0.1\% in the DM relic abundance. 
We conclude that using the Debye mass resummed rates over the whole range of momenta or $v_{\hbox{\scriptsize rel}}$, as we have done in figures~\ref{fig:boltz}, \ref{fig:boltznew} and~\ref{fig:boltznew_alpha},  
leads to results that are accurate for current measurements.

\section{Conclusions and outlook}
\label{sec:concl}
In this work, we have studied the impact of the Debye mass $m_\textrm{D}$, which is the inverse of the electric screening length of a thermal medium, on the dark matter relic density 
in a model where the dark sector is charged under an abelian U(1) gauge symmetry, dubbed U(1)$_{\textrm{DM}}$.
The Debye mass adds another energy scale to the system besides the temperature and the non-relativistic intrinsic scales of the dark matter bound states.  
We have studied the impact of the scale $m_\textrm{D}$ on the bound-state formation and dissociation, its interplay with the other energy scales and assessed its effect on the evolution of the dark matter population in the early universe. 
To the best of our knowledge, this has not been previously studied within the framework of a potential non-relativistic effective field theory.

We have shown that the bound-state formation cross section and dissociation width require the resummation of the thermal photon propagator in the temperature region where 
$T \gg m_\textrm{D} \sim \Delta E$, $\Delta E$ being the energy of the photon emitted or absorbed by the DM pair, as the effect of the resummation with respect to a perturbative fixed order calculation up to NLO is significant.
In particular, the fixed order result overestimates the bound-state formation cross section, bound-state dissociation and (de-)excitation width up to a factor of order three at high temperatures close to the freeze-out. 
For the case $m_\textrm{D} \sim \Delta E$, 
we have derived the observables in a closed form suitable for numerical evaluation. 
The results are given in eqs.~\eqref{bsf_final_hierarchy3} and~\eqref{bsd_final_hierarchy3}.
Analogous results hold for the bound-state to bound-state transition widths. 
The results can be extrapolated to analytical expressions, derived explicitly in this work as well, in the limits $m_\textrm{D} \gg \Delta E$ and $m_\textrm{D} \ll \Delta E$; 
in the case of the bound-state formation cross section, the limits are given in eqs. $\eqref{bsf_final_hierarchy1}$ and $\eqref{bsf_final_hierarchy2}$, respectively.

In figures~\ref{fig:bsf_comparison}, \ref{fig:bsd} and \ref{fig:btob}, we plot the thermally averaged cross sections and widths over the whole range of temperatures from the freeze-out till very late time, and low temperature $T =  10^{-4}M$. 
We observe that the correction at next-to-leading order as well as the correction due to the resummation of the Debye mass increase the rates drastically, by several orders of magnitude, in the vicinity of the freeze-out regime, i.e. $M/T \sim 10-50$, 
while the rates approach the leading order results at low $T$.
Nevertheless, the thermally averaged effective cross section, which is the quantity entering the effective Boltzmann equation~\eqref{Boltzmann_eq_eff},
gets modified only from a few percent up to about $25\%$ when considering NLO or Debye mass effects, cf. figure~\ref{fig:effcsnew}. 
This is due to the fact that the bound-state formation cross section and the bound-state dissociation width enter the effective cross section as a ratio, cf. eq.~\eqref{Cross_section_eff} in the no-transition limit, so that changes in one quantity are largely compensated by changes of the same size in the other one.
In particular, in the ionization limit at large $T$, the large corrections to the bound-state formation and bound-state dissociation process almost cancel each other in the ratio, while at small $T$ 
those corrections are small, correcting the thermally averaged cross section, 
which is dominated by the in-vacuum contribution, by only 1\%.

By numerically solving the effective Boltzmann equation, we have computed the present-day dark matter relic density.
As can be seen in the plots in figures~\ref{fig:boltznew} and~\ref{fig:boltznew_alpha}, the inclusion of NLO corrections to the rates monotonically decreases the present relic abundance up to $3.7-6.0\%$ when considering the ground state only, for couplings up to $\alpha(2M)=0.1$, $n_f={1, 2}$ dark light fermion species and DM masses ranging from 1 to 10 TeV, 
and up to $7.3-11.0$\% when including the first excited bound states with principal quantum number $n=2$ and allowing for transitions among them. 
The corrections become larger when adding more dark light fermion species. 
Performing instead the Debye mass resummation, which is a necessary requirement in the large temperature regime, results in up to $2.5-3.5$\% corrections on $\Omega_{\textrm{DM}}h^2$ when considering only the 1S state, and up to $4.8-6.3$\%  corrections when including $n\leq 2$ excited states beyond the no-transition limit, for the same values of the parameters considered above. The inclusion of the Debye mass resummation appears to give larger corrections to $\Omega_{\textrm{DM}}h^2$ when excited states are considered. 
Hence, our results may be also relevant when including a large number of excited states, see e.g.~\cite{Binder:2023ckj}, 
while keeping the full accounting of the bound-to-bound transitions and thermal scales at play. 
The parameter values have been chosen 
such that $T > m_{\textrm{D}}$ is fulfilled at any time. 
Larger values of the coupling, $\alpha(2M)>0.1$, and including more dark light fermion species beyond $n_f=2$, would spoil that condition and the 
effective field theory treatment would need to be adapted to the strongly coupled plasma case 
$T \sim m_\textrm{D}$.

We conclude with some general observations. 
For temperatures around the ultrasoft scale and below, the Debye mass and the NLO thermal corrections are very small and do not give sizeable contributions to the rates computed at leading order in refs.~\cite{vonHarling:2014kha,Biondini:2023zcz}. 
With increasing $T$ the NLO corrections are no longer suppressed but start to dominate over the LO results. 
This signals that the thermal part of the higher-order loop corrections needs to be resummed, which is the Debye mass resummation performed throughout the paper. 
It strongly increases the thermal rates in the vicinity of the freeze-out temperature, 
although the obtained rates turn out to be a factor $4-5$ smaller than the rates obtained at NLO without resummation, cf.~\cite{Binder:2020efn,Binder:2021otw}. 
At the level of the effective cross section and hence the effective Boltzmann equation, both the Debye mass resummation and the NLO corrections have a small effect, leading to corrections up to around 4\% and 11\% on the relic abundance, respectively. 
The number of dark light fermion species has a larger impact on the NLO treatment with increasing $\alpha$ compared to the correct (resummed) treatment of the Debye mass scale. 
In general, the correct treatment of the Debye mass scale gives smaller corrections to the relic abundance than the fixed order calculation, and,  surprisingly, the absolute value of these corrections is of the same order as the recoil corrections recently computed in ref.~\cite{Biondini:2024aan}, but with opposite sign.
Hence, when including the recoil effects in addition to the Debye-mass effects, we may expect a partial cancellation, 
such that the total correction to the energy density due to both effects may become smaller than the 1\% accuracy of the observed dark matter relic density.

In this work, a simple abelian model of the dark sector has been chosen to illustrate the effects due to the emergence of a Debye mass scale in the system, and we have established an approach to correctly account for those corrections in the rates as well as in the evolution equations. 
The approach is suited to be used in more complicated theories, e.g. in theories with non-abelian gauge fields 
or in spontaneously broken renormalizable gauge theories, whose degrees of freedom include massive thermal mediators.

\section*{Acknowledgments}
The work of S.B. is supported by the Swiss National Science Foundation (SNSF) under the Ambizione grant PZ00P2\_185783.
N.B., G.Q. and A.V. acknowledge support from the DFG (Deutsche Forschungsgemeinschaft, German Research Foundation) cluster of excellence ``ORIGINS'' under Germany's Excellence Strategy -  EXC-2094-390783311.  
A.D. acknowledges support by the DFG under Germany's Excellence Strategy – EXC 2121 ``Quantum Universe'' - 390833306.
N. B. acknowledges the European Union ERC-2023-ADG-Project EFT-XYZ.

\begin{appendices}

\section{Two-point functions in real-time formalism} 
\label{sec:app_A}
In the Schwinger--Keldysh representation, fields double into physical fields of type 1 located on the physical time axis and unphysical fields of type 2 shifted along the imaginary time axis with respect to the physical ones~\cite{Bellac:2011kqa}.
As a consequence, 
the dark photon two-point function in momentum space is a $2\times 2$ matrix of the form
\begin{equation}
    D_{\mu\nu}(q)= \left(\begin{matrix}
{D}^{11}_{\mu\nu}(q) & {D}^{12}_{\mu\nu}(q)\\
{D}^{21}_{\mu\nu}(q) &  {D}^{22}_{\mu\nu}(q)
\end{matrix}\right) = 
    \left(\begin{matrix}
{D}^{T}_{\mu\nu}(q) & D^<_{\mu\nu}(q)\\
D^>_{\mu\nu}(q) &  ({D}^{T}_{\mu\nu}(q))^*
\end{matrix}
\right) \, ,
\label{photon_prop_mom_space}
\end{equation}
where ${D}^{11}_{\mu\nu}(q)$ is the time-ordered two-point function in momentum space and ${D}^{22}_{\mu\nu}(q)$ the anti-time-ordered one.
The off-diagonal elements are the Wightman functions that describe correlators of mixed fields of type 1 and 2; in the case of thermal particles, they satisfy the Kubo--Martin--Schwinger relation
\begin{equation}
    D^<_{\mu\nu}(q) = e^{-q_0/T}D^>_{\mu\nu}(q) \,.
\label{kms_relation}
\end{equation}
The free thermal propagator in Coulomb gauge reads
\begin{equation}
    \begin{aligned}
    &D^{\textrm{LO}}_{00}(|\bm{q}|) 
    = \begin{pmatrix} \displaystyle \frac{i}{\bm{q}^2} && 0 \\ 0 && \displaystyle \frac{-i}{\bm{q}^2} \end{pmatrix} , \\
    &D^{\textrm{LO}}_{ij}(q)= \left(\delta_{ij} - \frac{q_iq_j}{|\bm{q}|^2}\right)
    \left[\begin{pmatrix} \displaystyle \frac{i}{q^2 +i\epsilon} && \theta(-q_0)2\pi \delta(q^2) \\ \theta(q_0)2\pi \delta(q^2) && \displaystyle \frac{-i}{q^2 -i\epsilon} \end{pmatrix}
      + 2\pi \delta(q^2)\,n_{\text{B}}(|q_0|)\begin{pmatrix} 1 && 1 \\ 1 && 1 \end{pmatrix}\right]  ,
  \end{aligned}
\label{XXpropGT}  
\end{equation}
where $n_{\text{B}}(E) = 1/(e^{E/T}-1)$ is the Bose--Einstein distribution and 
The 21 and time-ordered two-point functions can be derived from the retarded and advanced two-point functions via
\begin{equation}
\begin{aligned}
    D^>_{\mu\nu}(q) &= [1+n_B(q_0)][D^R_{\mu\nu}(q)-D^A_{\mu\nu}(q)] \\
    &= 2[1+n_B(q_0)]\textrm{Re}[D^R_{\mu\nu}(q)]\big|_{q_0>0} \, ,\\
    D^{11}_{\mu\nu}(q) &= D^R_{\mu\nu}(q)+D^<_{\mu\nu}(q)= D^A_{\mu\nu}(q)+D^>_{\mu\nu}(q) \\
    &= \frac{D^R_{\mu\nu}(q)+D^A_{\mu\nu}(q)}{2}+\left[\frac{1}{2}+n_B(q_0)\right] [D^R_{\mu\nu}(q)-D^A_{\mu\nu}(q)] \\ 
    &= \frac{1}{2}[D^R_{\mu\nu}(q)+D^A_{\mu\nu}(q) + D^S_{\mu\nu}(q)] \, ,
\end{aligned}
\label{LOprop_coulombgauge}
\end{equation}
where in the last line we have split the propagator into the sum of a symmetric, $D^S_{\mu\nu}(q)$,  and an anti-symmetric, $D^{AS}_{\mu\nu}(q)=D^R_{\mu\nu}(q)+D^A_{\mu\nu}(q)$, function, 
which are respectively real and imaginary. 
The free retarded and advanced propagators in Coulomb gauge are
\begin{equation}
\begin{aligned}
    D^{R/A}_{00,\textrm{LO}}(|\bm{q}|)&= \frac{i}{\bm{q}^2} \, ,\\
    D^{R/A}_{ij,\textrm{LO}}(q)&=\left(\delta_{ij}-\frac{q_i q_j}{\bm{q}^2}\right)\frac{i}{(q_0 \pm i\epsilon)^2-\bm{q}^2} = \left(\delta_{ij}-\frac{q_i q_j}{\bm{q}^2}\right)\frac{i}{q^2\pm i \sign(q_0)\epsilon} \\
    &\equiv \left(\delta_{ij}-\frac{q_i q_j}{\bm{q}^2}\right)\Delta^{R/A}_{\text{LO}}(q) \, .
\end{aligned}
\label{LO_retarded_advanced}
\end{equation}

At higher order in perturbation theory, the free dark photon propagator gets modified by loop corrections originating from the light dark fermions in the thermal bath.\footnote{
Loop corrections from the interaction with the heavy DM fermions are accounted for by the decoupling theorem.} 
The two-point function can be expanded in terms of the free propagator and the one-loop polarization tensor, $\Pi^{\alpha\beta}$,
\begin{equation}
   D_{\mu\nu}(q)=D^{\textrm{LO}}_{\mu\nu}(q)+D^{\textrm{LO}}_{\mu\lambda}(q)[i\Pi^{\lambda\rho}(q)]D^{\textrm{LO}}_{\rho\nu}(q)+ \dots \, ,
\end{equation}
and similarly for the retarded and advanced two-point functions. 
Under certain circumstances, the polarization tensor needs to be resummed, resulting in\footnote{
In appendix~\ref{sec:b3}, we discuss the HTL-resummation, needed for momenta close to the Debye mass.}
\begin{equation}
\begin{aligned}
    D^{R/A}_{00}(q)=& \frac{i}{\bm{q}^2+\Pi^{R/A}_{00}(q)} \, ,\\
    D^{R/A}_{ij}(q)=&\left(\delta_{ij}-\frac{q_i q_j}{\bm{q}^2}\right)\frac{i}{(q_0 \pm i\epsilon)^2-\bm{q}^2 + \Pi^{R/A}_{\text{trans}}(q)} \, ,
\label{HTL_resummation_general}
\end{aligned}
\end{equation}
where the transverse retarded/advanced polarization tensor is defined as $\Pi^{R/A}_{\text{trans}}(q) = (\delta^{ij} - q^i q^j/\bm{q}^2)\Pi^{R/A}_{ij}(q)/2$. 
The Wightman function in an expanded form is given as
\begin{equation}
\begin{aligned}
    D^>_{\mu\nu}(q) \!&=\! 2[1+n_B(q_0)]\textrm{Re}[D^{R,\textrm{LO}}_{\mu\nu}(q) + D^{R,\textrm{LO}}_{\mu \lambda}(q) [i\Pi_R^{\lambda\rho}(q)] D^{R,\textrm{LO}}_{\rho\nu}(q) + \dots]\big|_{q_0>0} \\
    &=\! 2[1+n_B(q_0)]\textrm{Re}[D^{R,\textrm{LO}}_{\mu\nu}(q)]\big|_{q_0>0} 
    \!+ 2[1+n_\text{B}(q_0)]\textrm{Im}[D^{R,\textrm{LO}}_{\mu \lambda}(q) \Pi_R^{\lambda\rho}(q) D^{R,\textrm{LO}}_{\rho\nu}(q)]\big|_{q_0>0} \\
    &\!\hspace{0.5cm} + \dots \\
    &\equiv\! D^{>,\textrm{LO}}_{\mu\nu}(q) + D^{>,\textrm{NLO}}_{\mu\nu}(q) + \dots \, .
\label{upper_Wightman_fkt}
\end{aligned}
\end{equation}
We compute the one-loop polarization tensor in the subsequent sections.

In pNRQED$_{\textrm{DM}}$, the thermal bosonic propagator of the DM heavy fermion-antifermion field $\phi$ reads
\begin{equation}
\begin{aligned}
    G(p_0) 
    &= \begin{pmatrix} \displaystyle \frac{i}{p_0 - H +i\epsilon} && 0 \\ 2\pi \delta(p_0 - H) && \displaystyle \frac{-i}{p_0 - H -i\epsilon} \end{pmatrix}
    + 2\pi \delta(p_0 - H)\,n_{\text{B}}(p_0)\begin{pmatrix} 1 && 1 \\ 1 && 1 \end{pmatrix} \\
    &\approx \begin{pmatrix} \displaystyle \frac{i}{p_0 - H +i\epsilon} && 0 \\ 2\pi \delta(p_0 - H) && \displaystyle \frac{-i}{p_0 - H -i\epsilon} \end{pmatrix} ,
\label{DM_pair_prop}
\end{aligned}
  \end{equation}
where $p_0$ is the energy of the dark fermion-antifermion pair. 
In the last line, recalling that $H = 2M + \dots$ and $n_{\text{B}}(H) \approx e^{-2M/T}$ for 
$M \gg T$, we have dropped terms that are exponentially suppressed in the heavy mass limit.
The real-time formalism is convenient when dealing with heavy fields, since in the heavy mass limit
the type 2 fermion-antifermion fields decouple from the type 1 fields and may be ignored~\cite{Brambilla:2008cx}.

As for the $n_f$ dark light fermionic fields $f_i$ with masses $m_i$, the corresponding free thermal Dirac fermion propagators are

\begin{equation}
\begin{aligned}
    S^{\textrm{LO}}_i(k) 
    &= (\slashed{k}+m_i)\left[\begin{pmatrix} \displaystyle \frac{i}{k^2-m_i^2 +i\epsilon} && 0 \\ 2\pi \sign(k_0)\delta(k^2-m_i^2) && \displaystyle \frac{-i}{k^2-m_i^2-i\epsilon} \end{pmatrix} \right. \\
    &\hspace{1cm}\left. - 2\pi \sign(k_0)\delta(k^2-m_i^2)n_{\text{F}}(k_0)\begin{pmatrix} 1 && 1 \\ 1 && 1 \end{pmatrix} \right] \,,
\label{XXpropT}
\end{aligned}
  \end{equation}
where $n_{\text{F}}(E) = 1/(e^{E/T}+1)$ is the Fermi--Dirac distribution. 
In practice, for temperatures $T\gg m_i$ we neglect the light fermion masses. 
Hence each of the $n_f$ light particles has the same thermal propagator. 
Similar relations among the different two-point functions as in~\eqref{LOprop_coulombgauge} for the bosonic case hold also here, 
except replacing the distribution function $n_{\text{B}}$ with $-n_{\text{F}}$. 
In particular, the symmetric massless fermion propagator is $S^S(k)=[1-2n_{\textrm{F}}(|k_0|)]2\pi \slashed{k}\delta(k^2)$.

\section{Retarded polarization tensor} 
\label{sec:app_B}
\begin{figure}[ht]
    \centering
    \includegraphics[scale=0.4]{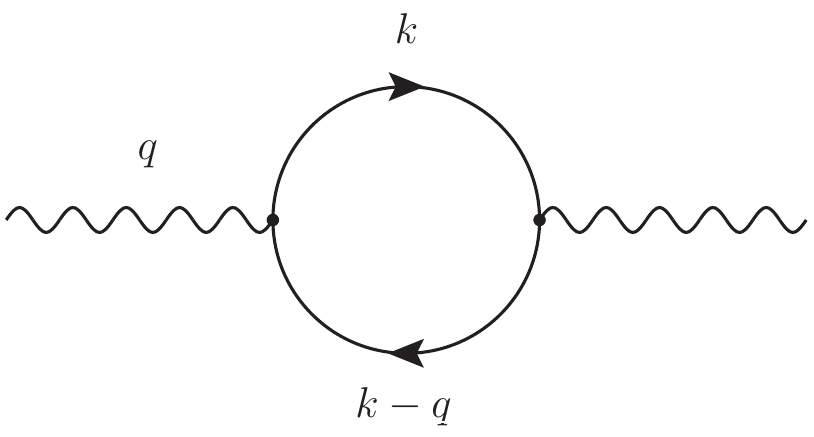}
    \caption{One-loop self-energy diagram of the dark photon propagator. 
    Only $n_f$ dark light fermions run in the loop.}
    \label{fig:self-energy}
\end{figure}

The retarded self-energy diagram shown in figure~\ref{fig:self-energy}, which enters in the NLO expression of the 21 dark photon propagator in~\eqref{upper_Wightman_fkt} with $n_f$ massless dark particles in the loop, reads~\cite{Carignano:2017ovz}
\begin{equation}
\begin{aligned}
    \Pi^{R}_{\mu\nu}(q) &= \Pi^{11}_{\mu\nu}(q) + \Pi^{12}_{\mu\nu}(q) \\
    &=-ig^2n_f\int \frac{d^4k}{(2\pi)^4}
    \left(\text{Tr}\left[\gamma_\mu S^{11, \text{LO}}(k-q) \gamma_\nu S^{11, \text{LO}}(k)\right]
    \right.\\
    &\hspace{7.5cm}
    \left. -\text{Tr}\left[\gamma_\mu S^{21, \text{LO}}(k-q) \gamma_\nu S^{12, \text{LO}}(k)\right]\right) \\
    &=-i\frac{g^2}{2}n_f\int \frac{d^4k}{(2\pi)^4}
    \left(\text{Tr}\left[\gamma_\mu S^{S, \text{LO}}(k-q) \gamma_\nu S^R(k)\right]
    +\text{Tr}\left[\gamma_\mu S^{A, \text{LO}}(k-q) \gamma_\nu S^{S, \text{LO}}(k)\right] \right. \\
    &\hspace{1cm} \left. +\text{Tr}\left[\gamma_\mu S^{A, \text{LO}}(k-q) \gamma_\nu S^{A, \text{LO}}(k)\right]
    +\text{Tr}\left[\gamma_\mu S^{R, \text{LO}}(k-q) \gamma_\nu S^{R, \text{LO}}(k)\right]\right) ,
\end{aligned}
\end{equation}
where the terms in the last line vanish after integration in $k_0$.
Shifting the momentum $k\to k+q$ in the first term and $k\to -k$ in the second term, and using the explicit form of the massless fermion propagators, we get
\begin{equation}
\begin{aligned}
        \Pi^R_{\mu\nu}(q)&=n_fg^2\int \frac{d^4k}{(2\pi)^3}   [1-2n_\textrm{F}(|k_0|)]  \text{Tr}[\gamma_\mu \slashed{k}\gamma_\nu (\slashed{k}+\slashed{q})] \frac{\delta(k^2)}{(k+q)^2+i \sign(k_0+q_0)\epsilon} \\
        &= \Pi^{R,T=0}_{\mu\nu}(q) +  \Pi^{R,T\neq 0}_{\mu\nu}(q) \, .
\label{retarded_pol_tensor}
\end{aligned}
\end{equation}
The vacuum polarization tensor stems from the first term in the square bracket. 
In the $\overline{\text{MS}}$ scheme, it reads 
\begin{equation}
    \Pi_{\mu \nu,\overline{\text{MS}}}^{R,T=0}(q)= (q_{\mu}q_{\nu} - g_{\mu \nu}q^2)\frac{n_f g^2}{12\pi^2} \left[\ln\left( \frac{(q_0+i\epsilon)^2-\bm{q}^2}{-\mu^2}\right)-\frac{5}{3}\right] .
\label{eq:zerotemppol}
\end{equation}

The retarded thermal polarization tensor reads
\begin{equation}
    \begin{aligned}
        \Pi^{R,T\neq 0}_{\mu\nu}(q) &= -2n_fg^2\int \frac{d^4k}{(2\pi)^3} n_\textrm{F}(|k_0|)  \text{Tr}[\gamma_\mu \slashed{k}\gamma_\nu (\slashed{k}+\slashed{q})] \frac{\delta(k^2)}{(k+q)^2+i \sign(k_0+q_0)\epsilon} \\
        &= -n_fg^2\int \frac{d^3k}{(2\pi)^3} \frac{n_\text{F}(|\bm{k}|)}{|\bm{k}|} \\
        &\hspace{0.5cm} \times \left(\frac{\text{Tr}[\gamma_\mu \slashed{k}\gamma_\nu (\slashed{q}+\slashed{k})]|_{k_0=|\bm{k}|}}{(q_0+|\bm{k}|)^2-|\bm{q}+\bm{k}|^2+i\epsilon_+}+
        \frac{\text{Tr}[\gamma_\mu \slashed{k}\gamma_\nu (\slashed{q}+\slashed{k})]|_{k_0=-|\bm{k}|}}{(q_0-|\bm{k}|)^2-|\bm{q}+\bm{k}|^2+i\epsilon_-} \right) ,
    \end{aligned}
\label{ret_pol_tensor1}
\end{equation}
where $\epsilon_{\pm}\equiv \sign(\pm |k|+q_0)\epsilon$. 
Alternatively, using the relation\footnote{
The equation follows from $S^{R}(x-y)=\theta(x_0-y_0)\left[ S^{>}(x-y)- S^{<}(x-y)\right]$.
}
\begin{equation}
    \frac{1}{(k+q)^2+i \sign(k_0+q_0)\epsilon}=\int \frac{dk_0'}{2\pi}\frac{\theta(k_0')-\theta(-k_0')}{k_0 + q_0 -k_0'+i\epsilon}2\pi\delta(k_0'^2-|\bm{k}+\bm{q}|^2) \, ,
\end{equation}
the thermal part of the retarded polarization tensor can be written as
\begin{equation}
        \Pi^{R,T\neq 0}_{\mu\nu}(q)=n_fg^2\int \frac{d^3k}{(2\pi)^3}   \frac{n_F(|\bm{k}|)}{2|\bm{k}+\bm{q}||\bm{k}|} \left(\sum_{\sigma_{1,2}=\pm 1}  \frac{\sigma_2 \text{Tr}[\gamma_\mu \slashed{k}\gamma_\nu (\slashed{q}+\slashed{k})]|_{k_0=\sigma_1|\bm{k}|}}{q_0+\sigma_1|\bm{k}|+\sigma_2|\bm{k}+\bm{q}|+i\epsilon}\right) ,
\label{therm_pol_tensor_alternative}
\end{equation}
which contains only poles of the first order. 
Next, we evaluate the trace appearing in the numerator, 
\begin{equation}
    T_{\mu\nu} \equiv \textrm{Tr}[\gamma_\mu \slashed{k}\gamma_\nu (\slashed{q}+\slashed{k})] 
    =4[k_{\mu}(q+k)_{\nu}+k_{\nu}(q+k)_{\mu}-g_{\mu\nu}k(q+k)] \, ,
\end{equation}  
which, in terms of its components, reads
\begin{equation}
    \begin{aligned}
         T_{00}&=4(k_{0}q_0+k^2_{0}+|\bm{k}||\bm{q}|\cos\varphi + |\bm{k}|^2) \, , \\
         T_{ij}&=4[2k_ik_j+k_iq_j+k_jq_i+\delta_{ij}(k^2+k_0q_0-|\bm{k}||\bm{q}|\cos\varphi)] \, ,\\
        T_{\text{trans}}&\equiv \frac{1}{2}\left(\delta^{ij}-\frac{q^iq^j}{\bm{q}^2}\right)T_{ij} 
        =4(k_0^2+k_0q_0-|\bm{k}||\bm{q}| \cos\varphi-\bm{k}^2 \cos^2\varphi) ,
\label{tr}
    \end{aligned}
\end{equation}
the angle $\varphi$ being the angle between the vectors $\bm{k}$ and $\bm{q}$: $\cos\varphi= \hat{\bm{k}}\cdot\hat{\bm{q}}$.

The longitudinal component of the retarded thermal polarization tensor in~\eqref{ret_pol_tensor1} reads
\begin{equation}
\begin{aligned}
        \Pi^{R,T\neq 0}_{00}(q)&=-4n_fg^2\int \frac{d^3k}{(2\pi)^3}   \frac{n_\text{F}(|\bm{k}|)}{|\bm{k}|}  \left[\frac{|\bm{k}|q_0+2\bm{k}^2+|\bm{k}||\bm{q}|\cos\varphi }{q_0^2+2q_0|\bm{k}|-\bm{q}^2-2|\bm{k}||\bm{q}|\cos\varphi+i\epsilon_+}\right. \\
        &\hspace{5cm}+\left.
        \frac{-|\bm{k}|q_0+2\bm{k}^2+|\bm{k}||\bm{q}|\cos\varphi }{q_0^2-2q_0|\bm{k}|-\bm{q}^2-2|\bm{k}||\bm{q}|\cos\varphi+i\epsilon_-} \right] \\
        &=n_f\frac{g^2}{\pi^2}\int^{\infty}_{0} d|\bm{k}| |\bm{k}|n_\textrm{F}(|\bm{k}|) \left[\frac{ 4|\bm{k}|^2+q^2+4q_0|\bm{k}|}{4|\bm{k}||\bm{q}|} \ln\left( \frac{q^2+2q_0|\bm{k}|-2|\bm{q}||\bm{k}|+i\epsilon_+}{q^2+2q_0|\bm{k}|+2|\bm{q}||\bm{k}|+i\epsilon_+}\right)\right. \\
        &\hspace{2cm}+\left. \frac{ 4|\bm{k}|^2+q^2-4q_0|\bm{k}|}{4|\bm{k}||\bm{q}|} \ln\left( \frac{q^2-2q_0|\bm{k}|-2|\bm{q}||\bm{k}|+i\epsilon_-}{q^2-2q_0|\bm{k}|+2|\bm{q}||\bm{k}|+i\epsilon_-}\right)+2\right] ,
\label{longitudinal_pol_tensor}
\end{aligned}
\end{equation}
and the transverse component reads
\begin{equation}
\begin{aligned}
        \Pi^{R,T\neq 0}_{\text{trans}}(q)&=-4n_fg^2\int \frac{d^3k}{(2\pi)^3}   \frac{n_\text{F}(|\bm{k}|)}{|\bm{k}|}  \left[\frac{\bm{k}^2+|\bm{k}|q_0-|\bm{k}||\bm{q}|\cos\varphi-\bm{k}^2\cos^2\varphi }{q_0^2+2q_0|\bm{k}|-\bm{q}^2-2|\bm{k}||\bm{q}|\cos\varphi+i\epsilon_+}\right. \\
        &\hspace{5cm}+\left.
        \frac{\bm{k}^2-|\bm{k}|q_0-|\bm{k}||\bm{q}|\cos\varphi-\bm{k}^2\cos^2\varphi }{q_0^2-2q_0|\bm{k}|-\bm{q}^2-2|\bm{k}||\bm{q}|\cos\varphi+i\epsilon_-} \right] \\
        &=n_f\frac{g^2}{\pi^2}\int^{\infty}_{0} d|\bm{k}| |\bm{k}|n_\textrm{F}(|\bm{k}|) \left[-\frac{q_0^2+\bm{q}^2}{\bm{q}^2} \right. \\
        &~~+  \frac{ 4\bm{q}^2(\bm{k}^2+|\bm{k}|q_0)+\bm{q}^4-(q_0^2+2q_0|\bm{k}|)^2}{8|\bm{k}||\bm{q}|^3} \ln\left( \frac{q^2+2q_0|\bm{k}|-2|\bm{q}||\bm{k}|+i\epsilon_+}{q^2+2q_0|\bm{k}|+2|\bm{q}||\bm{k}|+i\epsilon_+}\right) \\ 
        &~~+ \left. \frac{ 4\bm{q}^2(\bm{k}^2-|\bm{k}|q_0)+\bm{q}^4-(q_0^2-2q_0|\bm{k}|)^2}{8|\bm{k}||\bm{q}|^3} \ln\left( \frac{q^2-2q_0|\bm{k}|-2|\bm{q}||\bm{k}|+i\epsilon_-}{q^2-2q_0|\bm{k}|+2|\bm{q}||\bm{k}|+i\epsilon_-}\right) \right]  .
\label{transverse_pol_tensor}
\end{aligned}
\end{equation}
Finally, we split~\eqref{longitudinal_pol_tensor} and~\eqref{transverse_pol_tensor} into real and imaginary parts:
\begin{equation}
\begin{aligned}
        \text{Re}\left[\Pi^{R,T\neq 0}_{00}(q)\right]&=n_f\frac{g^2}{\pi^2}\int^{\infty}_{0} d|\bm{k}| |\bm{k}|n_\textrm{F}(|\bm{k}|) \left[\frac{ 4|\bm{k}|^2+q^2+4q_0|\bm{k}|}{4|\bm{k}||\bm{q}|} \ln\bigg| \frac{q^2+2q_0|\bm{k}|-2|\bm{q}||\bm{k}|}{q^2+2q_0|\bm{k}|+2|\bm{q}||\bm{k}|}\bigg|\right. \\
        &\hspace{2cm}+\left. \frac{ 4|\bm{k}|^2+q^2-4q_0|\bm{k}|}{4|\bm{k}||\bm{q}|} \ln\bigg| \frac{q^2-2q_0|\bm{k}|-2|\bm{q}||\bm{k}|}{q^2-2q_0|\bm{k}|+2|\bm{q}||\bm{k}|}\bigg|+2\right]  ,
\end{aligned}
\label{Re_long_pol_tensor}
\end{equation}
\begin{equation}
\begin{aligned}
        \text{Im}\left[\Pi^{R,T\neq 0}_{00}(q)\right]&=n_f\frac{g^2}{\pi}\int^{\infty}_{0} d|\bm{k}| |\bm{k}|n_\textrm{F}(|\bm{k}|) \\ 
        &\hspace{-2cm}\times \left[\frac{ 4|\bm{k}|^2+q^2+4q_0|\bm{k}|}{4|\bm{k}||\bm{q}|} \sign(\epsilon_+) \left[ \theta(2|\bm{k}|(q_0+|\bm{q}|)+\bm{q}^2)-\theta(2|\bm{k}|(q_0-|\bm{q}|)+\bm{q}^2)\right] \right. \\
        &\hspace{-1.5cm}+\left.\frac{ 4|\bm{k}|^2+q^2-4q_0|\bm{k}|}{4|\bm{k}||\bm{q}|} \sign(\epsilon_-)\left[\theta(2|\bm{k}|(q_0+|\bm{q}|)-\bm{q}^2)-\theta(2|\bm{k}|(q_0-|\bm{q}|)-\bm{q}^2)\right]\right] ,
\end{aligned}
\label{Im_long_pol_tensor}
\end{equation}
\\
\begin{equation}
\begin{aligned}
        \text{Re}\left[\Pi^{R,T\neq 0}_{\text{trans}}(q)\right]&=n_f\frac{g^2}{\pi^2}\int^{\infty}_{0} d|\bm{k}| |\bm{k}|n_\textrm{F}(|\bm{k}|) \left[-\frac{q_0^2+\bm{q}^2}{\bm{q}^2} \right. \\
        &~~+  \frac{ 4\bm{q}^2(\bm{k}^2+|\bm{k}|q_0)+\bm{q}^4-(q_0^2+2q_0|\bm{k}|)^2}{8|\bm{k}||\bm{q}|^3} \ln\bigg| \frac{q^2+2q_0|\bm{k}|-2|\bm{q}||\bm{k}|}{q^2+2q_0|\bm{k}|+2|\bm{q}||\bm{k}|}\bigg| \\ 
        &~~+ \left. \frac{ 4\bm{q}^2(\bm{k}^2-|\bm{k}|q_0)+\bm{q}^4-(q_0^2-2q_0|\bm{k}|)^2}{8|\bm{k}||\bm{q}|^3} \ln\bigg|\frac{q^2-2q_0|\bm{k}|-2|\bm{q}||\bm{k}|}{q^2-2q_0|\bm{k}|+2|\bm{q}||\bm{k}|}\bigg| \right] ,
\end{aligned}
\label{Re_trans_pol_tensor}
\end{equation}
\begin{equation}
\begin{aligned}
        \text{Im}\left[\Pi^{R,T\neq 0}_{\text{trans}}(q)\right]&=n_f\frac{g^2}{\pi}\int^{\infty}_{0} d|\bm{k}| |\bm{k}|n_\textrm{F}(|\bm{k}|) \\ 
        &\hspace{0.5cm}\times \left[\frac{ 4\bm{q}^2(\bm{k}^2+|\bm{k}|q_0)+\bm{q}^4-(q_0^2+2q_0|\bm{k}|)^2}{8|\bm{k}||\bm{q}|^3} \sign(\epsilon_+)  \right.\\
        &\hspace{2cm}\times\left[ \theta(2|\bm{k}|(q_0+|\bm{q}|)+\bm{q}^2)-\theta(2|\bm{k}|(q_0-|\bm{q}|)+\bm{q}^2)\right] \\
        &\hspace{1cm}+\frac{ 4\bm{q}^2(\bm{k}^2-|\bm{k}|q_0)+\bm{q}^4-(q_0^2-2q_0|\bm{k}|)^2}{8|\bm{k}||\bm{q}|^3} \sign(\epsilon_-)\\
        &\hspace{2cm}\times  \left[\theta(2|\bm{k}|(q_0+|\bm{q}|)-\bm{q}^2)-\theta(2|\bm{k}|(q_0-|\bm{q}|)-\bm{q}^2)\right]\bigg]  .
\end{aligned}
\label{Im_trans_pol_tensor}
\end{equation}
The advanced thermal polarization tensor can be obtained from $\textrm{Re}[\Pi_{\mu\nu}^R]=\text{Re}[\Pi_{\mu\nu}^A]$ and $\textrm{Im}[\Pi_{\mu\nu}^R]=-\text{Im}[\Pi_{\mu\nu}^A]$.

\subsection{Hierarchy \texorpdfstring{$T \sim |\bm{q}| \gg q_0$}{T ~ q >> q0}}
In the thermal polarization tensor, the loop momentum $\bm{k}$ is of the order of the temperature~$T$. 
Hence, for the particular hierarchy $T \sim |\bm{q}| \gg q_0$, which is relevant for interactions mediated by  
space-like dark photons, 
we may expand eqs.~\eqref{Re_long_pol_tensor}--\eqref{Im_trans_pol_tensor} for small $q_0$, resulting in
\begin{equation}
        \text{Re}\left[\Pi^{R,T\neq 0}_{00}(q)\right]=n_f\frac{g^2}{\pi^2}\int^{\infty}_{0} d|\bm{k}| |\bm{k}|n_\textrm{F}(|\bm{k}|) \left[2-\frac{4\bm{k}^2-\bm{q}^2}{2|\bm{k}||\bm{q}|} \ln\left| \frac{|\bm{q}|-2|\bm{k}|}{|\bm{q}|+2|\bm{k}|}\right|\right]  ,
\label{Re_long_pol_tensor_hierarchy1}
\end{equation}
\begin{equation}
        \hspace{-5.3cm}\text{Im}\left[\Pi^{R,T\neq 0}_{00}(q)\right]=2n_f\frac{g^2}{\pi}\frac{q_0}{|\bm{q}|}\int^{\infty}_{|\bm{q}|/2} d|\bm{k}| |\bm{k}|n_\textrm{F}(|\bm{k}|) \,,
\label{Im_long_pol_tensor_hierarchy1}
\end{equation}
\begin{equation}
        \text{Re}\left[\Pi^{R,T\neq 0}_{\text{trans}}(q)\right]=n_f\frac{g^2}{\pi^2}\int^{\infty}_{0} d|\bm{k}| |\bm{k}|n_\textrm{F}(|\bm{k}|) \left[\frac{4\bm{k}^2+\bm{q}^2}{4|\bm{k}||\bm{q}|} \ln\left| \frac{|\bm{q}|-2|\bm{k}|}{|\bm{q}|+2|\bm{k}|}\right|-1\right] ,
\label{Re_trans_pol_tensor_hierarchy1}
\end{equation}
\begin{equation}
        \hspace{-5.4cm}\text{Im}\left[\Pi^{R,T\neq 0}_{\text{trans}}(q)\right]=n_f\frac{g^2}{\pi}\frac{q_0}{|\bm{q}|}\int^{\infty}_{|\bm{q}|/2} d|\bm{k}| |\bm{k}|n_\textrm{F}(|\bm{k}|) \, .
\label{Im_trans_pol_tensor_hierarchy1}
\end{equation}
Equations~\eqref{Re_long_pol_tensor_hierarchy1} and~\eqref{Im_long_pol_tensor_hierarchy1} are in agreement with the abelian analog of the expressions in~\cite{Brambilla:2008cx, Brambilla:2013dpa}. 
Eventually, the longitudinal and transverse symmetric polarizations can be written as
\begin{equation}
\begin{aligned}
\Pi^{S,T\neq 0}_{00}(q)&=[1+2n_\text{B}(q_0)]\left[\Pi^{R,T\neq 0}_{00}(q)-\Pi^{A,T\neq 0}_{00}(q)\right]=2i[1+2n_\text{B}(q_0)]\text{Im}\left[\Pi^{R,T\neq 0}_{00}(q)\right] \\
&=8in_f\frac{g^2}{\pi}\frac{T}{|\bm{q}|}\int^{\infty}_{|\bm{q}|/2} d|\bm{k}| |\bm{k}|n_\textrm{F}(|\bm{k}|) = 2\Pi^{S,T\neq 0}_{\text{trans}}(q) \, ,
\end{aligned}
\label{symm_pol_tensor_hierarchy1}
\end{equation}
where we have expanded the Bose--Einstein distribution up to leading order, making explicit the Bose enhancement $n_\text{B}(q_0) = T/q_0 + \dots$~.

\subsection{Hierarchy \texorpdfstring{$T \gg q_0, |\bm{q}|$}{T >> q0, q}}
\label{sec:b2}
The hierarchy $T \gg  q_0, |\bm{q}|$ is of relevance for interactions mediated by non-thermal photons.
In this case, we expand eqs.~\eqref{Re_long_pol_tensor}--\eqref{Im_trans_pol_tensor} up to leading order in $|\bm{q}|$ and $q_0$, and perform the integral over $|\bm{k}|$. 
The analytic expressions read
\begin{equation}
        \hspace{-3cm}\text{Re}\left[\Pi^{R,T\neq 0}_{00}(q)\right]= m_\text{D}^2 \left[1+\frac{q_0}{2|\bm{q}|} \ln\left| \frac{q_0-|\bm{q}|}{q_0+|q|}\right| \right] ,
\label{Re_long_pol_tensor_hierarchy2}
\end{equation}
\begin{equation}
        \hspace{-5cm}\text{Im}\left[\Pi^{R,T\neq 0}_{00}(q)\right]=\pi m_\text{D}^2\frac{q_0}{2|\bm{q}|}\theta(-q^2) \, ,
\label{Im_long_pol_tensor_hierarchy2}
\end{equation}
\begin{equation}
        \text{Re}\left[\Pi^{R,T\neq 0}_{\text{trans}}(q)\right]=-m_\text{D}^2\frac{q_0^2}{2\bm{q}^2} \left[1-\frac{q_0}{2|\bm{q}|}\left(1-\frac{\bm{q}^2}{q_0^2}\right) \ln\left| \frac{q_0+|\bm{q}|}{q_0-|\bm{q}|}\right| \right] ,
\label{Re_trans_pol_tensor_hierarchy2}
\end{equation}
\begin{equation}
        \hspace{-2.6cm}\text{Im}\left[\Pi^{R,T\neq 0}_{\text{trans}}(q)\right]=-\pi m_\text{D}^2\frac{q_0^3}{4|\bm{q}|^3} \left(1-\frac{\bm{q}^2}{q_0^2}\right) \theta(-q^2) \, ,
\label{Im_trans_pol_tensor_hierarchy2}
\end{equation}
where we have used $\displaystyle \int^{\infty}_{0} d|\bm{k}|\, |\bm{k}| \, n_\text{F}(|\bm{k}|)=\pi^2T^2/12$ and 
defined the Debye mass as $m_\text{D}^2\equiv n_fg^2T^2/3$. 
The retarded thermal polarization tensor develops an imaginary part only for space-like dark photons.
The results are in agreement with the abelian analogs in refs.~\cite{Brambilla:2008cx, Brambilla:2013dpa, Brambilla:2010vq,Carrington:1997sq}.
Finally, the longitudinal and transverse symmetric polarizations are
\begin{equation}
\hspace{-2.2cm}\Pi^{S,T\neq 0}_{00}(q)=  2i\pi m_\text{D}^2 \frac{T}{ |\bm{q}|} \theta(-q^2) \, ,
\label{symm_long_pol_tensor_hierarchy2}
\end{equation}
\begin{equation}
        \Pi^{S,T\neq 0}_{\text{trans}}(q)= -i\pi m_\text{D}^2\frac{T q_0^2}{|\bm{q}|^3} \left(1-\frac{\bm{q}^2}{q_0^2}\right) \theta(-q^2) \, .
\label{symm_trans_pol_tensor_hierarchy2}
\end{equation}

\subsection{Hierarchy \texorpdfstring{$T \gg |\bm{q}| \gg q_0$}{T >> q >> q0}}
\label{sec:b3}
The hierarchy $T \gg |\bm{q}| \gg q_0$ is a special case of the one considered in~\ref{sec:b2}, hence we expand the expressions in~\eqref{Re_long_pol_tensor_hierarchy2}--\eqref{Im_trans_pol_tensor_hierarchy2} up to first order in $q_0 \ll |\bm{q}|$,
\begin{equation}
        \hspace{-0.9cm}\text{Re}\left[\Pi^{R,T\neq 0}_{00}(q)\right]= m_\text{D}^2  \, , \qquad\text{Re}\left[\Pi^{R,T\neq 0}_{\text{trans}}(q)\right]=m_\text{D}^2 \times \mathcal{O}\left(\frac{q_0^2}{\bm{q}^2}\right)  ,
\label{Re_pol_tensor_hierarchy3}
\end{equation}
\begin{equation}
        \text{Im}\left[\Pi^{R,T\neq 0}_{00}(q)\right]=\pi m_\text{D}^2\frac{q_0}{2|\bm{q}|} \, ,\qquad \text{Im}\left[\Pi^{R,T\neq 0}_{\text{trans}}(q)\right]=\pi m_\text{D}^2\frac{q_0}{4|\bm{q}|} .
\label{Im_pol_tensor_hierarchy3}
\end{equation}
If the momentum of the dark photon is of the order of the Debye mass, $T \gg |\bm{q}|\sim m_\text{D} \gg q_0$, then the thermal loop corrections to the dark photon propagator need to be resummed, which is called the \textit{hard thermal loop} (HTL) resummation. 
For the hierarchy considered in this particular section, only the longitudinal polarization tensor does not vanish at leading order. 
Resummation according to eq.~\eqref{HTL_resummation_general} leads to the following \textit{dressed} longitudinal retarded/advanced propagators 
\begin{equation}
    D_{00}^{R/A}(q) = \frac{i}{\bm{q}^2+m_\text{D}^2 \pm i \pi m_\text{D}^2\frac{q_0}{2|\bm{q}|}} = \frac{i}{\bm{q}^2+m_\text{D}^2} \pm \frac{\pi}{2}\frac{q_0}{|\bm{q}|}\frac{m_\text{D}^2}{(\bm{q}^2+m_\text{D}^2)^2}+\mathcal{O}\left(\frac{q_0^2}{\bm{q}^2}\right) ,
\end{equation}
where the Debye mass plays the role of an effective thermal mass. 
Using the relations in~\eqref{kms_relation} and~\eqref{LOprop_coulombgauge}, the resummed longitudinal dark photon propagator becomes
\begin{equation}
 D_{00}(|\bm{q}|) =  \frac{i}{\bm{q}^2+m_\text{D}^2}\left(\begin{matrix}
1 & 0\\
0 &  -1
\end{matrix}\right)+\pi \frac{T}{|\bm{q}|}\frac{m_\text{D}^2}{(\bm{q}^2+m_\text{D}^2)^2} \left(\begin{matrix}
1 & 1\\
1 &  1
\end{matrix}
\right) ,
\label{resummed_prop}
\end{equation}
and the resummed longitudinal symmetric propagator reads
\begin{equation}
    D^S_{00}(q) = \frac{-i\Pi_{00}^S(q)}{(\bm{q}^2+\Pi_{00}^{R}(q))(\bm{q}^2+\Pi_{00}^{A}(q))}  = 2\pi \frac{T}{|\bm{q}|}\frac{m_\text{D}^2}{(\bm{q}^2+m_\text{D}^2)^2} \, .
\label{resummed_symm_prop}
\end{equation}

In case of the less strict hierarchy considered in section~\ref{sec:b2}, if also the dark photon energy becomes of the order of the Debye mass, such that $q_0, |\bm{q}| \sim  m_\text{D}$, both the longitudinal and transverse polarization tensors in~\eqref{Re_long_pol_tensor_hierarchy2}--\eqref{Im_trans_pol_tensor_hierarchy2} need to be resummed and one obtains
\begin{equation}
\begin{aligned}
    D^{R/A}_{00}(q) &=  \frac{i}{\bm{q}^2 + m_\text{D}^2 \left[1+\frac{q_0}{2|\bm{q}|} \ln\left| \frac{q_0-|\bm{q}|}{q_0+|q|}\right| \right] \pm i\pi m_\text{D}^2\frac{q_0}{2|\bm{q}|}\theta(-q^2)} \\
    &= \frac{i}{\bm{q}^2+m_\text{D}^2\left[1+\frac{q_0}{2|\bm{q}|}\ln\left(\frac{q_0-|\bm{q}|\pm i\epsilon}{q_0+|\bm{q}|\pm i\epsilon}\right)\right] } \, ,
\label{resummed_longitudinal_prop}
\end{aligned}
\end{equation}
\begin{equation}
\begin{aligned}
    &D^{R/A}_{ij}(q) \\
    &=\frac{i\left(\delta_{ij}-\frac{q_i q_j}{\bm{q}^2}\right)}{(q_0 \pm i\epsilon)^2-\bm{q}^2 -m_\text{D}^2\frac{q_0^2}{2\bm{q}^2} \left[1-\frac{q_0}{2|\bm{q}|}\left(1-\frac{\bm{q}^2}{q_0^2}\right) \ln\left| \frac{q_0+|\bm{q}|}{q_0-|\bm{q}|}\right| \right] \mp i\pi m_\text{D}^2\frac{q_0^3}{4|\bm{q}|^3} \left(1-\frac{\bm{q}^2}{q_0^2}\right) \theta(-q^2)} \\
    &=\left(\delta_{ij}-\frac{q_i q_j}{\bm{q}^2}\right)\frac{i}{q^2 \pm i\sign(q_0)\epsilon -\frac{m_\text{D}^2}{2} \left[\frac{q_0^2}{\bm{q}^2}-\frac{q_0}{2|\bm{q}|^3}q^2 \ln\left( \frac{q_0+|\bm{q}| \pm i\epsilon}{q_0-|\bm{q}| \pm i\epsilon}\right) \right]}\, ,
\label{resummed_transversal_prop}
\end{aligned}
\end{equation}
from which one can compute the dressed symmetric propagator and the full matrix-valued dark photon two-point function.

\section{Bound-state formation at NLO}
\label{sec:app_C}
\begin{figure}[ht]
    \centering
    \includegraphics[scale=0.85]{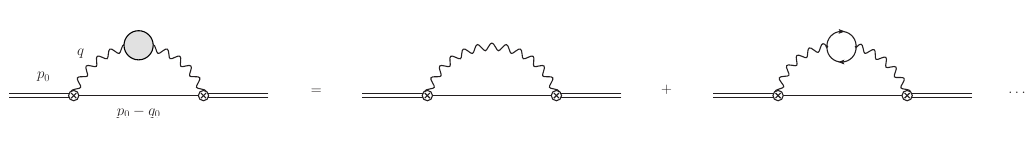}
    \caption{Scattering-state self-energy diagrams contributing to the bound-state formation cross section.
    Symbols are as in figure~\ref{fig:bsf_different}.
    The curly line with the shaded loop represents the resummed dark electric field correlator.
    On the right-hand side of the equality, we display the self-energies at LO and NLO in the weak-coupling expansion.}
    \label{fig:sed}
\end{figure}

The bound-state formation cross section at leading order in $r$ can be determined from the self-energy diagram of the scattering state in figure~\ref{fig:sed}.\footnote{
In the case of dissociation and bound-to-bound transitions, one has just to invert the double- and single-line heavy pair propagators in figure~\ref{fig:sed}, or consider only single-line propagators, respectively. 
The electric correlator and the vertices remain unchanged.}
Using the Kobes--Semenoff's cutting rules at finite $T$~\cite{Kobes:1985kc}, for the particular self-energy diagram under consideration, 
we obtain the following relation between the 11-component and the off-diagonal components in the time path index space,
\begin{equation}
    \text{Im}[\Sigma^{11}]=-\frac{1}{2i}(\Sigma^> + \Sigma^<) \approx  \frac{i}{2}\Sigma^> \, ,
\label{cutrulesexp}
\end{equation}
where we have used that the 12-component $\Sigma^<$ is exponentially suppressed,  cf.~\eqref{DM_pair_prop}. 
Then from the optical theorem, it follows
\begin{equation}
(\sigma_{\text{bsf}} \, v_{\hbox{\scriptsize rel}})(\bm{p})
= -2 \langle \,\bm{p}|  {\rm{Im}}[\Sigma^{11}(p_0)] | \bm{p} \, \rangle = \langle\,\bm{p}|  [-i\Sigma^{21}(p_0)] | \bm{p} \, \rangle \, .
  \label{bsf_projected}
\end{equation}

In the center-of-mass frame of the incoming scattering state with energy $p_0$ and relative momentum $\bm{p}$, such that $p_0 = 2M+\bm{p}^2/M$, we can freely choose the center-of-mass coordinate to be at the origin, $\bm{R}=0$. 
In dimensional regularization, the physical 11-component of the self-energy reads
\begin{equation}
\begin{aligned}
        \Sigma^{11}(p_0) &=-ig^2 \mu^{4-D}\int_0^\infty dt ~r^i e^{it(p_0-H)}  r^j \langle E^i (t,0) E^j (0,0)\rangle^{11} \\  
        &= -ig^2 \frac{\mu^{4-D}}{D-1}\int_0^\infty dt ~r^i e^{it(p_0-H)}  r^i\langle \bm{E} (t,0) \bm{E} (0,0)\rangle^{11} \\
        &= -ig^2 \frac{\mu^{4-D}}{D-1}\int_0^\infty dt ~r^i e^{it(p_0-H)} r^i \int \frac{d^Dq}{(2\pi)^D} e^{-iq_0t} \langle \bm{E}  \bm{E} \rangle^{11}(q) \\
        &=-ig^2 \frac{\mu^{4-D}}{D-1} \int \frac{d^D q}{(2\pi)^D} r^i \frac{i}{p_0-q_0-H+i\epsilon}r^i\langle \bm{E}  \bm{E} \rangle^{11}(q) \, ,
\label{scattering_self_energy}
\end{aligned}
\end{equation}
where the time-ordered DM propagator enters as $i[p_0-q_0-H+i\epsilon]^{-1}$. 
The dark electric field correlator can be written in terms of the matrix-valued dark photon two-point function as
\begin{equation}
\begin{aligned}
    \langle {\rm T} E_i (x) E_i (0)\rangle
    &= -\partial_0^2 D_{ii} (x)-{\bm{\nabla}}^2 D_{00} (x) \, ,
\end{aligned}
\end{equation}
or equivalently in momentum space
\begin{equation}
\begin{aligned}
    \langle {E_i}  {E_i} \rangle(q) \equiv 
    \int \frac{d^Dq}{(2\pi)^D}\,e^{-iq\cdot x}\langle {\rm T} E_i (x) E_i (0)\rangle
    &= q_0^2 D_{ii} (q)+{\bm{q}}^2 D_{00} (q) \, ,
\end{aligned}
\label{EEform}
\end{equation}
where the dark photon two-point function in momentum-space has been defined in appendix~\ref{sec:app_A}. 
Projecting the imaginary part of equation~\eqref{scattering_self_energy} onto a scattering state $|\bm{p}\rangle$ and inserting a complete set of bound states with quantum numbers $n$ in between the DM propagator and one of the quantum-mechanical relative position operators $r^i$, 
one can determine the bound-state formation cross section. 
Equivalently, according to equation~\eqref{bsf_projected}, it may be extracted from the 21-component of the scattering-state self-energy, leading to\footnote{
The advantage of using the 21-component of the self-energy, $\Sigma^{21}$, 
rather than the imaginary part of the 11-component, is in the simplification of the integrals, 
as we don't need to worry about a possible contribution from the imaginary part of the DM pair propagator,
$$
    \Sigma^> (p_0)  \propto \int \frac{d^ Dq}{(2\pi)^D} G^{21}(p_0-q_0) \langle \bm{E}  \bm{E} \rangle^{>}(q) 
    \propto \int  \frac{d^ Dq}{(2\pi)^D} \delta(p_0 - q_0 - H)\langle \bm{E}  \bm{E} \rangle^{>}(q) \, ,
$$
whereas when computing the imaginary part involves dealing with the (anti-)symmetric electric correlator,
$$
\begin{aligned}
    \text{Im}[\Sigma^{11}] \propto & \int  \frac{d^ Dq}{(2\pi)^D} \text{Re}[G^{11}(p_0-q_0)] \text{Re}[\langle \bm{E}  \bm{E} \rangle^{>}(q)]
    +\int  \frac{d^ Dq}{(2\pi)^D}\text{Im}[G^{11}(p_0-q_0) ]\text{Im}[\langle \bm{E}  \bm{E} \rangle^{>}(q)] \\ 
    \propto & \int  \frac{d^ Dq}{(2\pi)^D} \delta(p_0 - H - q_0)\, \frac{1}{2}\langle \bm{E}  \bm{E} \rangle^{S}(q)
    +\int  \frac{d^ Dq}{(2\pi)^D} \textrm{P} \left[\frac{1}{p_0 - H - q_0}\right]\, \frac{1}{2i}  \langle \bm{E}  \bm{E} \rangle^{AS}(q) \, .
\end{aligned}
$$
In this work, we use both ways. 
For the determination of the bound-state formation cross section up to NLO, 
we compute the 21-component of the scattering-state self-energy. 
When exploiting the hierarchy of energy scales and performing a HTL-resummation to obtain the appropriate cross section, we compute the imaginary part of $\Sigma^{11}$.}
\begin{equation}
\begin{aligned}
(\sigma_{\text{bsf}} \, v_{\hbox{\scriptsize rel}})(\bm{p}) = g^2 \frac{\mu^{4-D}}{D-1}\sum_{n}|\langle n|\bm{r}|\bm{p}\rangle|^2 \int \frac{d^{D-1} q}{(2\pi)^{D-1}}  \langle \bm{E}  \bm{E} \rangle^{>}(\Delta E^p_n , \bm{q})  \, .
\end{aligned}
\label{bsf_final_form}
\end{equation}
The bound-state formation cross section depends on the quantum-mechanical electric dipole matrix element, whose general expression can be taken, e.g., from~\cite{Biondini:2023zcz}, and the $D-1$ dimensional integral of the 21-component of the dark electric correlator at energy $\Delta E^p_n = \bm{p}^2/M + M\alpha^2/(4n^2) \geq 0$. 
We deal with this latter quantity in the following paragraphs. 

At leading order in the coupling $\alpha$ and using eq. \eqref{XXpropGT}, we obtain\footnote{
The electric correlator is gauge invariant and, therefore, independent of the specific gauge chosen for the photon propagator.
}
\begin{equation}
\begin{aligned}
\langle \bm{E}  \bm{E} \rangle^{>}_{\text{LO}}(\Delta E^p_n , \bm{q})&= (\Delta E^p_n)^2 D^{>,\text{LO}}_{ii}(\Delta E^p_n , \bm{q})+{\bm{q}}^2 D^{>,\text{LO}}_{00} (\Delta E^p_n , \bm{q}) \\
&= 4\pi (\Delta E^p_n)^2 \delta[(\Delta E^p_n)^2 - \bm{q}^2][1+n_{\text{B}}(\Delta E^p_n)] \, ,
\end{aligned}
\end{equation}
which, plugged in into~\eqref{bsf_final_form}, gives \cite{vonHarling:2014kha,Biondini:2023zcz}
\begin{equation}
\begin{aligned}
(\sigma_{\text{bsf}} \, v_{\hbox{\scriptsize rel}})^{\text{LO}}(\bm{p}) =\sum_n (\sigma^n_{\text{bsf}} \, v_{\hbox{\scriptsize rel}})^{\text{LO}}(\bm{p})= \frac{4}{3}\alpha(\mu)\sum_n |\langle n|\bm{r}|\bm{p}\rangle|^2 (\Delta E^p_n)^3 [1+n_{\text{B}}(\Delta E^p_n)] \, .
\end{aligned}
\label{bsf_final_LO}
\end{equation}
At next-to-leading order, one takes into account the NLO expression of the 21-component of the dark photon propagator in terms of the retarded propagator at LO and the retarded polarization tensor, cf.~\eqref{upper_Wightman_fkt}. 
Hence, the 21-electric correlator at NLO becomes
\begin{equation}
\begin{aligned}
\langle \bm{E}  \bm{E} \rangle^{>}_{\text{NLO}}(q)&= q_0^2 D^{>,\text{NLO}}_{ii} (q)+{\bm{q}}^2 D^{>,\text{NLO}}_{00} (q) \\
&\hspace{-1cm} = 2[1+n_\text{B}(q_0)]\textrm{Im}\left[q_0^2D^{R,\textrm{LO}}_{i \lambda}(q) \Pi_R^{\lambda\rho}(q) D^{R,\textrm{LO}}_{\rho i}(q) + \bm{q}^2D^{R,\textrm{LO}}_{0 \lambda}(q) \Pi_R^{\lambda\rho}(q) D^{R,\textrm{LO}}_{\rho 0}(q)\right] \\
&\hspace{-1cm}\equiv \langle \bm{E}  \bm{E} \rangle^{>,n_\text{F}=0}_{\text{NLO}}(q)+\langle \bm{E}  \bm{E} \rangle^{>,n_\text{F}\neq0}_{\text{NLO}}(q)\, ,
\label{NLO_el_corr}
\end{aligned}
\end{equation}
where $q_0 = \Delta E^p_n$ and in the last line we have written the electric correlator as the sum of a term retaining the retarded polarization tensor in vacuum ($n_\text{F}(q_0)=0$) and a term retaining its thermal part ($n_\text{F}(q_0)\neq 0$) according to eq.~\eqref{retarded_pol_tensor}.\footnote{
Note that, while we distinguish contributions to the 21-electric correlator coming from either a vanishing or non-vanishing Fermi--Dirac distribution in $\Pi_R^{\lambda\rho}(q)$, the Bose--Einstein distribution $n_\text{B}(q_0)$ is included in $\langle \bm{E}  \bm{E} \rangle^{>,n_\text{F}=0}_{\text{NLO}}(q)$ as well as in $\langle \bm{E}  \bm{E} \rangle^{>,n_\text{F}\neq 0}_{\text{NLO}}(q)$.} 

Using the expressions in~\eqref{LO_retarded_advanced} and~\eqref{eq:zerotemppol}, the $n_\text{F}(q_0)=0$ part in dimensional regularization and in the $\overline{\text{MS}}$ scheme becomes
\begin{equation}
\begin{aligned}
\langle \bm{E}  &\bm{E} \rangle^{>,n_\text{F}=0}_{\text{NLO}}(q) \\
&= 2[1+n_\text{B}(q_0)]\textrm{Im}\left[q_0^2D^{R,\textrm{LO}}_{i \lambda}(q) \Pi_{R,\overline{\text{MS}}}^{\lambda\rho,T=0}(q) D^{R,\textrm{LO}}_{\rho i}(q) + \bm{q}^2D^{R,\textrm{LO}}_{0 \lambda}(q) \Pi_{R,\overline{\text{MS}}}^{\lambda\rho,T=0}(q) D^{R,\textrm{LO}}_{\rho 0}(q)\right] \\
&=[1+n_\text{B}(q_0)]\frac{n_fg^2}{6\pi^2}\textrm{Im}\left\{\frac{q^2(\bm{q}^2-3q_0^2)}{((q_0+i\epsilon)^2-\bm{q}^2)^2}\left[\ln{\left(\frac{(q_0+i\epsilon)^2-\bm{q}^2}{-\mu^2}\right)}-\frac{5}{3}\right]\right\} \, .
\label{NLO_el_corr_vac}
\end{aligned}
\end{equation}
Inserting this expression into~\eqref{bsf_final_form} and integrating over $\bm{q}$ by means of the residue theorem, we obtain the $n_\text{F}(q_0)=0$ part of the bound-state formation cross section at NLO,\footnote{
This contribution does not need to be resummed to all orders as long as $n_f \alpha/\pi \ll 1$.}
\begin{equation}
\begin{aligned}
(\sigma_{\text{bsf}} \, v_{\hbox{\scriptsize rel}})^{\text{NLO}}_{n_\text{F}=0}(\bm{p}) = \sum_n(\sigma^n_{\text{bsf}} \, v_{\hbox{\scriptsize rel}})^{\text{LO}}(\bm{p})\frac{n_f}{3\pi}\alpha
\left[\ln{\left(\frac{4(\Delta E^p_n)^2}{\mu^2}\right)}-\frac{10}{3}\right] \, .
\end{aligned}
\label{bsf_final_NLO}
\end{equation}
Adding up~\eqref{bsf_final_LO} and~\eqref{bsf_final_NLO} yields
\begin{equation}
\begin{aligned}
(\sigma_{\text{bsf}} \, v_{\hbox{\scriptsize rel}})^{\text{LO}+\text{NLO}}_{n_\text{F}=0}(\bm{p}) &\equiv (\sigma_{\text{bsf}} \, v_{\hbox{\scriptsize rel}})^{\text{LO}}(\bm{p}) + (\sigma_{\text{bsf}} \, v_{\hbox{\scriptsize rel}})^{\text{NLO}}_{n_\text{F}=0}(\bm{p}) \\
&= \sum_n(\sigma^n_{\text{bsf}} \, v_{\hbox{\scriptsize rel}})^{\text{LO}}(\bm{p})\bigg\{1+\frac{n_f}{3\pi}\alpha
\left[\ln{\left(\frac{4(\Delta E^p_n)^2}{\mu^2}\right)}-\frac{10}{3}\right]\bigg\} \, ,
\end{aligned}
\label{bsf_final_NLO_adding}
\end{equation}
where the scale dependence at NLO cancels against the $\mu$ dependence of the coupling in the LO expression  $(\sigma_{\text{bsf}} \, v_{\hbox{\scriptsize rel}})^{\text{LO}}(\bm{p})$, according to
\begin{equation}
\frac{d}{d\ln\mu} \left[ \alpha(\mu) + \frac{n_f}{3\pi}\alpha^2\ln{\left(\frac{4(\Delta E^p_n)^2}{\mu^2}\right)}\right]    = -\frac{\alpha^2}{2\pi}\beta_0 + \frac{n_f}{3\pi}\alpha^2\,(-2) + {\cal O}(\alpha^3)= {\cal O}(\alpha^3)\, ,
\label{eq:muinvariance}
\end{equation}
$\beta_0=-4n_f/3$ being the first coefficient of the beta function in QED.

Next, we compute the $n_\text{F}(q_0)\neq0$ part in~\eqref{NLO_el_corr}, which depends on the thermal retarded polarization tensor given in~\eqref{therm_pol_tensor_alternative}, and obtain
\begin{equation}
\begin{aligned}
\langle \bm{E}  &\bm{E} \rangle^{>,n_\text{F}\neq 0}_{\text{NLO}}(q) \\
&= 2[1+n_\text{B}(q_0)]\textrm{Im}\left[2q_0^2\Delta^{R}_{\textrm{LO}}(q) \Pi_\text{trans}^{R,T\neq0}(q) \Delta^{R}_{\textrm{LO}}(q) + \bm{q}^2D^{R,\textrm{LO}}_{00}(q) \Pi_{00}^{R,T\neq0}(q) D^{R,\textrm{LO}}_{00}(q)\right] \\
&=n_f g^2[1+n_\text{B}(q_0)]  \int \frac{d^3k}{(2\pi)^3}\frac{n_\text{F}(|\bm{k}|)}{|\bm{k}||\bm{k}+\bm{q}|}\textrm{Im}\left[\sum_{\sigma_{1,2}=\pm 1}  \frac{\sigma_2 T_{00}/\bm{q}^2|_{k_0=\sigma_1|\bm{k}|}}{q_0+\sigma_1|\bm{k}|+\sigma_2|\bm{k}+\bm{q}|+i\epsilon} \right.\\
&\hspace{4.cm}\left. +\frac{1}{((q_0+i\epsilon)^2-\bm{q}^2)^2}\left(\sum_{\sigma_{1,2}=\pm 1}  \frac{\sigma_2 2q_0^2T_{\text{trans}}|_{k_0=\sigma_1|\bm{k}|}}{q_0+\sigma_1|\bm{k}|+\sigma_2|\bm{k}+\bm{q}|+i\epsilon}\right)\right],
\label{NLO_el_corr_therm}
\end{aligned}
\end{equation}
where $\Delta^{R}_{\textrm{LO}}(q)$, $\Pi_{\text{trans}}^{R,T\neq0}(q)$, $T_{00}$ and $T_{\text{trans}}$ are given in appendix~\ref{sec:app_A} and \ref{sec:app_B}. 
We insert~\eqref{NLO_el_corr_therm} into~\eqref{bsf_final_form}, integrate first over the momentum $|\bm{q}|$ using the residue theorem,\footnote{
Equation~\eqref{NLO_el_corr_therm} has two single poles at $|\bm{q}|_{\pm}=-|\bm{k}|\cos\varphi \pm \sqrt{\bm{k}^2\cos^2\varphi+ q_0^2+2\sigma_1 q_0|\bm{k}|}$, which contribute to bound-state formation via bath-particle scattering as well as to the off-shell decay of the emitted dark photon into a light dark particle and antiparticle. 
Moreover, equation~\eqref{NLO_el_corr_therm} has a double pole at $|\bm{q}|=q_0$, which contributes to bound-state formation via on-shell emission.} 
then add up the individual terms and integrate over the angular coordinate $\varphi =\sphericalangle(\bm{k},\bm{q})$. 
We end up with a single integral expression in $|\bm{k}|$ that is finite and numerically solvable,\footnote{
Following the arguments in~\cite{Binder:2020efn, Binder:2021otw}, potential collinear divergences from the individual terms in~\eqref{bsf_final_NLO_therm} cancel each other in the sum.}
\begin{equation}
\begin{aligned}
(\sigma_{\text{bsf}} \, v_{\hbox{\scriptsize rel}})^{\text{NLO}}_{n_\text{F}\neq0}(\bm{p}) &= \sum_n(\sigma^n_{\text{bsf}} \, v_{\hbox{\scriptsize rel}})^{\text{LO}}(\bm{p})\frac{n_f\alpha}{\pi (\Delta E^p_n)^3} \int_0^{\infty} d|\bm{k}| ~2n_\text{F}(|\bm{k}|)\Bigg[-2|\bm{k}|\Delta E^p_n \\
&\hspace{0cm}+2|\bm{k}|\Delta E^p_n\ln{\left|\frac{\bm{k}^2-(\Delta E^p_n)^2}{(\Delta E^p_n)^2}\right|}+[2|\bm{k}|^2+(\Delta E^p_n)^2]\ln{\left|\frac{|\bm{k}|+\Delta E^p_n}{|\bm{k}|-\Delta E^p_n}\right|}\Bigg].
\end{aligned}
\label{bsf_final_NLO_therm}
\end{equation}
The complete bound-state formation cross section up to NLO in the coupling is given by the sum of~\eqref{bsf_final_NLO_adding} and~\eqref{bsf_final_NLO_therm}, 
\begin{equation}
\begin{aligned}
(\sigma_{\text{bsf}} \, v_{\hbox{\scriptsize rel}})^{\text{LO}+\text{NLO}}(\bm{p}) &= (\sigma_{\text{bsf}} \, v_{\hbox{\scriptsize rel}})^{\text{LO}+\text{NLO}}_{n_\text{F}=0}(\bm{p}) + (\sigma_{\text{bsf}} \, v_{\hbox{\scriptsize rel}})^{\text{NLO}}_{n_\text{F}\neq0}(\bm{p}) \\
&= \sum_n(\sigma^n_{\text{bsf}} \, v_{\hbox{\scriptsize rel}})^{\text{LO}}(\bm{p})\bigg\{1+\frac{n_f}{\pi}\alpha
\left[\mathcal{X}_1(\Delta E^p_n,\mu
)
+\mathcal{X}_2(\Delta E^p_n/T)\right]\bigg\} \, ,
\end{aligned}
\label{bsf_final_fixed_NLO}
\end{equation}
where $\mathcal{X}_1$ can be read off from the NLO expression in~\eqref{bsf_final_NLO_adding} and $\mathcal{X}_2$ can be written as an integral over the dimensionless variable $t\equiv |\bm{k}|/T$~\cite{Binder:2020efn, Binder:2021otw}:
\begin{equation}
   \mathcal{X}_2(x)=\frac{2}{x^3}\int_0^{\infty}  \frac{dt}{e^{t}+1}\left[(2t^2+x^2)\ln{\left|\frac{t+x}{t-x}\right|}+2tx\ln{\left|\frac{t^2-x^2}{x^2}\right|}-2xt\right] .
   \label{dimless_x2}
\end{equation}
While the vacuum loop corrections are quite small for small couplings, the thermal loop corrections can become large if the temperature exceeds the ultrasoft energy scale $\Delta E^p_n$,
in which case they need to be resummed, symbolized by the gray loop in figure~\ref{fig:sed}. 
We have scrutinized the effects of the resummation in section~\ref{sec:el_dipole_transitions}.

\end{appendices}

\newpage

\bibliographystyle{JHEP}
\bibliography{bibl}

\end{document}